\documentclass[a4paper,11pt]{article}
\pdfoutput=1
\usepackage{slashed}
\usepackage[affil-sl,auth-sc]{authblk}
\usepackage{amssymb,amsmath,amsbsy}
\usepackage{mathrsfs}
\usepackage{mathbbol}
\usepackage{bm}
\usepackage{appendix}
\usepackage{hyperref}
\usepackage[top=1.2in, bottom=1.2in,left=0.9in,includefoot]{geometry}               
\usepackage{graphicx}
\usepackage{cite}
\usepackage{float}
\usepackage{caption}
\usepackage{wasysym}
\usepackage{float}
\usepackage{caption}
\usepackage[utf8]{inputenc}

\usepackage{subcaption}
\usepackage{amsthm}

\newif\ifContLineOne
\newif\ifContLineTwo
\newif\ifContLineThree

\def\conC#1{\vbox{\ialign{##\crcr
  \ifContLineThree\hrulefill\else\vphantom{\hrulefill}\fi\crcr
  \noalign{\kern3.2pt\nointerlineskip}
  \ifContLineTwo\hrulefill\else\vphantom{\hrulefill}\fi\crcr
  \noalign{\kern3.2pt\nointerlineskip}
  \ifContLineOne\hrulefill\else\vphantom{\hrulefill}\fi\crcr
  \noalign{\nointerlineskip}
  $\hfil\textstyle{\vbox to 14pt{}#1}\hfil$\crcr}}}

\def\DrawLeg#1#2{
  \kern-.2pt              % back up half width of leg
  \dimen2 =#1             % =height of whatever is underneath leg
  \advance\dimen2 by 2pt  % 2pt space below bottom of leg
  \dimen3 = 10.6pt        % base value of height of top of leg
  \dimen4 =3.6pt          % add this much time 1 2 or 3 to base value
  \advance\dimen3 by -\dimen2 
  \multiply\dimen4 by #2
  \advance\dimen3 by \dimen4
  \raise\dimen2 \hbox{\vrule height\dimen3 width .4pt} % draw it
  \kern-.2pt}             % and back up half width of line

\def\begC#1#2{\setbox0 =\hbox{$\textstyle{#2}$}
  \dimen0=.5\wd0 \dimen1=\ht0
  \conC{\hskip\dimen0}
  \count255=#1
  \ifnum\count255 =1 \ContLineOnetrue\else
  \ifnum\count255 =2 \ContLineTwotrue\else
  \ifnum\count255 =3 \ContLineThreetrue\fi\fi\fi
  \DrawLeg{\dimen1}{\count255}
  \conC{\hskip\dimen0}
  \kern-\dimen0\kern-\dimen0 \box0}

\def\endC#1#2{\setbox0 =\hbox{$\textstyle{#2}$}
  \dimen0=.5\wd0 \dimen1=\ht0
  \conC{\hskip\dimen0}
  \count255=#1
  \ifnum\count255 =1 \ContLineOnefalse\else
  \ifnum\count255 =2 \ContLineTwofalse\else
  \ifnum\count255 =3 \ContLineThreefalse\fi\fi\fi
  \DrawLeg{\dimen1}{\count255}
  \conC{\hskip\dimen0}
  \kern-\dimen0\kern-\dimen0 \box0}

\theoremstyle{definition}

\theoremstyle{remark}

\def\theequation{\arabic{section}.\arabic{equation}}
\def\eq#1{eq.~(\ref{#1})}
\def\Eq#1{Eq.~(\ref{#1})}

\def\eqs#1#2{eqs.~(\ref{#1}) and (\ref{#2})}
\def\eqss#1#2#3{eqs.~(\ref{#1}), (\ref{#2}) and (\ref{#3})}
\def\Eqsss#1#2#3{Eqs.~(\ref{#1}), (\ref{#2}) and (\ref{#3})}

\def\Ref#1{ref.~\cite{#1}}

\def\Refs#1{refs.~\cite{#1}}
\newcommand{\ie}{{\it i.e., }}
\newcommand{\eg}{{\it e.g., }}

\newcommand{\bea}{\begin{eqnarray}}
\newcommand{\eea}{\end{eqnarray}}

\newcommand{\GeV}{\; \mathrm{GeV}}
\newcommand{\TeV}{\; \mathrm{TeV}}

\usepackage{color}

\definecolor{orange}{rgb}{0.9,0.2,0}
\definecolor{brown}{rgb}{0.7,0.3,0.2}
\definecolor{fuxia}{rgb}{1,0,1}
\definecolor{skyblue}{rgb}{0,0.1,0.9}
\definecolor{violetred}{rgb}{0.8,0.13,0.56}
\definecolor{deeppink}{rgb}{1.00,0.08,0.5}
\definecolor{pink}{rgb}{1.00,0.75,0.80}
\definecolor{orchid}{rgb}{0.85,0.44,0.84}
\definecolor{lightpink}{rgb}{1.00,0.71,0.76}
\definecolor{bluish}{rgb}{0,0.6,0.8}

\newcommand{\vev}[1]{\left\langle #1\right\rangle}

\title{\bf Effective Theory for \\ Electroweak  Doublet  Dark Matter}

\author{A.  Dedes$^{1}$\footnote{email: {\tt adedes@cc.uoi.gr}}, ~
 D. Karamitros$^{1}$\footnote{email: {\tt dkaramit@cc.uoi.gr}}  
   ~and V. C. Spanos$^{2}$\footnote{email: {\tt vspanos@phys.uoa.gr}}}
\affil{\small $^{1}$Department of Physics, Division of Theoretical
  Physics, \\ University of Ioannina, GR 45110, Greece}
\affil{\small $^{2}$Section  of Nuclear {\rm \&} Particle Physics, Department  of Physics,     \\
   National and Kapodistrian University of Athens, GR--15784 Athens, Greece}

\date{\today}

\begin{document}

\maketitle

\begin{abstract}

We perform a detailed study of an effective field theory
which includes the  Standard Model particle content extended by
a pair of Weyl fermionic
SU(2)-doublets with opposite hypercharges.
A discrete symmetry guarantees that a linear combination
of the doublet components is stable and can act as a candidate
particle for Dark Matter.
The dark sector fermions interact with the Higgs and gauge bosons
through renormalizable $d=4$ operators, and non-renormalizable
$d=5$ operators that appear after integrating out extra degrees
of freedom  above the TeV scale.
We study collider, cosmological and  astrophysical probes for
this effective theory of Dark Matter.
We find that a WIMP with a mass nearby to the electroweak scale,
and thus observable  at  LHC,
is consistent with collider and astrophysical data only when
fairly large magnetic dipole moment transition operators with
the gauge bosons exist, together with moderate
Yukawa interactions.

\end{abstract}

%%%%%%%%%%%%
%\tableofcontents
%\listoffigures
%\listoftables
%%%%%%%%%%%%

%%%%%%%%%%%%%%%%%%%%%%%%%%%%%%%%%
\section{Introduction and Motivation}
\setcounter{equation}{0}
\label{sec:intro}
%%%%%%%%%%%%%%%%%%%%%%%%%%%%%%%%%

There is  convincing evidence for the existence of Dark Matter (DM) 
from observation of gravitational effects at astrophysical and 
cosmological scales but not yet confirmed at Earth's colliders, 
 where interactions between the hypothetical Weakly Interacting 
DM particle (WIMP) is probed through its interactions with the 
Standard Model particles (for recent reviews see ~\cite{Bertone:2004pz,Gelmini:2015zpa,Lisanti:2016jxe,DeSimone:2016fbz}).
Out of all energy density in the universe, approximately  $25\%$ seems to
consist of DM,  probably in the form of WIMPs, with its relic
density today with respect to the critical density, to be
precisely known by the  Planck  collaboration~\cite{Ade:2013zuv,Agashe:2014kda}:
%%%%%%%%%%%%%%%%%%%%
\begin{equation}
\Omega \, h^2=0.1198 \pm 0.0026\;.
\label{Planck}
\end{equation}
%%%%%%%%%%%%%%%%%%%%%%
 Out of many WIMP candidates one of the most studied is the 
lightest higgsino particle~\cite{Roszkowski:1991ng,Drees:1996pk},
a fermion which is a linear combination of the neutral components 
of the $SU(2)_L$-bi-doublet superpartners of the  Minimal Supersymmetric Standard Model (MSSM) scalar 
Higgs doublets. A higgsino WIMP fulfilling  the constraint of \eq{Planck}, which concurrently  escapes  the 
direct DM search bounds, 
must be  heavier than the TeV scale, and therefore 
difficult to be reached at the Large Hadron Collider (LHC). 

 In this article, we shall consider a ``higgsino like''
DM sector of the Standard Model (SM) gauge structure,  with mass 
\emph{as close to the electroweak scale as possible,}  supplied also by  related 
effective operators of dimension less than or equal to five.  Since
$SU(2)_L$ fermionic doublets are not singlets under the SM gauge group, 
there are  important interactions already at the renormalizable
level, providing annihilation processes of WIMP to SM particles or interactions
between  the WIMP and the nucleons. 
Other, what we call ``Earth" detectable  effects, include contributions to the Electroweak (EW)
 parameters,  to the Higgs boson decay into diphotons, and to other LHC processes,
like mono-jets, mono-$Z$, etc.~\cite{Crivellin:2015wva}.

 Apart from  MSSM and its variants, 
there are many simple models for DM that contain bi-doublets,\footnote{By the name  ``bi-doublets''   we mean two Weyl fermion 
 $SU(2)_L$-doublets with opposite hypercharge.}
in their low energy spectrum.
For instance, there are  models with $SU(2)_L$
doublets+singlet(s)~\cite{Carena:2004ha,Mahbubani:2005pt,D'Eramo:2007ga,
Cohen:2011ec,Joglekar:2012vc,Abe:2014gua,Banerjee:2016hsk,Calibbi:2015nha}
or doublets+triplet~\cite{Dedes:2014hga,Freitas:2015hsa}.
% or triplet+quadraplets~\cite{Tait:2016qbg}.  
For EW scale DM at work in most of these models, the need of low energy cut-off, of the order of 1 TeV,
is sometimes unavoidable.\footnote{It has been shown in \Ref{Dedes:2014hga} that for EW scale DM particle mass one needs relatively large Yukawa couplings
between the extra vector-like fermions and the Higgs boson.  
These lead in turn to vacuum instabilities of the Higgs potential~\cite{ArkaniHamed:2012kq}, that arise
already at the TeV scale, depending on the largeness of the Yukawa couplings and the particle content of the model.}
In addition, recent attempts to investigate low energy DM-models 
arising from Grand Unified Theories (e.g., from an SO(10) GUT), seem 
to incorporate bi-doublets, often in association with other particles, in their low energy particle 
content~\cite{Arbelaez:2015ila,Nagata:2015dma,Boucenna:2015sdg}. This low energy content, may also
be part of a non-GUT extension of the Standard Model,
 as for instance  a subgoup of SO(10), such as the left-right symmetric model~\cite{Garcia-Cely:2015quu}.   
There are also Effective Field Theory (EFT) approaches with the SM+$\chi$, or simply $SM_\chi$,
where $\chi$ is the SM-singlet, up to dimension six effective operators~\cite{D'Eramo:2014aba,Matsumoto:2016hbs}.
One should remark  however,  that a light singlet fermionic dark matter is not favoured by SO(10)--GUT constructions
consistent with a unification  and  intermediate symmetry breaking scale at the TeV scale~\cite{Nagata:2015dma,Mambrini:2013iaa}.
 
 Motivated by
all the above we would like to study the phenomenology
 of a SM with  $SU(2)_L$-bi-doublets with electroweak mass. In terms of physical masses,
this  model contains
a charged Dirac fermion and two Majorana (or  Pseudo-Dirac) neutral
fermions with their masses splitted with mass differences in the vicinity of tens of GeV
due to the presence of $d=5$ non-renormalizable operators.  We study the
implications of all the related to dark matter $d=5$ operators for the relic abundance,
for direct as well as indirect searches. 
A general study of Majorana fermionic dark matter based on SM-extensions of the bi-doublets
has been discussed in \Ref{Chua:2015ixi}. Our EFT can be viewed as a decoupling limit of all extra fermion states
but not those arising from  the $SU(2)_L$ bi-doublet system.

The EFT at hand, generalizes the phenomenology of Standard Models with additional $SU(2)_L$ multiplets, sometimes called 
Minimal Dark Matter models~\cite{Cirelli:2005uq,DelNobile:2009st,Kim:2006af}. The most basic  of these models is just a Dirac mass term,
 {\it c.f.} \eq{eq:md},  for the bi-doublet fermion multiplet.
However,  without the imposition of a symmetry  the WIMP will not  be stable (although higher spin SU(2)-reps will be ``accidentally'' stable).  
We  discuss in the next section available symmetries that 
not only protect the WIMP  for decaying, like a $Z_2$ or lepton number, but also forbid potentially  dangerous couplings to 
the $Z$ boson like  charge conjugation  or custodial symmetry. 

A similar to our  EFT,
 has been studied in~\Ref{Nagata:2014wma} for  higgsino DM scenario in high scale supersymmetry breaking, 
using a mass splitting of $\mathcal{O}(\lesssim 1 \GeV)$ originated through $d=5$ Yukawa interactions  and radiative corrections.
 For higgsino mass parameter $\lesssim \mathcal{O}(1) \TeV$,
 the parameter space is constrained from direct detection  and Electric Dipole Moment searches.
The EFT employed here is complementary to~\Ref{Nagata:2014wma}. 
 We assume that the cut-off scale is of order $\Lambda = \mathcal{O}(1 \TeV)$ and 
for this reason, we introduce a complete set of $d=5$ operators, \ie 
Yukawa and dipole transition operators. We later use all these operators to calculate different observables and constrain the parameter
space accordingly. Furthermore,  the Yukawa couplings are not restricted
by supersymmetry. This, in turn, allows us to focus on larger mass differences and therefore different phenomenology.

As we show in this article, a  viable WIMP with mass nearby the electroweak scale acquires fairly large non-zero magnetic dipole moments.
Magnetic dipole interacting DM has already been studied  in  \Refs{Chang:2010en,Weiner:2012cb,Weiner:2012gm}, a scenario called 
 Magnetic Inelastic Dark Matter (MiDM). 
In MiDM,   the WIMP ($\chi$) is supplemented by a  ``excited WIMP state'',  ($\chi^{\star}$), with $m_{\chi^{\star}}-m_{\chi} = \mathcal{O}(100)~\mathrm{KeV}$. 
A consequence of this, is a large nucleus-WIMP cross-section, comparable  to experimental limits for inelastic nucleus-WIMP scattering. Moreover, in \Ref{Weiner:2012gm}, a connection between direct detection and Gamma-ray line signals  pointed out, for such small mass splitting.
Our work is more general than this scenario, simply because  the fermions we introduce are doublets under the $SU(2)_{L}$. 
Apart from this, we focus on relatively large mass difference, of order $\mathcal{O}(1-10) \GeV$, between the two neutral fermion states. 
These facts lead to qualitatively different phenomenology. In particular, the direct detection scattering, in our case, is elastic. Also, due to a symmetry 
 the lightest fermion does not interact directly with  $Z$-boson
 and the dominant annihilation channels in the early universe are different. 
Although the EFT studied here is more general from the one suggested previously in the literature, 
the dipole moments that are responsible for  the observed DM relic abundance,  provide also  
enough monochromatic photon flux from the center of our galaxy, to bound considerably (but not to exclude)
the parameter space of the model.  It is therefore understood that our model  
could provide an explanation for a possible  signal in the near future. 

The outline of the article is the following: 
in section~\ref{sec:eft} we describe    the   effective theory and associated possible accidental symmetries  and in  \ref{sec:ope} we list
 the effective $d=5$ and $d=6$ operators, that may be present in this  extension of the SM. 
In section~\ref{sec:pheno} we describe the interactions and the mass spectrum.  
Consequently,  in section~\ref{sec:earth} various collider and direct DM detection constraints are  analysed.
In addition, in section~\ref{sec:astro}   the DM relic density    is calculated, and we
study the corresponding cosmological constraints.  Moreover, we discuss the phenomenology 
of  possible indirect signals for DM searches, from   gamma-rays,  and briefly, from neutrino fluxes. 
 In section~\ref{sec:LHC}  we study possible signals of this model  at LHC  at  $8$ and $13 \TeV$.
Finally, in section~\ref{sec:concl} we summarise our findings.

%%%%%%%%%%%%%%%%%%%%%%%%%%%%%%%%%
\section{Symmetries and the effective theory}
\setcounter{equation}{0}
\label{sec:eft}
%%%%%%%%%%%%%%%%%%%%%%%%%%%%%%%%%

 In the SM particle content we add a fermionic bi-doublet, that is a pair of Weyl fermion  $SU(2)$-doublets 
 with opposite hypercharges, ${\mathbf{{D}_{1}}}$, 
 that transform under $(SU(3),SU(2)_L)_Y$ like $ \mathbf{(1^{c}, 2)_{-1}} $ and 
${\mathbf{{D}_{2} }}$, that transform as $ \mathbf{(1^{c}, 2)_{+1}} $. The
 doublet $\mathbf{D_{2}}$ has  exactly the same gauge quantum numbers as the SM Higgs field
$\mathbf{H} $,  while   $\mathbf{D}_{1}$ carries  the   quantum
numbers of the SM lepton doublet but not necessarily sharing lepton number.    
 Then the  model under study includes  gauge invariant kinetic terms like\footnote{Throughout this paper, 
 we  adopt the convenient  two-component Weyl spinor notation of~\Ref{Dreiner:2008tw}.}
$D^{\dagger\, xa} \bar{\sigma}^{\mu} \mathscr{D}_{\mu} D_{xa}$, with $( x=1,2$) the number of doublets and $(a=1,2)$
their $SU(2)_L$-quantum numbers. These fields have 
 renormalizable couplings with the SM electroweak gauge bosons  through  $ \mathscr{D}_{\mu}$,  the
covariant derivative for the SM gauge group $SU(2)_L  \times U(1)_Y$.

%%%%%%%%%%%%%%%%%%%%
\subsection{Custodial symmetry} 
%%%%%%%%%%%%%%%%%%%%% 
 
In addition to gauge invariant kinetic term, an invariant Dirac-type mass term for the  bi-doublets is
 %%%%%%%%%%%%%%%%%%%
\begin{eqnarray}
\mathscr{L}_{\mathrm{DM}} \ \supset \  - M_{D}\: \epsilon^{ab}  {D}_{1\,a} {D}_{2\, b} 
\ + \ {\rm H.c.} \ =\  -M_D \: \det \mathcal{D} \ + \ {\rm H.c.}\;,
\label{eq:md}
\end{eqnarray}
where $\epsilon^{ab}$ is the antisymmetric tensor, with 
$\epsilon^{12}=-\epsilon^{21}=1$ and, for later notational use, 
we define $D_{1a} \equiv ( D_{1}^0, D_{1}^-)^T \;, \; D_{2a} \equiv (D_{2}^+,D_{2}^0)^T$.
%%%%%%%%%%%%%%%%%
In order to make things clearer below, 
in the second equality of \eq{eq:md} we used the definition of the determinant  to write the matrix
\begin{eqnarray}
\mathcal{D}_{x a} \ = \  (D_{1a} \quad D_{2a} ) \ =\  \left ( \begin{array}{cc} D_1^0  & D_2^+ \\ D_1^- & D_2^0  \end{array} \right ) \;.
\label{eq:det}
\end{eqnarray} 
%%%%%%%%%%%%%%%%%
Written in this form it is now transparent that $\mathcal{D}$ is invariant not only under the $SU(2)_L$  but also 
under another $SU(2)$, say   $SU(2)_R$. The transformation rule under $SU(2)_L\times SU(2)_R$ with corresponding 
unitary matrices $U_L$ and $U_R$ is 
\begin{equation}
\mathcal{D} \rightarrow U_L \, \mathcal{D} \, U_R \;,
\end{equation} 
%%%%%%%%%%%%%%%%
where $U_L$ acts on the rows and $U_R$ acts on the columns of $\mathcal{D}$, respectively.
On the other hand, it is  well known~\cite{Sikivie:1980hm} that, the SM Higgs sector is also invariant under 
a global $SU(2)_R$ symmetry. In this case we can write the Higgs field in $\mathbf{(2,2)}$ form of $SU(2)_L\times SU(2)_R$
as
%%%%%%%%%%%%%%%%%
\begin{equation}
\mathcal{H}_{a x} \ =\ (H_a^* \quad H_a) \ = \ \left ( \begin{array}{cc} 
-\Phi^{0*} & \Phi^+ \\ \Phi^- &  \Phi^0 
\end{array} \right )\;.
\label{eq:H}
\end{equation}
%%%%%%%%%%%%%%%%%%
Similarly, the Higgs field is invariant under $SU(2)_L\times SU(2)_R$ 
with  a transformation law $\mathcal{H} \rightarrow U_L \, \mathcal{H} \, U_R$. Obviously, we can now write down a  
 $SU(2)_L\times SU(2)_R$ non-renormalizable $d=5$ Yukawa operator  as 
 %%%%%%%%
 \begin{equation}
 \mathscr{L} \supset \frac{y}{\Lambda} \, [\mathrm{Tr}(\mathcal{H}^\dagger \mathcal{D})]^2 \ + \  \mathrm{H.c.}
 \label{eq:cus}
 \end{equation}
 %%%%%%%%%%
where $\Lambda$ is the scale of masses that are being integrated out. EW symmetry breaking 
breaks $SU(2)_L\times SU(2)_R$ down to its diagonal subgroup,  $SU(2)_{L+R}$. The latter symmetry 
is the well known \emph{custodial} symmetry~\cite{Sikivie:1980hm}. Most pronouncedly it is broken by the difference in magnitude between the top 
and bottom Yukawa couplings and by the $U(1)_Y$ gauge symmetry but, importantly, keeps radiative EW corrections
under control. One of our study benchmarks below arises from \eq{eq:cus}.

%%%%%%%%%%%%%%%%%%%%
\subsection{Charge conjugation symmetry}
%%%%%%%%%%%%%%%%%%%%

The new $\mathbf{D_1}$- and $\mathbf{D_2}$-fermion fields  form
 a pseudo-real representation of SU(2). 
In order  to make the presentation transparent, we  redefine the Weyl fields as 
%%%%%%%%%%%%%%%%%%
\begin{equation}
\xi^b = \epsilon^{ab} \: D_{1a} \;, \qquad \eta_b = D_{2b}\;,
\end{equation}
%%%%%%%%%%%%%%
where we can easily arrive at a Dirac fermion field Lagrangian written in terms of the 
two, two-component Weyl spinor fields, $\xi$ and $\eta$, as
%%%%%%%%%%%%%%
\begin{equation}
\mathscr{L}_{\mathrm{DM}} \ =  \ i \xi^{\dagger}_{a} \bar{\sigma}^\mu \mathscr{D}_{\mu} \xi^a + i \eta^{\dagger a} \bar{\sigma}^\mu \mathscr{D}_{\mu} \eta_a - M_D \: (\eta_a \xi^a + \eta^{a \dagger} \xi_{a}^\dagger) \;.
\label{eq:MD2}
\end{equation}
%%%%%%%%%%%%%%%
The   bi-doublets-mass term, $M_D$, can be taken real and positive.  
In \eq{eq:MD2}, we have suppressed  all spinor indices, but have left the gauge group indices
intact to show our covariant notation (to be used below). Now, it is well known that
  the Lagrangian \eqref{eq:MD2}, beyond $SU(2)_L$ symmetry, accommodates 
a $O(2)$-symmetry which, apart from 
making  the usual phase invariance transformation $SO(2)\sim U(1)$ group
 $\xi\to e^{-i\theta} \xi$ and  $\eta \to e^{i\theta} \eta$, it contains a discrete symmetry under which
%%%%%%%%%%%%
\begin{equation}
 C^{-1}\, \eta_a \, C \ = \ \xi^a \;.
\label{cc1}
\end{equation}
%%%%%%%%%%%%
This discrete symmetry  is a  \emph{charge conjugation symmetry (c.c.)},
 associated to  the charge conjugation operator   $C$ with $C^2=(C^{-1})^2=I$.
This symmetry simply exchanges the two Weyl fields $\xi \leftrightarrow \eta$ 
or to a ``free'' notation, $D_1 \leftrightarrow D_2$. 
There is a similar symmetry in the Higgs sector, where another explicit bi-doublet mass term exists, that of the Higgs field.
Then the  corresponding charge conjugation symmetry for the Higgs field, which leaves invariant the kinetic terms 
as well as the Higgs potential in the Standard Model, reads accordingly as, 
%%%%%%%%%%%%
\begin{equation}
C^{-1} \, H_a \, C \ = \  H^{\dagger \, a}\;,
\label{cc2}
\end{equation}     
%%%%%%%%%%%%%%
where $H_a$ is the SM Higgs doublet, $H_a \equiv (\Phi^+, \Phi^0)^T$. 
What basically c.c. symmetry does, is to exchange the columns of matrices $\mathcal{D}$ and $\mathcal{H}$ 
in \eqs{eq:det}{eq:H},  respectively.
For the Higgs field, charge conjugation becomes somewhat trivial for the following reason.
In order to read physical masses we have to expand the Lagrangian in terms of 
fields that vanish at the minimum. There are many $SU(2)_L\times U(1)_Y$ equivalent Higgs
representations, but the most known is the so-called Kibble parametrization~\cite{Kibble:1967sv},
%%%%%%%%%%%%%%%%%%%5
\begin{equation}
\mathbf{H}  \ =\ \mathbf{U} \; \mathbf{H_0} \ = \ \mathbf{U} \; \left ( \begin{array}{c} 0 \\  v + \frac{h}{\sqrt{2}}   \end{array} \right ) \;,
\label{ugauge}
\end{equation}
%%%%%%%%%%%%%%%%%%%5
where $\mathbf{U}$ is any $2\times 2$ unitary matrix describing a unitary gauge transformation, 
$v$ is the vacuum expectation value (vev) [{\it c.f.} \eq{eq:vev}], and $h$ is the \emph{real}-valued Higgs field. 
The matrix $\mathbf{U}$ is absorbed in gauge boson, lepton, quark field redefinitions, \emph{and, in particular}
 model at hand, in the dark sector fields $\xi$ and $\eta$ (or $D_1$ and $D_2$). Therefore, 
c.c.  symmetry, \eqref{cc2},  has no effect on $\mathbf{H_0}$. On the other hand,
the discrete c.c. symmetry in \eqref{cc1}, acts in a non-trivial way in the dark sector of the model after EW symmetry breaking. 
We will \emph{assume} that this is 
a symmetry of the Lagrangian and examine implications from this hypothesis. 

%%%%%%%%%%%%%%%%%%%%%%%%
\subsection{The discrete $Z_2$-symmetry}
%%%%%%%%%%%%%%%%%

Unfortunately, the c.c. or the custodial symmetries alone can not account  for the stability of DM lightest particle
and an extra discrete $Z_2$-symmetry 
 that distinguishes SM-particles from DM-particles is needed. 
To ``throw away"  dangerous  $d=4$ operators that are responsible for WIMP decay, 
like  $D_1 H^\dagger \bar{e}$  or higher [see \ref{sec:ope} for assignments and in particular \eq{leptonDM}],   
 it could be enough  to impose a lepton number  symmetry for example.
It is safer however, to impose an external $Z_2$-discrete symmetry under which the SM fermions are odd while the 
dark matter fermions and the Higgs boson are even eigenstates.
Such a discrete symmetry, or equivalently, its variant known from  MSSM as  R-parity, is preserved in SO(10) with the Higgs field 
in a {\bf{126}} representation~\cite{Kibble:1982ae}  and  are common in Grand Unified Theories (GUTs)  with low mass  dark 
matter particles~\cite{Frigerio:2009wf,Arbelaez:2015ila,Nagata:2015dma,Boucenna:2015sdg}. We shall therefore
assume such a $Z_2$-symmetry in what follows.

%%%%%%%%%%%%%%%%%%%%%%%%%%%%%%%
\subsection{Symmetric limits used in the analysis}
%%%%%%%%%%%%%%%%%%%%%%%%%%%%%

Our model, is based on an  effective theory described by  the  following Lagrangian:
%%%%%%%%%%%%%%%
\begin{eqnarray}
\mathscr{L} = \mathscr{L}_{\rm SM} + \mathscr{L}_{\rm DM} + 
\mathscr{L}^{d=5}_{\rm SM+DM} \;.
\label{eq:larg}
\end{eqnarray}
%%%%%%%%%%%%%%%%
$\mathscr{L}_{\rm SM}$ is the SM renormalizable Lagrangian, $\mathscr{L}_{\rm DM}$ is the DM  sector renormalizable Lagrangian 
given by \eq{eq:MD2} and $\mathscr{L}^{d=5}_{\rm SM+DM}$ is the Lagrangian that contains the dimension-5 operators relevant to
DM interactions.   
We assume that higher dimensional  operators ($d\ge 6$) are suppressed
and  throughout this article we are focusing on up-to $d=5$ effective operators.
For  the sake of completeness, however,  in  \ref{sec:ope} we construct all relevant operators for both  dimensionalities
 $d=5$ and  $d=6$. 
 
  We show below that by using the  the  c.c.  symmetry of \eq{cc1}, or the custodial symmetry  or  just the $U(1)$ phase symmetry 
  we can arrive   at four distinct choices in the parameter space. Moreover, 
  this is very convenient for the phenomenological study that follows. First, $\mathscr{L}^{d=5}_{\rm SM+DM}$ contains 
effective operators that after spontaneous EW symmetry breaking split the masses of the neutral
particles from their original common mass $M_D$. The most general, linearly independent  set of operators, is 
\begin{eqnarray}
-\mathscr{L}^{d=5}_{\rm SM+DM} \ \supset  & & \frac{y_1}{2\Lambda} \, (H_a \xi^a) \, (H_b\xi^b) \ + \
\frac{y_2}{2\Lambda} \, (H^{\dagger \, a} \eta_a) \, (H^{\dagger \, b} \eta_b) \ - \ \frac{y_{12}}{\Lambda} \, (H_a \xi^a) \, (H^{\dagger \, b} \eta_b) 
\nonumber \\[2mm]
&+& \frac{\xi_{12}}{\Lambda} \, (\xi^a \eta_a) \, (H^{\dagger\, b} H_b) \ + \ \mathrm{H.c.}
\label{d5mass}
\end{eqnarray}
where $\Lambda$ is the cutoff of the effective, SM+bi-doublet, theory.\footnote{In \eq{eq:op1} we give examples of what sort of 
heavy particle mass the $\Lambda$ might be.} 
If the c.c. symmetry \eqref{cc1} is imposed the last two terms of \eq{d5mass} are unaffected, but the 
first two terms must be the same. This means that under c.c. symmetry the relation
%%%%%%%%%%%%%%%%%
\begin{equation}
y_1\  =  \  y_2 \ \equiv \ y \;,
\label{eq:y}
\end{equation}
%%%%%%%%%%%%%%%%
holds. We always follow this symmetry condition in the analytical expressions as well in the numerical results throughout this article.
Even more, one can write the independent  c.c. symmetry invariant $d=5$  operators  
%%%%%%%%
\begin{equation}
 y\, (H_a \xi^a  - H^{\dagger a} \eta_a)^2  \quad \mathrm{or} \qquad   y\,(H_a \xi^a  + H^{\dagger a} \eta_a)^2  \;,
\end{equation}
%%%%%%%%%%%%%
in addition to the  operators multiplying $y_{12}$ and $\xi_{12}$  in \eq{d5mass}. 
Based on symmetries discussed above, there are  additional restrictions on Yukawa couplings 
%%%%%%%%%
\begin{equation}
(1) ~y = y_{12}\;, \qquad  (2) ~ y =  - y_{12} \; , \qquad (3) ~ y_{12}=0\;, \qquad   (4) ~y = y_{12} = 0 \;,  \quad  \forall y \;, \quad  \forall \xi_{12} \,.  \label{eq:cases}
\end{equation}
%%%%%%%%%%%
%for every value of $\xi_{12}$,  since it is always self invariant under \eqref{cc1}. 
Cases (1) and (2) above,  may correspond to the $SU(2)_L\times SU(2)_R$
symmetry limit of \eq{eq:cus}.
Case (3)  is not really supported by any symmetry consideration, in fact it violates the custodial symmetry,
 and is only adopted here
for covering the mass spectrum phenomenology  ({\it c.f.} Fig.~\ref{spectra}).
In choosing the benchmark for case (4) we are motivated by the following:
 in a full  gauge invariant theory, 
$y_{12}$ and $\xi_{12}$ may have certain relations with $y$. For example,
 in the fermionic doublet-triplet DM model of  \Ref{Dedes:2014hga}  one finds $\xi_{12}=-2 y = 2 y_{12}$ after decoupling
the heavy triplet in the custodial limit.
If the continuous $U(1)$-phase symmetry is employed (or if the two $SU(2)_R$ symmetries for $\mathcal{D}$ and $\mathcal{H}$  are
different) then $y=0$ for all $y_{12}$ and $\xi_{12}$. In this case there are two, mass degenerate, 
 Dirac fermions in the spectrum: one neutral and one charged.
This completes our study benchmark points which are mostly based 
upon the underlying global symmetries of the model rather on a random choice of the model parameters.

There are also  $d=5$  magnetic and electric dipole operators related to
the dark sector particles. A detailed form of these operators is given in \ref{sec:ope}. In this article we
shall focus on the magnetic dipole operators
\begin{equation}
-\mathscr{L}^{d=5}_{\rm SM+DM} \ \supset
 \frac{d_{\gamma}}{\Lambda}\: \xi^a \: \sigma^{\mu\nu}  \: \eta_{a}\: 
 B_{\mu\nu}  \ + \
\frac{d_{W}}{\Lambda}\: \xi^b \: \sigma^{\mu\nu} \: (\tau^{A})_{b}^{\ c} \: \eta_c\: 
 W_{\mu\nu}^{A} \ + \ \mathrm{H.c.}\;,
\end{equation}
where $B_{\mu\nu}$ and $W_{\mu\nu}^A$ are  the $U(1)_Y$ and $SU(2)_L$ field strength tensors respectively and
$\tau^A$  the Pauli matrices with $A=1,2,3$ and $\sigma^{\mu\nu} \equiv \frac{i}{4} (\sigma^\mu\bar{\sigma}^\nu - \sigma^\nu\bar{\sigma}^\mu)$ .
These operators are invariant under \eqref{cc1} since $C^{-1} \, \xi \sigma^{\mu\nu} \eta \, C = \eta \sigma^{\mu\nu} \xi = - \xi \sigma^{\mu\nu} \eta$, 
 $C^{-1} W^{A}_{\mu\nu} C =(-1)^{2-A} W^{A}_{\mu\nu}$ (no sum in $A$) and 
$C^{-1} B_{\mu\nu} C =- B_{\mu\nu}$.  We shall see below that both moments  $d_\gamma$ and $d_W$,
play an important role in achieving the correct relic density.

As promised earlier in this section, the new, beyond the SM parameters needed to describe the dark sector are   the following six: 
\begin{equation}
M_D \;, \quad \Lambda\;, \quad y\;, \quad \xi_{12}\;, \quad d_\gamma\;, \quad d_W \;.
\label{params}
\end{equation}
Throughout, we assume them all to be real.
More importantly, we \emph{assume} that the mass $M_{D}$  is around or 
below the EW-scale, that is  of the order of $\mathcal{O}(100) $ GeV.
The mass scale $\Lambda$   for  extra  scalars and fermions,    are far 
above the EW scale, possibly at the TeV-scale.  As a result, we assume that  this 
EFT contains three   (but two distinct) mass scales, 
%%%%%%%%%%%%%%%%%
\begin{equation}
M_{D}\simeq v \simeq 174 ~\mathrm{GeV} 
\;, \qquad \Lambda \simeq \mathcal{O}(1)~\mathrm{TeV}\;.
\label{eq:vev}
\end{equation}
%%%%%%%%%%%%%%%%

\section{Phenomenology}
\setcounter{equation}{0}
\label{sec:pheno}

%%%%%%%%%%%%%%%%%%%%%%%%%
\subsection{Mass Spectrum}\label{Spectrum-section}
%%%%%%%%%%%%%%%%%%%%%%%%%

%Apart from the tree-mass $M_{D}$ of \eqref{eq:md} there are also 
%mass terms arising at the vacuum from operators in \eqref{eq:op2}.
After electroweak symmetry breaking and the shift of the neutral component of the
Higgs field $H_{0} = (0, v + h/\sqrt{2})^T$, in \eqs{eq:MD2}{d5mass}, we obtain 
%%%%%%%%%%%%%%%%%%%%%%% 
\begin{align}
\mathscr{L}^{\rm DM}_{\rm  (mass)} 
 = -  m_{\chi^{\pm}}\: \chi^{-} \: \chi^{+} \ - \ 
\frac{1}{2} \sum_{i=1}^{2} \: m_{\chi^{0}_{i}} \: 
\chi_{i}^{0}  \chi_{i}^{0} \ + \ {\rm H.c.} \;,
\label{Lmass}
\end{align}    
%%%%%%%%%%%%%%%%%%%
where,   under the c.c. symmetry restrictions  \eqref{eq:y}, the physical fields are two neutral Majorana fermions ($\chi_1^0, \chi_2^0$) and one 
pair of Dirac charged  fermions ($\chi^\pm$)
%%%%%%%%%%%%%%%%%
\begin{subequations}
\begin{align}
\chi_{1}^{0} = \frac{1}{\sqrt{2}} \: ( D_{1}^{0} + D_{2}^{0} ) \;, \quad &
\chi_{2}^{0} = -\frac{i}{\sqrt{2}} \: ( D_{1}^{0} - D_{2}^{0} ) \;,  \label{eq:chi0} \\[1mm]
\chi^{+} = i \: D_{2}^{+} \;, \qquad  &\chi^{-} = i \: D_{1}^{-} \;,
\label{eq:chipm}
\end{align}
\end{subequations}
%%%%%%%%%%%%%%%%%%%%
with   masses,
%%%%%%%%%%%%%%%%%
\begin{subequations}
\begin{align}
m_{\chi^{\pm}} \ &= \ M_{D}\ + \ \xi_{12} \: \omega \;, 
 \\[1mm]
m_{\chi_{1}^{0}}  \ &= \ m_{\chi^{\pm}} \ + \ \omega\:  (y - y_{12}) \;,  \qquad \omega \equiv \frac{v^{2}}{\Lambda} \;,
 \\[1mm]
m_{\chi_{2}^{0}} \ &= \ m_{\chi^{\pm}}\ - \ \omega \: (y + y_{12}) \;. 
\end{align}
\label{eq:spec}
\end{subequations}
%%%%%%%%%%%%%%%%%%
Without loss of generality, our natural choice for field redefinitions is such that $M_{D} >0$.
Under the c.c. symmetry  the state $\chi_1^0$ is even,  while the states $\chi_2^0,\chi^\pm$ are odd, \ie
\begin{align}
C^{-1} \, \chi_1^0 \, C &= + \chi_1^0 \;, \qquad C^{-1} \, \chi_2^0 \, C = -\chi_2^0 \;, \label{cchi0} \\[2mm]
C^{-1} \, \chi^+ \, C &=- \chi^- \;, \qquad C^{-1} \, \chi^- \, C = -\chi^+ \;. \label{cchipm}
\end{align}
However, in general and far from custodial symmetry limits,
 only $\chi^+$ and $\chi^-$ are particle-antiparticle states with common mass, $m_{\chi^\pm}$. 

%Please note that we  append the complete spectrum. 
In what follows, we sort  the masses so that the lightest particle is $\chi_{1}^{0}$. 
Also, we assume $M_D + \xi_{12} \, \omega >0$, 
for otherwise the contribution from  $d=5$ operators to the masses, \ie the term $\xi_{12} \, \omega$
would be unnaturally large,  in order to satisfy the  LEP bound~\cite{LEP_SUSY:2001,LEP:2001qw,Abdallah:2003xe} 
  $m_{\chi^{\pm}}\gtrsim 100 \; \mathrm{GeV}$.
%%%%%%%%%%%%%%%%%%%%%%%%%%%%%%
\begin{figure}[t]
   \centering
   \includegraphics[height=2.0in]{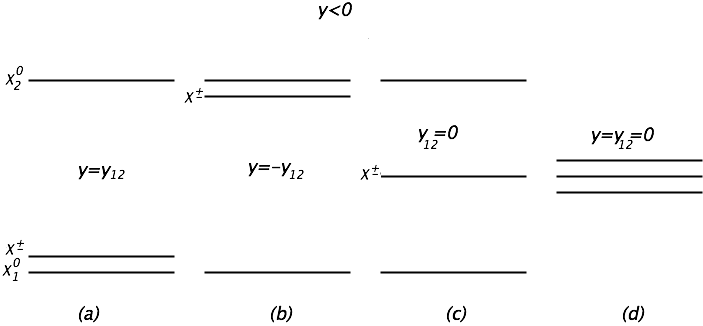} % requires the graphicx package
   \caption{\em Mass hierarchies of the dark  fermions   $\chi_1^0$, $\chi^\pm$ and $\chi_2^0$ (bottom to top)
    for   $y<0$, following  the c.c. symmetry  of \eq{eq:cases} for  the  cases (a) $y=y_{12}$,  (b) $y=-y_{12}$, (c) $y_{12}=0$ and (d) $y=y_{12}=0$.
    The mass spectrum for $y>0$ is obtained  from this figure by exchanging  $\chi_1^0 \leftrightarrow \chi_2^0$.}
   \label{spectra}
\end{figure}
%%%%%%%%%%%%%%%%%%%%%%%%%%%%%%
There are two equivalent set of mass spectra: one with
 $y \le 0$ where $m_{\chi_1^0} \le m_{\chi_2^0}$ and the other $y \ge 0$ where  $m_{\chi_2^0} \le m_{\chi_1^0}$. 
 In Fig.~\ref{spectra},
we show the spectrum for the $y\le 0$ case. 
The mass spectrum for $y>0$ is exactly the same after exchanging $\chi_1^0 \leftrightarrow \chi_2^0$.
We note that the mass hierarchies  between $\chi^\pm$, $\chi_{1}^0$ and $\chi_2^0$ 
displayed  in Fig.~\ref{spectra} do not depend on $M_D$ and $\xi_{12}$,   although their central mass values  
 are all shifted uniformly upon their variation. 
 Therefore, following  \eq{eq:cases}, we  distinguish four mass spectra:

{\bf (a) } $y=y_{12}<0$ : the lightest neutral DM fermion $\chi_1^0$ is almost degenerate 
with the charged one $\chi^\pm$ (see Fig.~\ref{spectra}a)  with 
\begin{equation}
m_{\chi_1^0} = m_{\chi^\pm}\;, \quad m_{\chi_2^0} = m_{\chi^\pm} + 2 \omega |y| \;.
\end{equation}

{\bf (b)} $y=-y_{12}<0$ : the heavy neutral fermion $\chi_2^0$ is degenerate with the charged fermion
 $\chi^\pm$ (Fig.~\ref{spectra}b) with
\begin{equation}
m_{\chi_1^0} = m_{\chi^\pm} - 2 \omega |y| \;,  \quad m_{\chi_2^0} = m_{\chi^\pm}\  \;.
\end{equation}

{\bf (c) } $y_{12}=0\;, \forall y<0$ : all $\chi_1^0$ and  $\chi_2^0$ are split from  $\chi^\pm$ 
by an equal amount $\omega |y|$ with (Fig.~\ref{spectra}c)
\begin{equation}
m_{\chi_1^0} = m_{\chi^\pm} -  \omega |y| \;,  \quad m_{\chi_2^0} = m_{\chi^\pm}  +  \omega |y|  \;.
\end{equation}

{\bf (d) } $y_{12}=y=0$ : all four particles and antiparticles are degenerate in mass
(Fig.~\ref{spectra}d)
\begin{equation}
m_{\chi_1^0} = m_{\chi^\pm}  = m_{\chi_2^0}   \;.
\end{equation}
This case describes two Dirac fields: one  neutral and one  charged.  It can be viewed as a limit of case (c)
 when $y\to 0$.
All these mass relations  have been derived at  tree level. However,  it  is  known that these 
mass differences are altered by a finite piece of $\mathcal{O}$(100 -- 1000 MeV), when radiative corrections
are taken into account~\cite{Hisano:2014kua}.    Even in the custodial symmetry limit, 
these corrections should be proportional to the $U(1)_Y$ gauge coupling.
They are small compared to  $\omega |y|$ contributions
to the masses from the $d=5$ operators when the scale $\Lambda$  is low,  e.g.,  $\mathcal{O}(1)$ TeV.
 As a result, the mass hierarchies depicted in Fig.~\ref{spectra} will survive  beyond tree level in all cases apart from case (d).

%%%%%%%%%%%%%%%%%%%%%%%
\subsection{Dark Matter Particle Interactions}
\label{inter}
%%%%%%%%%%%%%%%%%%%%%%%%%%
Our notation follows closely 
that  of \Ref{Dedes:2014hga}.
We calculate the  Higgs interactions with the extra fermions from \eq{d5mass}. 
We find,
%%%%%%%%%%%%%%%%%%%%
\begin{equation}
\begin{aligned}
\mathscr{L}_{\rm Y (int)}^{\rm DM} = & - Y^{h\chi^{-} \chi^{+}} \: h \: \chi^{-} \: \chi^{+}
\  - \ \frac{1}{2}\: Y^{h\chi_{i}^{0} \chi_{j}^{0}} \: h \: \chi_{i}^{0} \: \chi_{j}^{0} \  \\ 
 & - \ \frac{1}{2}\: Y^{hh\chi^{-} \chi^{+}} \: h \: h \: \chi^{-} \: \chi^{+}  - \ \frac{1}{4}\: Y^{hh\chi_{i}^{0} \chi_{j}^{0}} \: h \: h \: \chi_{i}^{0} \: \chi_{j}^{0} \ +  \ {\rm H.c.} \; , 
 \end{aligned} 
 \label{Lhchi}
 \end{equation}   
 %%%%%%%%%%%%%%%%%%
where
%%%%%%%%%%%%%%%%%%%%
%
%%%%%%%%%%%%%%%%%%%%%%%%%%%%%%%%
\begin{subequations}
\begin{align}
 Y^{h\chi^{-} \chi^{+}}  =  \sqrt{2}\: \xi_{12} \: \frac{\omega}{v}, &&  Y^{hh\chi^{-} \chi^{+}}  =   \xi_{12} \: \frac{\omega}{v^2} , \label{yhpm}\\
 Y^{h\chi_{1}^{0} \chi_{1}^{0}}  = \frac{\sqrt{2} \, \omega}{v} \: (\xi_{12} + y - y_{12} ),  &&  Y^{hh\chi_{1}^{0}\chi_{1}^{0}}  =  \frac{\omega}{v^2} \: (\xi_{12} + y - y_{12} ), \label{yhx1x1} \\
 Y^{h\chi_{2}^{0} \chi_{2}^{0}}  = \frac{\sqrt{2}\, \omega}{v} \: (\xi_{12} - y - y_{12} ), &&Y^{hh\chi_{2}^{0}\chi_{2}^{0}}  =  \frac{\omega}{v^2} \: (\xi_{12} - y - y_{12} ),\label{yhx2x2}\\
 Y^{h\chi_{1}^{0} \chi_{2}^{0}}  = 0,   &&  Y^{hh\chi_{1}^{0}\chi_{2}^{0}}  = 0 .\label{yhx1x2}
 \end{align}
 \end{subequations}
%%%%%%%%%%%%%%%%%%%%%%%%%%%%%%%%

The 4-point $h^2\chi^2$ vertices are proportional to 3-point $h\chi^2$ vertices. 
Interestingly enough, off-diagonal couplings to $h$ in \eqref{yhx1x2}, vanish identically 
due to the c.c. symmetry of  \eqs{cchi0}{cchipm},  using  that   $C^{-1}\,  h \, C =  h$.

Since $D_{1}$ and $D_{2}$ carry  $SU(2)_{L}\times U(1)_{Y}$ quantum numbers, there are 
renormalizable interactions involving gauge bosons and the   dark fermions, $\chi_{1,2}^0$  and $\chi^\pm$. For instance, 
the interaction between $\chi^{\pm}$ and the photon reads
 %%%%%%%%%%%%%%%
 \begin{equation}
 \mathscr{L}_{\rm KIN (int)}^{\gamma - \chi^{\pm}} = - 
 (+e) \:(\chi^{+})^{\dagger} \bar{\sigma}^{\mu}
 \chi^{+} \: A_{\mu} - (-e)\: (\chi^{-})^{\dagger} \bar{\sigma}^{\mu}
 \chi^{-} \: A_{\mu} \;,
 \end{equation} 
 %%%%%%%%%%%%%%%%%%%%%
 where $A_{\mu}$ is the photon field and $(-e)$ 
 the electron electric charge.  Similarly, the $Z$-gauge boson couplings to 
 charged and neutral dark fermions are 
 %%%%%%%%%%%%%%%%%%%
 \begin{equation}
 \mathscr{L}_{\rm KIN (int)}^{Z - \chi} = \frac{g}{c_{W}} O^{\prime \, L} \: (\chi^{+})^{\dagger}
\: \bar{\sigma}^{\mu}\: \chi^{+} \:Z_{\mu} -
\frac{g}{c_{W}} O^{\prime \, R} \: (\chi^{-})^{\dagger}
\: \bar{\sigma}^{\mu}\: \chi^{-} \:Z_{\mu}  +
\frac{g}{c_{W}} O_{ij}^{\prime\prime \, L} \: (\chi_{i}^{0})^{\dagger}
\: \bar{\sigma}^{\mu}\: \chi_{j}^{0} \:Z_{\mu} \;, 
\label{Zxx}
 \end{equation} 
 %%%%%%%%%%%%%%%%%%%%%%%
 where
 %%%%%%%%%%%%%%%%%%
 \begin{subequations}
 \begin{align}
O^{\prime\, L} &= O^{\prime \, R} = -\frac{1}{2}(1-2 s_{W}^{2}) \;, \label{eq:OLp}\\
O_{ij}^{\prime\prime\, L} &= \frac{1}{2}\: \left ( O_{2i}^{*} \: O_{2j} - O_{1i}^{*} \: O_{1j} \right ) \;, 
\label{gZxx}\\[2mm]
O &=\frac{1}{\sqrt{2}} \left(\begin{array}{cr}1 & i \\1 & -i\end{array}\right)\;, \qquad
O^{\prime\prime\, L} = -\frac{i}{2}\left(\begin{array}{rc}0 & 1 \\-1 & 0\end{array}\right)  \,. 
\label{OLpp}
\end{align}
\end{subequations}
%%%%%%%%%%%%%%%%%%%%%%%%
With $s_{W}$ ($c_{W}$) we denote the $\sin \theta_W$ ($\cos \theta_W $)
 of the weak mixing angle  and with $g$ the $SU(2)_{L}$ gauge
coupling. The  coupling 
$Z\, \chi_i^{0}\,\chi_j^0$ is non-zero only for $i \neq j$
due to the c.c. symmetry 
with  $C^{-1} Z_\mu C = - Z_\mu$.
The  $O^{\prime\prime \, L}$ is an antisymmetric matrix due to 
  the Majorana nature of $\chi_i^0$ fermions and the hermiticity of the Lagrangian.

Interactions between $\chi$'s  and $W$--bosons 
are described by the following terms
%%%%%%%%%%%%%%%%
\begin{align}
\mathscr{L}_{\rm KIN (int)}^{W^{\pm}-\chi^{0}-\chi^{\mp}} \ &= \ 
g\: O_{i}^{L} \: (\chi_{i}^{0})^{\dagger} \: \bar{\sigma}^{\mu} \: \chi^{+} \: W_{\mu}^{-}
-g \: O_{i}^{R} \: (\chi^{-})^{\dagger} \: \bar{\sigma}^{\mu} \: \chi^{0}_{i} \: W_{\mu}^{-} 
\nonumber \\[2mm]
& + g \: O_{i}^{L*} \: (\chi^{+})^{\dagger} \: \bar{\sigma}^{\mu}\: \chi^{0}_{i}\: W_{\mu}^{+}
-g\: O_{i}^{R*} \: (\chi_{i}^{0})^{\dagger} \: \bar{\sigma}^{\mu}\: \chi^{-} \: W_{\mu}^{+} \;,
\label{Wxx}
\end{align}
%%%%%%%%%%%%%%%%%
where the mixing column matrices $O^{L}$ and $O^{R}$ are given by 
%%%%%%%%%%%%%%%
\begin{subequations}
\begin{align}
O_{i}^{L} &=  \ \frac{i}{\sqrt{2}} \: O_{2i}^{*}  = \frac{1}{2} \left(\begin{array}{r}i \\-1\end{array}\right) \;, 
\\[2mm]
O_{i}^{R} &=  \frac{i}{\sqrt{2}} \: O_{1i}  = \frac{1}{2} \left(\begin{array}{r}i \\-1\end{array}\right)  \;,
\end{align}
\label{eq:OLOR}
\end{subequations}
%%%%%%%%%%%%%%%%
with the identity $O_{i}^{R} =  O_{i}^{L}$  being again a  consequence  of  the c.c. symmetry.
Using the same matrices we can write the three-point dipole 
interactions of \eq{dipoles} in the diagonal basis~\footnote{We are not concerned here about CP-violating phenomena and we set $e_{\gamma,W}=0$ .}
% %%%%%%%%%%%%%%%%%%%%%%%%%%%%%%%%%%
\begin{align}
\mathscr{L}^{3-\text{point}}_{\text{dipole}}\ =& -\frac{\omega}{v^2}\, (d_{\gamma}\, s_{W} \,+ \,d_W \,c_{W}) \,  O^{\prime \prime L}_{ij} \, \chi_{i}^{0} \, \sigma_{\mu\nu} \, \chi_{j}^{0} \, F_{Z}^{\mu\nu}  \, - \,  \frac{\omega}{v^2}\, (d_{\gamma} \,s_{W} \, - \, d_W \,c_{W}) \, \chi^{-} \, \sigma_{\mu\nu} \, \chi^{+} \, F_{Z}^{\mu\nu} \nonumber\\ 
& + \frac{\omega}{v^2}\, (d_{\gamma} \,c_{W} \, - \, d_W \,s_{W}) \,  O^{\prime \prime L}_{ij} \, \chi_{i}^{0} \, \sigma_{\mu\nu} \, \chi_{j}^{0} \, F_{\gamma}^{\mu\nu} \, + \, \frac{\omega}{v^2}\, (d_{\gamma} \,c_{W} \, + \, d_W \,s_{W}) \, \chi^{-} \, \sigma_{\mu\nu} \, \chi^{+} \, F_{\gamma}^{\mu\nu} \nonumber\\
& - 2\frac{\omega}{v^2}\,  d_W \,  O^{R\, *}_{i} \, \chi^{-} \, \sigma_{\mu\nu} \, \chi_{i}^{0} \,  F_{W^{+}}^{\mu\nu}   \, + \,  2\frac{\omega}{v^2}\,  d_W \,  O^{L}_{i} \, \chi^{+} \, \sigma_{\mu\nu} \, \chi_{i}^{0} \,  F_{W^{-}}^{\mu\nu} \ + \  \mathrm{H.c.} , \label{diag_dipole-3point}  
\end{align}
%%%%%%%%%%%%
where $F^{\mu\nu}_{V}=\partial^{\mu} V^{\nu}-\partial^{\nu} V^{\mu}$, $V=Z,A \text{ and }W^{\pm}$. 
Interestingly enough, EFT dipole  $d=5$ operators, generate  photon interactions with the neutral dark particles,
 with a coupling that vanishes in the limit  $d_{\gamma} c_{W}  \simeq  d_W s_{W}$. There  is also  an alignment of
couplings in \eq{diag_dipole-3point} with the those  in \eqs{Zxx}{Wxx},  that  is important for achieving a ``natural'' cancellation 
of two different contributions in the cross-section  for $\chi_1^0 \chi_1^0 \rightarrow VV$, where $V$ can be $Z$,$W$ or $\gamma$.
% as we will show below in section~\ref{sec:astro}.
Moreover, the four-point  interactions involving dipole operators are 
%%%%%%%%%%%%%%%%%%%%%%%%%%%%%%%
\begin{align}
\mathscr{L}^{4-\text{point}}_{\text{dipole}}\ =& - 2\, i\, g\, \frac{\omega}{v^2}\,  d_W \,  O^{\prime \prime L}_{ij} \, \chi_{i}^{0} \, \sigma_{\mu\nu} \, \chi_{j}^{0} \, W^{+ \mu} \, W^{- \nu} + 2\, i\, g\, \frac{\omega}{v^2}\,  d_W  \, \chi^{-} \, \sigma_{\mu\nu} \, \chi^{+} \, W^{+ \mu} \, W^{- \nu}  \nonumber \\
& + 4\, i\, g\, \frac{\omega}{v^2}\,  d_W  c_{W}\,O^{R\, *}_{i} \, \chi^{-} \, \sigma_{\mu\nu} \, \chi_{i}^{0} \, W^{+ \mu} Z^{\nu}  \, + \,  4\, i\, g\, \frac{\omega}{v^2}\,  d_W  c_{W} \,O^{L}_{i} \, \chi^{+} \, \sigma_{\mu\nu} \, \chi_{i}^{0} \, W^{- \mu} Z^{\nu} \nonumber \\
& + 4\, i\, g\, \frac{\omega}{v^2}\,  d_W  s_{W}\,O^{R \, *}_{i} \, \chi^{-} \, \sigma_{\mu\nu} \, \chi_{i}^{0} \, W^{+ \mu} A^{\nu}  \, + \,  4\, i\, g\, \frac{\omega}{v^2}\,  d_W  s_{W}\,O^{L}_{i} \, \chi^{+} \, \sigma_{\mu\nu} \, \chi_{i}^{0} \, W^{- \mu} A^{\nu} \nonumber \\ \ & + \ \mathrm{H.c.} \label{diag_dipole-4point}
\end{align}
% \end{equation}
%%%%%%%%%%%%%%%%%%%%%%%%%%%%%%%%%%%%%%%%%

In section~\ref{sec:astro}, we will see that these EFT dipole interactions   
are  important  for making  the annihilation and coannihilation cross section 
 of  the WIMP dark matter particle $\chi_1^0$,  compatible with the measurement \eqref{Planck} for the DM relic density.

%%%%%%%%%%%%%%%%%%%%%%%%%%
\section{``Earth'' constraints in the Dark Sector}\label{Earth constraints}
\setcounter{equation}{0}
\label{sec:earth}
%%%%%%%%%%%%%%%%%%%%%%%%%%

 In this section we  study   constraints  imposed on  the parameter space, 
from WIMP($\chi_1^0$)-nucleon scattering experiments searching directly for DM,
from  direct and oblique  LEP  electroweak observables and from the  LHC data for the Higgs boson decay 
to two photons.

%%%%%%%%%%%%%%%%%%%%%%%%%%%%
\subsection{Nucleon-WIMP direct detection experimental bounds} 
%%%%%%%%%%%%%%%%%%%%%%%%%%%%
In the limit that the  DM particle $\chi_1^0$ is much heavier  than  nucleon,  the spin independent (SI) and  spin dependent  (SD) cross
sections are given by~\cite{Cheung:2012qy}
%%%%%%%%%%%%%%%%%
\begin{equation}
\sigma_{\rm SI} =  8\times 10^{-45}\, \mathrm{cm}^{2} \, 
\left (\frac{Y^{h\chi_{1}^{0}\chi_{1}^{0}}}{0.1} \right )^{2}\;, \qquad 
\sigma_{\rm SD} =  3\times 10^{-39}\, \mathrm{cm}^{2} \, 
\left (\frac{g^{Z\chi_{1}^{0}\chi_{1}^{0}}}{0.1} \right )^{2}\;.
\end{equation}
From the interactions in  \eq{Zxx},
 we see  that $g^{Z\chi_{1}^{0}\chi_{1}^{0}} = 0$ at tree level
 and therefore
$\sigma_{\rm SD} \approx 0$. For the SI cross-section
the current  bound  from LUX~\cite{Akerib:2015rjg,Akerib:2013tjd}  
is  $\sigma_{\rm SI} \simeq \{1 - 3.5 \}  \times 10^{-45} ~\mathrm{cm}^{2}$, for 
$   m_{\mathrm {DM}}  \simeq  \{100 -  500 \} \GeV$, respectively.
This gives 
%%%%%%%%%%%%%%%%
\begin{equation}
|Y^{h\chi_{1}^{0} \chi_{1}^{0}}| \ \lesssim \ \{0.04,0.06\}   \, ,
\label{dircon}
\end{equation}
%%%%%%%%%%%%%%%
 which through the r.h.s. of  \eq{yhx1x1} yields  
a  constraint for the combination $\xi_{12} + y - y_{12}$. The
 one-loop contributions  have been calculated in \Refs{Dedes:2014hga,Freitas:2015hsa}. We have worked out
the  formula given in the Appendix A of \Ref{Dedes:2014hga},   for zero Yukawa couplings  
and  $M_D \gg M_{W,Z}$, and we find 
%%%%%%%%%%%%%%%%
\begin{equation}
\delta Y^{h\chi_1^0\chi_1^0} = \left (\frac{3 g}{8\pi} \right )^2 \frac{\sqrt{2} M_Z}{v} \, \frac{M_D-M_Z}{M_D} \simeq 4\times 10^{-3}\, .
\end{equation}
%%%%%%%%%%%%%%
This  is an order of magnitude smaller\footnote{It is shown  in~\Ref{Hisano:2015rsa} that 
for next to leading order corrections  in $\alpha_s$, the  SI cross-section is even smaller.} than the current LUX bound in \eq{dircon}. 
In the limit  $M_D\to \infty$, the model exhibits  a non-decoupling behaviour,   as expected from the EFT analysis of
\Refs{Hill:2013hoa,Cirelli:2005uq,Hisano:2011cs}. On the other hand, for $M_D\to 0$ the one-loop contribution vanishes.

Based upon \eqs{yhx1x1}{dircon} we obtain the inequality 
%%%%%%%%%%%%%%%%%%%%%%%%
\begin{equation}
%-\frac{1}{\sqrt{2}} \, \left (\frac{\Lambda}{v} \right ) \, Y^{\mathrm{bound}}(m_{\chi_1^0}) \,  \  \le \
| \xi_{12} + y - y_{12} |  \ \le \ 
\frac{1}{\sqrt{2}} \, \left (\frac{\Lambda}{v} \right )\, Y^{\mathrm{bound}}(m_{\chi_1^0})  \;,
\label{dlim}
\end{equation}
%%%%%%%%%%%%%%%%%%%%%%%%%%%
where $Y^{\mathrm{bound}}(m_{\chi^{0}_{1}})$ 
is the bound of \eq{dircon}.  \Eq{dlim} sets strong bounds 
on the couplings $\xi_{12}$ and/or $y$. Relevant to the cases depicted in Fig.~\ref{spectra}
we obtain,   for $\Lambda=1~\mathrm{TeV}$ and $m_{\chi_1^0} \sim 100$ GeV the following constraints:
%%%%%%%%%%%%%%%%%%%
\begin{align}
& \mathrm{(a,d)} : \quad  -0.16  \lesssim \ \xi_{12} \ \lesssim 0.16    \;, \nonumber \\[2mm]
& \mathrm{(b)} : \quad- \frac{16 + 200\: y}{100}  \lesssim \ \xi_{12} \  
\lesssim \frac{16  - 200\: y}{100}\;,
\label{xicon} \\[2mm]
& \mathrm{(c)} : \quad- \frac{16  + 100\: y}{100}  \lesssim \ \xi_{12} \  
\lesssim \frac{16  - 100\: y}{100}\;.  \nonumber 
\end{align}
%%%%%%%%%%%%%%%%%%%%
%%%%%%%%%%%%%%%%%%%%%%%%%%%%%%%%%
\begin{figure}[ht]
  \centering
   \includegraphics[width=0.58\textwidth]{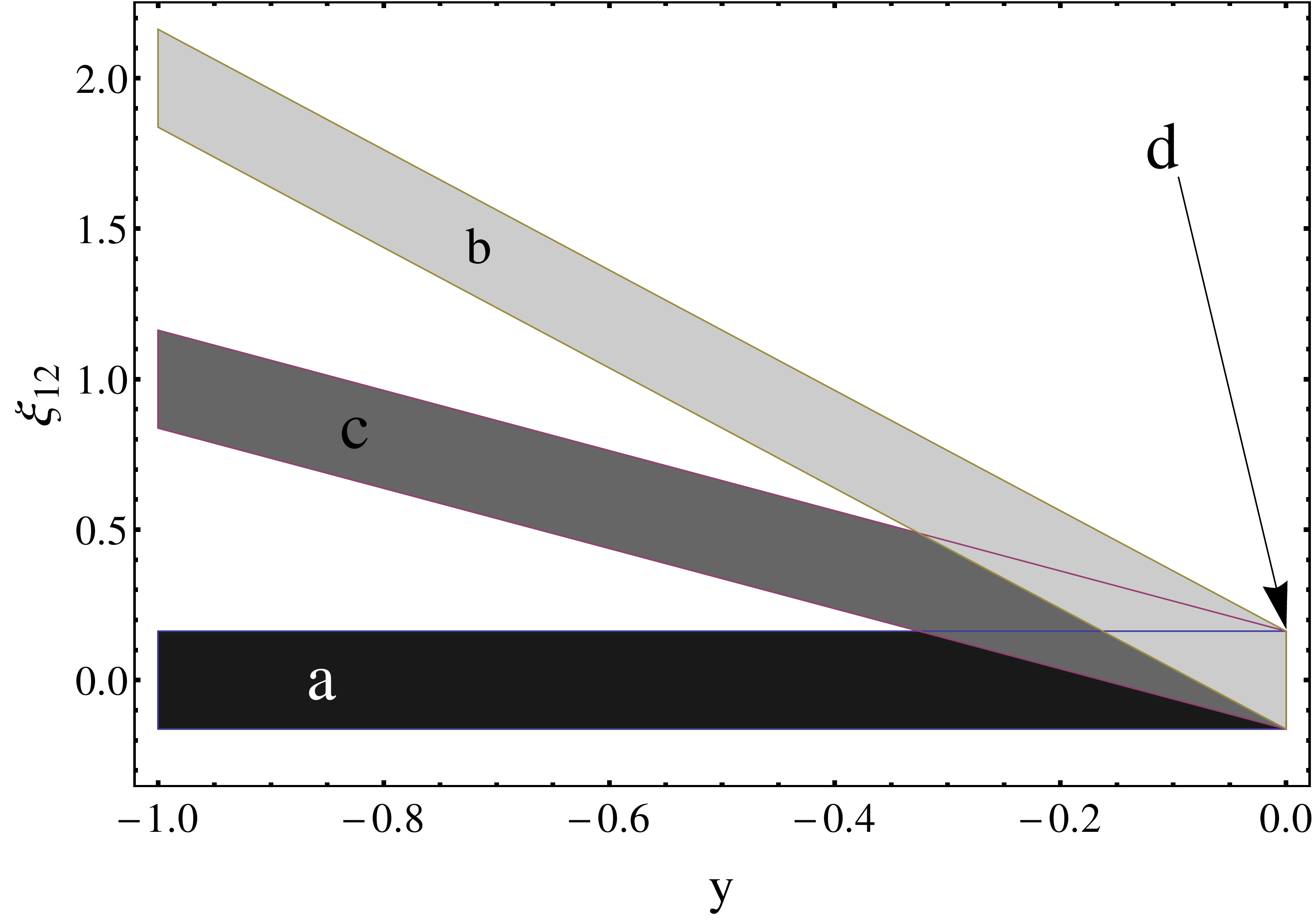}
        \caption{\em Yukawa couplings $y$ vs. $\xi_{12}$  compatible with
        the bound  of \eq{dircon}   
        related to the LUX DM detection experiment,
              for the four  cases of the mass spectrum and $\Lambda=1 \; \mathrm{TeV}$. 
%         {\bf Dimitri: can you make labels bigger here?}
        }
        \label{DirDet-region}
    \end{figure}
%%%%%%%%%%%%%%%%%%%%%%%%%%%%%%%%%%

Therefore, {\em for $y \sim -1$, the parameter $\xi_{12}$ is always positive} with a small variation 
band of about 10\% w.r.t. the $|y|$ value due to the LUX bound. An example for the case (c) is shown 
in Fig.~\ref{DirDet-region}. For even bigger values of $|y|$, we obtain, $\xi_{12}$ as big as $|y|$
for case (c) or as big as $2 |y|$ in case (b). The band of the allowed values for $\xi_{12}$, e.g., 
the shaded area in Fig.~\ref{DirDet-region},
expands  if we increase $\Lambda$. 

Apparently,   from \eq{eq:spec},  if  $y=0$ we get $m_{\chi_1^0}=m_{\chi_2^0}$.
In addition,   if dipole operators of  \eq{diag_dipole-3point} are present,
severe bounds on $d_\gamma$ and $d_W$ can be set based on contribution to WIMP-nucleon cross section 
from  $\gamma \, , Z$-exchange  graphs~\cite{Barger:2010gv,Banks:2010eh,Fortin:2011hv}.
In our case these bounds are avoided because we choose always 
 $m_{\chi_2^0}-m_{\chi_1^0} \gtrsim 2\GeV$~\cite{Chang:2010en}. 

It is worth repeating here,   that $\xi_{12}$ is in principle   positive everywhere for the cases (b,c), 
which means that essentially the charged particle $\chi^\pm$ is behaving  as an extra lepton 
circulating in the  $h \to \gamma \gamma$ loop decay process. Therefore,  we expect that 
$R_{ h \to \gamma \gamma }$ will be in general   smaller than in the  
SM.

%%%%%%%%%%%%%%%%%%%%%%%%
\subsection{LEP bounds}
\label{LEPcon}
%%%%%%%%%%%%%%%%%%%%%%%%%
Next we  examine  constrains from LEP, that although   have been  derived particularly for the  MSSM, 
 they can  easily be adapted to this model.
From Fig.~\ref{spectra} we observe that always, 
the next-to-lightest particle is the charged dark fermion  $\chi^\pm$ 
with mass $m_{\chi^{\pm}}=M_D + \xi_{12} \, \omega$ that, as explained
before,  is assumed to be positive. 

Depending on the mass difference between the  lightest neutral particle $m_{\chi_1^0}$
and the charged one  $m_{\chi^\pm}$,
the bound on $m_{\chi^\pm}$ varies within 
$\sim 90 \; \mathrm{GeV} $ to $\sim 100 \; \mathrm{GeV}$~\cite{LEP_SUSY:2001,LEP:2001qw,Abdallah:2003xe}. 
We will use the most conservative choice
%%%%%%%%%%%%%%%%%%%%%%%%%%%%%%
\begin{equation}
  m_{\chi^{\pm}}  \ \gtrsim \ 100  \; \mathrm{GeV} \;,
  \label{LEP-Bound}
\end{equation} 
%%%%%%%%%%%%%%%%%%%%%%%%%%%
which in terms of $\xi_{12}$, $\omega$ and $M_D$ becomes:
\begin{equation}
 \xi_{12} \ \gtrsim  \ \frac{100-M_{D}}{\omega} , \quad \forall \, y_{12} \;. \label{LEP-bound}
\end{equation} 
%%%%%%%%%%%%%%%%%%%%%%%%%%%%%%
As we have seen, the bound  from direct detection experiments implies a positive 
value on $\xi_{12}$ for the cases (b,c). Thus, the LEP bound (\ref{LEP-bound}) is 
always satisfied if $M_D \gtrsim 100$ GeV.  
In the case where $M_D \lesssim 100$ GeV one may evade the LEP bound with a large positive $\xi_{12}$.
 For example,  for $\Lambda=1~\mathrm{TeV}$ and $M_D=50$ GeV,
we need, $\xi_{12} \gtrsim 1.7$. Interestingly, this may be  compatible with  \eqref{xicon} 
only in  cases (b) and (c)  with  $\Lambda=1$ TeV and certain values of $y$.

%%%%%%%%%%%%%%%%%%%%%%%%%%%%%%
\subsection{$h\to \gamma\gamma$}
%%%%%%%%%%%%%%%%%%%%%%%%%%%%%%%%%
\label{sect:htogg}

For the model under study,  the ratio $R_{h\to \gamma\gamma}
\equiv 
\frac{\Gamma(h\to \gamma\gamma)}{\Gamma(h\to \gamma\gamma)_{\mathrm{(SM)}}}$ 
is given by~\cite{Dedes:2014hga}
%%%%%%%%%%%%%
\begin{equation}
R_{h\to \gamma\gamma} \ = \  
\: \biggl |\: 1 \ + \  
\frac{1}{A_{\mathrm{SM}}} \:   
 \frac{\sqrt{2}\: Y^{h\chi^{-}\chi^{+}}\, v}{m_{\chi^{\pm}}} \: A_{1/2}(\tau)\: \biggr |^{2}\;,
\label{hi2g}
\end{equation}
%%%%%%%%%%%%%%%%%%
where $A_{\mathrm{SM}} \simeq -6.5$ for $m_{h}=125$ GeV.
This  is the SM result dominated by the $W$-loop,
with $\tau \equiv m_{h}^{2}/4 m_{\chi^{\pm}}^{2}$ and $A_{1/2}$ is the well known function
given for example in \Ref{Djouadi:2005gi}.
%\footnote{In our notation,  
%the Higgs-fermion vertex is parametrized here
%as $\mathcal{L} \supset - Y_{f} f \bar{f} \, h$ and therefore, for the top-quark Yukawa coupling, 
%we obtain: $Y_{i} \to Y_{t}/\sqrt{2}$ while for the new charged fermions 
%$Y \to Y^{h\chi^{-}\chi^{+}}$.}
The ratio $R$  is currently under experimental scrutiny   at LHC. The current combined 
value is
$R_{h\to \gamma\gamma} = 1.15^{+0.28}_{-0.25}$~\cite{Aad:2015zhl}.  
Note that the  gluon fusion channel  $g g \to h$, involved in the Higgs boson production at LHC,
is not affected in the context of this  model, 
 since $\chi_i^0,\chi^\pm$ are uncoloured particles. In principle, there are
 $d=6$ operators, such as  $H^\dagger H \, G_{\mu\nu}\, G^{\mu\nu}$, 
  but  we assume that these are quite suppressed in comparison   to the SM contribution.
An analogous operator exists in the case of $h\to \gamma\gamma$, just replacing $G^{\mu\nu}$ with 
the photon field strength tensor, $F^{\mu\nu}$ when integrated out heavy (of order bigger than $\Lambda$) 
particles. These operators arise at loop level and are suppressed by the scale $\Lambda$.
 Therefore, for the process $h\to \gamma\gamma$, the effect is  dominated  by the SM charged particles and the new
$\chi^\pm$ circulating  in the  triangle diagram.

Below we study  the ratio $R$ in two complementary regions for  $M_D$:
a)  $M_D \lesssim 100 \GeV$ and  b)  $M_D \gtrsim 100 \GeV$.

%%%%%%%%%%%%%%%%%%%%%%%%%%%%%
\subsubsection{$M_D \lesssim 100 \; \mathrm{GeV}$}
%%%%%%%%%%%%%%%%%%%%%%%%%%%

From \eq{yhpm} we expect that $\xi_{12}$ would be restricted to small values from the 
loop induced $h\to \gamma \gamma$ bound, 
where we should also expect that, for $M_D$ below $100~\mathrm{GeV}$,
the bound from LEP will be important   as we explained previously in section~\ref{LEPcon}.
%
%%%%%%%%%%%%%%%%%%%
\begin{figure}[t]
\mbox{ %\hspace{2mm}
\includegraphics[width=7.4cm]{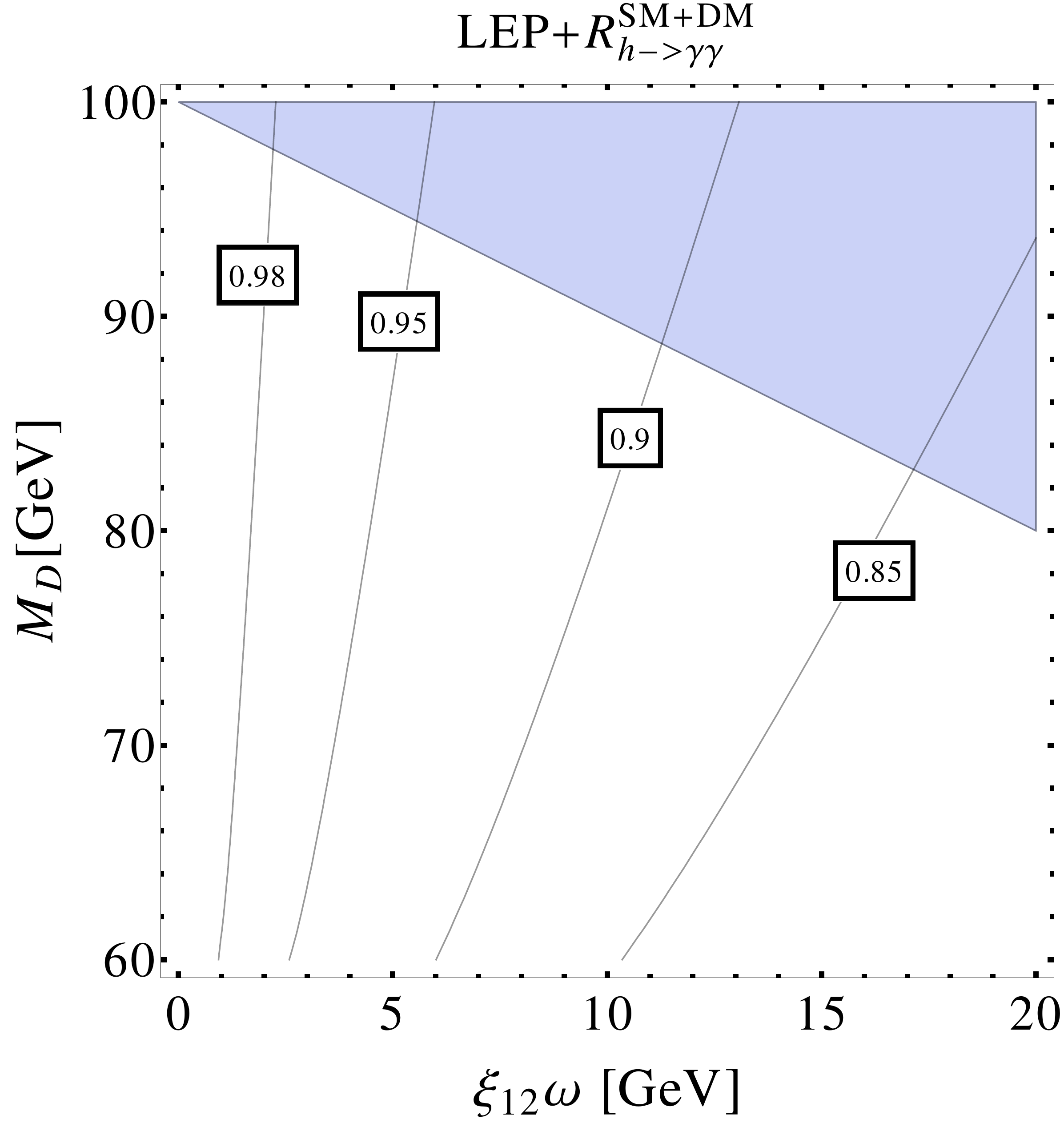}
\hspace{3mm}
\hfill
\includegraphics[width=7.4cm]{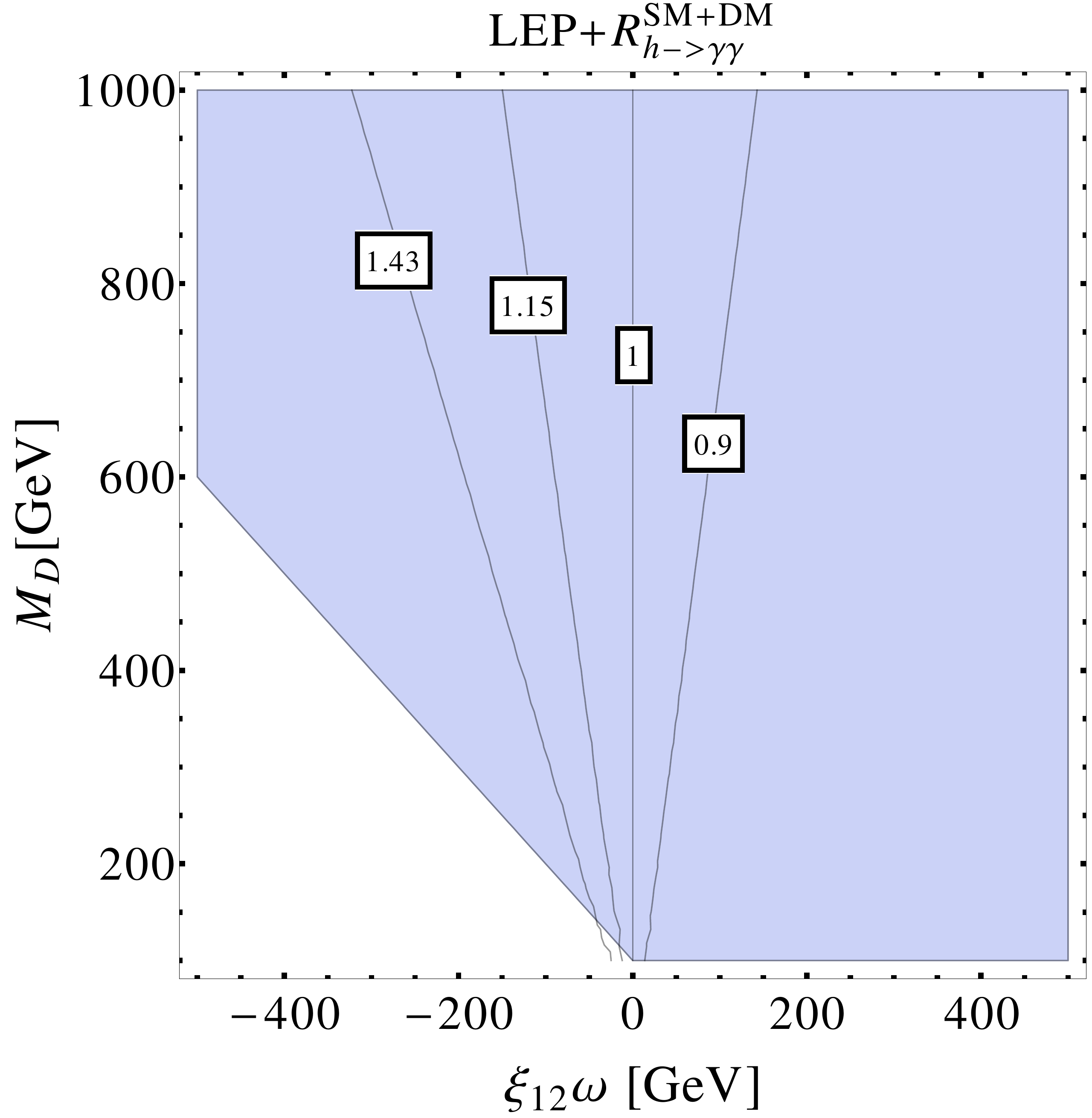}
}
\vspace*{-0.5cm}
\caption{\em  Combination of  the constraints from negative LEP ``chargino''  searches applied to $\chi^\pm$ 
  and the ratio $R_{h\to\gamma\gamma}$ from LHC (Run~I), 
  for (a) $M_D \lesssim 100  \GeV$  and (b) $M_D \gtrsim 100  \GeV$.
  The shaded region is  compatible to the   LEP  $\chi^\pm$ bound,  
  while the contours show the values of $R_{h\to\gamma\gamma}$ from \eq{hi2g} on  $\xi_{12}\, \omega-M_D$ plane.}
     \label{LEP-Rbsm-Region-MD_100}
\end{figure}
%
%%%%%%%%%%%%%%%%
When $M_D \lesssim 100~\mathrm{GeV}$, 
$\xi_{12}$ should  always be positive or zero in order to satisfy the LEP bound 
\eqref{LEP-bound}. Then the charged fermion behaves as an extra lepton 
and lowers the ratio  $R$. This is clear from \eqs{yhpm}{hi2g}.
In addition, one can easily observe that, 
since LEP restricts $m_{\chi^{\pm}}$ to be above $100~\mathrm{GeV}$, 
the function $A_{1/2}(\tau=   {m_{h}^{2}}/{4 m_{\chi^{\pm}}})$ lies within
the interval $\sim 1.5 \;-\; 1.3$. 
These observations lead us to another improved bound 
between $\xi_{12}$, $\omega$ and $M_D$, for the combined constraints from LEP and $R_{h\to\gamma\gamma}$: 
% {\bf Dimitri: what bound on R did you use here? 1$\sigma$, $2\sigma$?}
%%%%%%%%%%%%%%%%%%%%%%%%
  \begin{equation}
  100-M_{D} \ \lesssim \  \xi_{12} \: \omega \ \lesssim  \ 0.1\, M_D \;, 
  \qquad \forall  \, y_{12}. \label{LEP_Rbsm-bound-MD<100-xi>0}
\end{equation}
 %%%%%%%%%%%%%%%%%%%%%%%%%%%% 
 % 
Therefore,  if $M_D \lesssim 100 \; \mathrm{GeV}$, 
we obtain a minimum allowed value for $M_D$,  which is around $90 \; \mathrm{GeV}$, as
illustrated  in  Fig.~\ref{LEP-Rbsm-Region-MD_100}(a). As a consequence, the  case
 $M_D \lesssim 100$ GeV is disfavoured.

 %%%%%%%%%%%%%%%%%%%%%%%%%%%%%%%
\subsubsection{$M_D \gtrsim 100 \; \mathrm{GeV}$}
%%%%%%%%%%%%%%%%%%%%%%%%%%%%%%
If $\xi_{12}>0$, then  \eq{LEP_Rbsm-bound-MD<100-xi>0} still holds.
 The only difference from the previous case arises 
 from \eq{LEP-bound}, which now allows $\xi_{12}$ to  be also negative.
 Consequently, for  $\xi_{12}<0$, the $R_{h\to \gamma\gamma}$ can be greater than unity and we obtain
%\ie the charged fermion does not behaves as an extra SM-like lepton,
%and the corresponding bound becomes
%%%%%%%%%%%%%%%%%%%%%%%%%%
 \begin{equation}
  \xi_{12} \, \omega \ \gtrsim \  -0.3 \, M_{D} \;, \quad \text{ or } \quad  \xi_{12} \, \omega
   \ \gtrsim \ 100-M_{D} \;,  \quad \forall y_{12} \text{ and } \xi<0.
    \label{LEP_Rbsm-bound-MD>100-xi<0}
\end{equation}
%%%%%%%%%%%%%%%%%%%%%%%
Therefore, the combined result for $M_D \gtrsim 100 \; \mathrm{GeV}$ is:
%%%%%%%%
 \begin{equation}
    -0.3\,  M_D \ \lesssim \  \xi_{12}\, \omega \  \lesssim \ 0.1\, M_D \quad  \text{ and } \quad 
    \xi_{12}\, \omega \ \gtrsim \ 100-M_{D}  \;, \quad \forall y_{12}. 
    \label{LEP_Rbsm-bound-MD>100}
\end{equation}
%%%%%%%%%%%%%%%%%%%%%%%%%%%%
This inequality is illustrated  in  Fig.~\ref{LEP-Rbsm-Region-MD_100}(b).  
We notice that    \eq{LEP_Rbsm-bound-MD>100} results in
 a very weak bound for $\xi_{12}<0$ compared to the constraints from direct detection  experiments, as can be seen 
 in Fig.~\ref{DirDet-region}. 
  %%%%%%%%%%%%%%%%%%%%%%%%%  
 \Eq{LEP_Rbsm-bound-MD>100} may nicely 
be combined in terms of the physical charged fermion mass $m_{\chi^\pm}$ 
and the ``doublet'' mass $M_D$ as
%%%%%%%%%%%%%%%%
 \begin{equation}
0.7 \, M_D \ \lesssim \ m_{\chi^\pm} \ \lesssim \ 1.1\, M_D \;.
\end{equation}
%%%%%%%%%%%%%%%%%%

%%%%%%%%%%%%%%%%%%
 \begin{figure}[H]
    \centering
    \includegraphics[width=.6\linewidth]{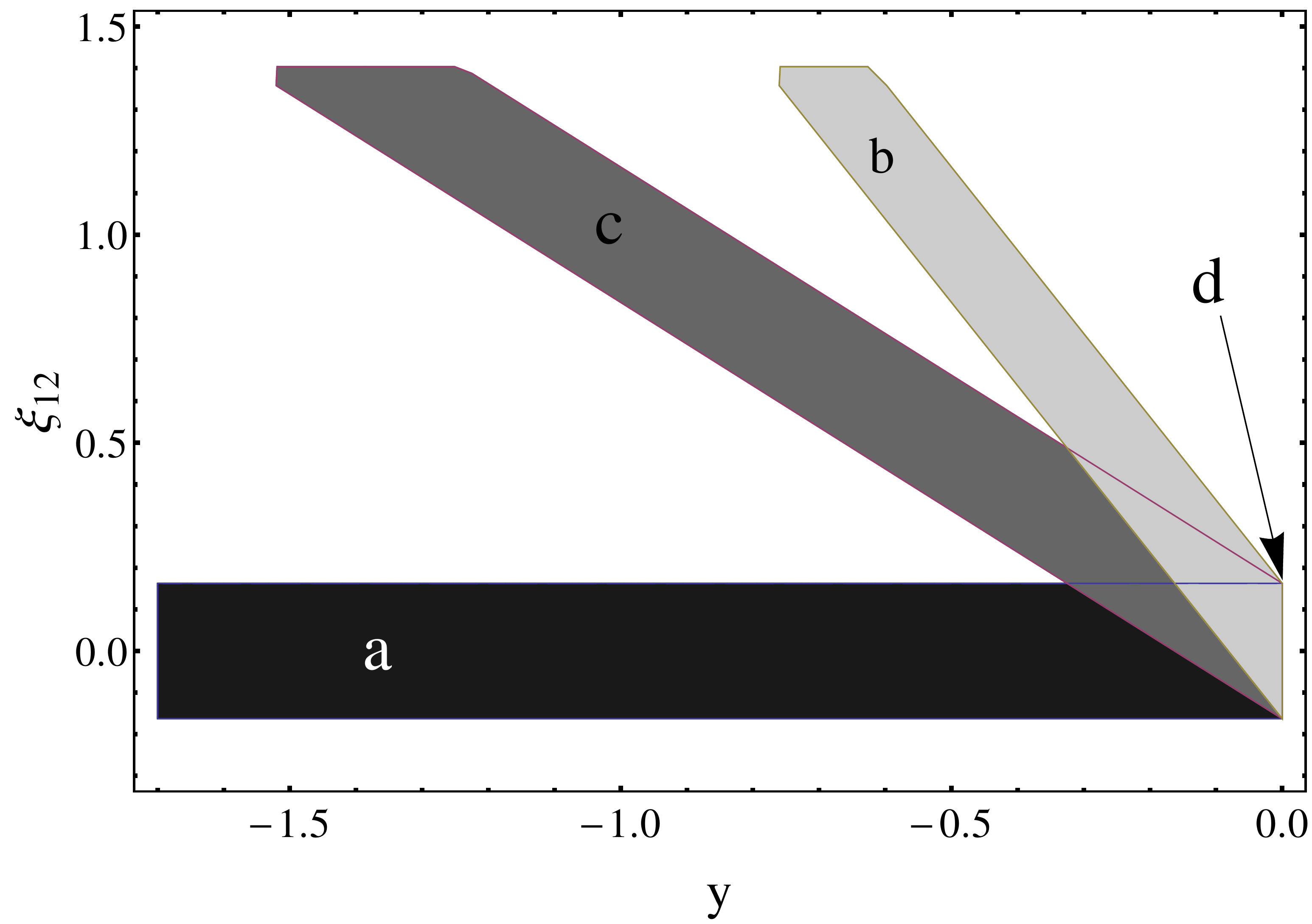}
  \caption{\em 
  $y$ vs. $\xi_{12}$ regions allowed by combining  LEP, $R_{h \to \gamma \gamma}$ and DM direct detection constraints,
   for $\Lambda =1  \TeV$ and $M_D=300 \GeV$ for  the four cases studied. Notice that case (d) is the intersection of the three other cases.}
    \label{Combined}
  \end{figure}
%%%%%%%%%%%%%%%%%%

Before moving on  to the calculation of the relic density, we summarize 
  the phenomenological constraints  imposed to this model by  LEP $\chi^\pm$ 
  searches, the $h \to \gamma \gamma$ decay   and the direct DM detection experiments.
As can be seen from Fig.~\ref{Combined}, 
these  constraints  confine the parameters $y$ and $\xi_{12}$ in small regions for given $M_D$ and the 
cut-off of the theory. As discussed previously, $M_D$ is always   $\gtrsim 90 \; \mathrm{GeV}$ which is independent of the cut-off.   
A general comment is that the bound imposed by the direct detection experiments in \eq{dircon} binds $y$ and $\xi_{12}$ together 
(and also forces $\xi_{12}$ to be mostly positive).

%%%%%%%%%%%%%%%%%%%%%%%%%%%%%
\subsection{Electroweak oblique corrections}  
%%%%%%%%%%%%%%%%%%%%%%%%%%%%%%

In general,  when one adds new matter into the SM particle content, with non-trivial 
gauge quantum numbers, severe  bounds arise  from the so-called oblique electroweak corrections. 
These loop  corrections to electroweak precision observables are commonly parametrised by three parameters,  
$S$,  $T$ and $U$,  introduced long ago in \Refs{Peskin:1991sw,Barbieri:2004qk}. 
Even though the new matter fields $\mathbf{D_1}$ and $\mathbf{D_2}$ have common,
vectorlike, mass $M_D$ from  \eq{eq:md}, there are mass splittings amongst the two doublets as well
amongst their components themselves. These mass splittings arise from $d=5$ operators in \eq{d5mass} 
as discussed  in the previous section.

\begin{figure}[t]
    \centering
    \includegraphics[width=.8 \linewidth]{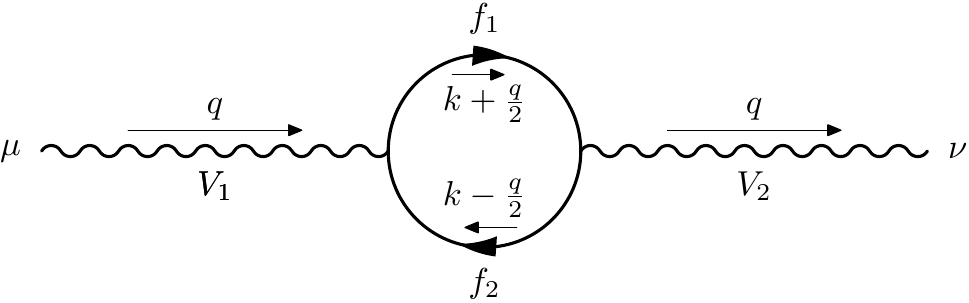}
  \caption{\em The Feynman diagram  contributing to the oblique parameters.
   $V_{1,2}$ represent the gauge bosons $Z, W$,  or  $\gamma$,  while 
   $f_{1,2}$ are $\chi_{1,2}^0$ and/or $\chi^{\pm}$.}
     \label{VacPol}
  \end{figure}

In order to calculate the $S,T$ and $U$ parameters in the EFT at hand,  we need to calculate vacuum polarization diagrams
like the one depicted in Fig.~\ref{VacPol},  for all relevant interactions arisen from $d=4$ and $d=5$ operators  given  in section~\ref{inter}.
%
%%%%%%%%%%%%%%%%%%%%%%%  
%%%%%%%%%%%%%%%%%%%%%%%%%%  
%
The general form of this diagram is  
%%%%%%%%%%%%%%%%%%%%%
  \begin{align}
 i \,  \Pi^{\mu \nu}_{V_1 V_2} & = \int d^d k \mu^{\epsilon} \dfrac{(-1)}{(2 \pi)^d} \dfrac{1}{\left[(k+\dfrac{q}{2})^2-m_1 ^2 \right]  \left[(k-\dfrac{q}{2})^2-m_2 ^2\right]} \times  \nonumber   \\
   &    \mathrm{Tr}\left[ \left(a_{21} \gamma^\mu + \dfrac{b_{21}}{4\Lambda} [\slashed{q},\gamma^\mu] \right)  \left(\slashed{k}
   +\dfrac{\slashed{q}}{2}+ m_1 \right)   \left(a_{12} \gamma^\nu 
           - \dfrac{b_{12}}{4\Lambda} [\slashed{q},\gamma^\nu] \right)  \left(\slashed{k}-\dfrac{\slashed{q}}{2}+ m_2 \right)\right],   
   \label{eq:VC}
   \end{align}  
% %%%%%%%%%%%%%%%%%
%
%
where $\mu$ is the renormalization scale, $\epsilon \equiv 4-d$, $a_{12,21}$ and $b_{12,21}$ are the gauge and the dipole couplings for every possible $\{f_{1,2}  \; , \; V_{1,2}\}$ combination, where  $V_{1,2}$ can be  the gauge bosons $Z, W$,  or  $\gamma$
and  $f_{1,2}$ are $\chi_{1,2}^0$ and/or $\chi^{\pm}$. 
 
If we express the  fermion masses circulating in the loop  as 
$m_{1,2} = M_D+ c_{1,2} {\, v^2}/{\Lambda}$ and expand \eq{eq:VC} up to the order  $\mathcal{O}(\Lambda^{-1})$~\footnote{ 
By doing so, one avoids the introduction of  involved $d=6$ operators.
%These operators  occur in this model and, of course,  in  the SM~\cite{Grzadkowski:2010es}.
Their inclusion would lead to weak bounds on the corresponding Wilson coefficients  
(a related discussion can be found in \Ref{Mebane:2013zga}).},
the term  proportional to $g^{\mu \nu}$  and its derivative w.r.t. $q^2$ at $q^2 = 0$,  read as
%
%%%%%%%%%%%%%%%%
\begin{subequations}
\begin{align}
&\Pi_{ V_1 V_2}(q^2 =0) = \epsilon \;\frac{a_{12}a_{21}}{8\pi^2 \Lambda}(c_1+c_2) v^2 \;M_D \; \log\left(\dfrac{M_D ^2}{\mu^2}\right)  \xrightarrow{\epsilon \rightarrow 0} 0
\;,  \\[3mm]
&\dfrac{d}{dq^2}\Pi_{ V_1 V_2}(q^2 =0) =  \frac{a_{12}a_{21}}{12\pi^2} \left [\log\left(\dfrac{M_D ^2}{\mu^2}\right)+\dfrac{2}{\epsilon}-\gamma+\log(4\pi)\right]\\
%&+ \frac{4 a_{12}a_{21} (c_1 + c_2) v^2 +6M_D ^2 \left( a_{21}b_{12}+a_{12}b_{21} \right)  
%   \left(  \log\left(\dfrac{M_D ^2}{\mu^2}\right)+\dfrac{2}{\epsilon}-\gamma+\log(4\pi) \right)   }
%   {48\pi^2 \Lambda M_D}  \nonumber
&+ \frac{1}{48\pi^2 \, \Lambda M_D} 
\left\{ 4 a_{12}a_{21} (c_1 + c_2) v^2 +6M_D ^2 \left( a_{21}b_{12}+a_{12}b_{21} \right)  
   \left[  \log\left(\dfrac{M_D ^2}{\mu^2}\right)+\dfrac{2}{\epsilon}-\gamma+\log(4\pi) \right]   \right\}.
    \nonumber  
\end{align}
\label{eq:PIS}
\end{subequations}
%%%%%%%%%%%%%%%%%%%%%%
 
Using these  equations  and substituting  for every combination of $\{f_{1,2},V_{1,2}\}$, the
 $a_{12}, a_{21}$, $b_{12},b_{21}$ and $c_{1},c_{2}$  
in the expressions for the parameters $S,T$, and $U$~\cite{Peskin:1991sw},
with the interactions   given in section~\ref{inter},  one obtains  up to terms of $\mathcal{O}(1/\Lambda^2)$,  that
%%%%%%%%%%%%%%%%%%%%%%%%%%
\begin{subequations}
\begin{align}
%&S= - \frac{2 v^2 y_{12}}{3 \pi \Lambda M_D}
\label{eq:S}
& S = - \frac{2}{3 \pi} \: \frac{v^{2} \, y_{12}}{\Lambda \: M_{D}}\;, \qquad  T=0 \;, \qquad U=0 \;. 
\end{align}
\label{STU}
\end{subequations}
%%%%%%%%%%%%%%%%%%%%%%%%
%
These results have been checked  independently using the analytical expressions of  \Ref{Dedes:2014hga} 
 and interactions  from section~\ref{inter}
keeping terms  up to $1/\Lambda$. In addition, they 
have been verified numerically by taking the decoupling limit of the fermion triplet mass  $M_T \gg M_D$ in  \Ref{Dedes:2014hga}.

The parameter $S$ measures the size of  the new fermion sector 
\ie the number of the  extra $SU(2)_L$ irreducible representations that have been added in the model.
In general, the contribution of degenerate fermions to the $S$-parameter is
 \begin{equation}
S\sim  \sum_{\text{new fermions}} (T^3 _{(R)} - T^3 _{(L)})\, ,
\end{equation}
where $T^3_{(L,R)}$ is the isospin of the left- and right-handed fermions. 
So, in a case similar to ours, where the fermions are nearly degenerate, the $S$-parameter takes the form
 \begin{equation}
S\sim  \sum_{\text{new fermions}} (T^3 _{(R)} - T^3 _{(L)}) +  f(m_{\chi_1 ^0},m_{\chi_2 ^0},m_{\chi ^+}) \, ,
\end{equation}
where $ f(m_{\chi_1 ^0},m_{\chi_2 ^0},m_{\chi ^+}) $ is a function that vanishes if the three masses are equal. 
Therefore, in our case, the $S-$parameter for two vector-like doublets 
would  arise  only from the mass differences, which  means
that  S-parameter is proportional to the 
Yukawa couplings~\footnote{The coupling $\xi_{12}$ is 
just a universal shift to $M_D$ and thus it does not contribute 
to the mass difference. Also, as it turns out, $y$ does not appear in 
$\dfrac{d}{dq^2}\Pi_{V_1V_2}(q^2 =0)$ (for every $V_{1}$ and $V_{2}$ combination).
Only $y_{12}$ contributes to the oblique EW parameters at the  approximation in $1/\Lambda$.}.
After performing the calculation, it turns out that
 \begin{equation}
f(m_{\chi_1 ^0},m_{\chi_2 ^0},m_{\chi ^+}) \propto \dfrac{\Delta m_{1^+}+\Delta m_{2^+}}{M_{D}}  \, ,
\end{equation}
where $\Delta m_{i^+}\equiv m_{\chi_i ^0}-m_{\chi^{\pm}}$.
 This is proportional to  $y_{12}$, as can been seen  in eq.~\eqref{eq:S}. 
 Furthermore, no magnetic dipole parameters $d_\gamma$ or $d_W$ are involved in $S$-parameter in \eqref{eq:S}
up to $\mathcal{O}(1/\Lambda^2)$,  as also expected from dimensional arguments.
% {\bf $\Delta m$ here depends on $y$ and not on $y_{12}$ that we find in $S$ parameter. 
% What is going on?}

The $U-$parameter, on the other hand, measures the size  of the 
isospin breaking contribution from the new fermions. So, it should be suppressed due to the c.c. (or custodial) 
symmetry (which limits the isospin breaking) and the 
fact that we are keeping only terms up to ${1}/{\Lambda}$.
Up to this order, the parameter $T$ is zero too, because  $\Pi_{V_1 V_2} (q^2 =0)=0$,
a result which is  independent of the symmetric limits for $y_{12}$.
Usually, the parameters $T$ and  $U$ are  proportional 
to the ratio $\dfrac{\Delta m^2}{M_{Z}^2 \text{ or } M_{D}^2}$, where  $\Delta m^2$ is some  mass-squared difference arising from isospin breaking.
In our model  this should be the case  when $~y\ne 0 \text{ and }y_{12}\ne 0$,  
which means that higher order terms could give a non-vanishing (but suppressed by terms $\propto \Lambda^{-2}$) contribution.
%
%
%%%%%%%%%%%%%%%%%%%%%%%%%%%%%%%%% 
 \begin{figure}[t]
    \centering
    \includegraphics[width=.58\linewidth]{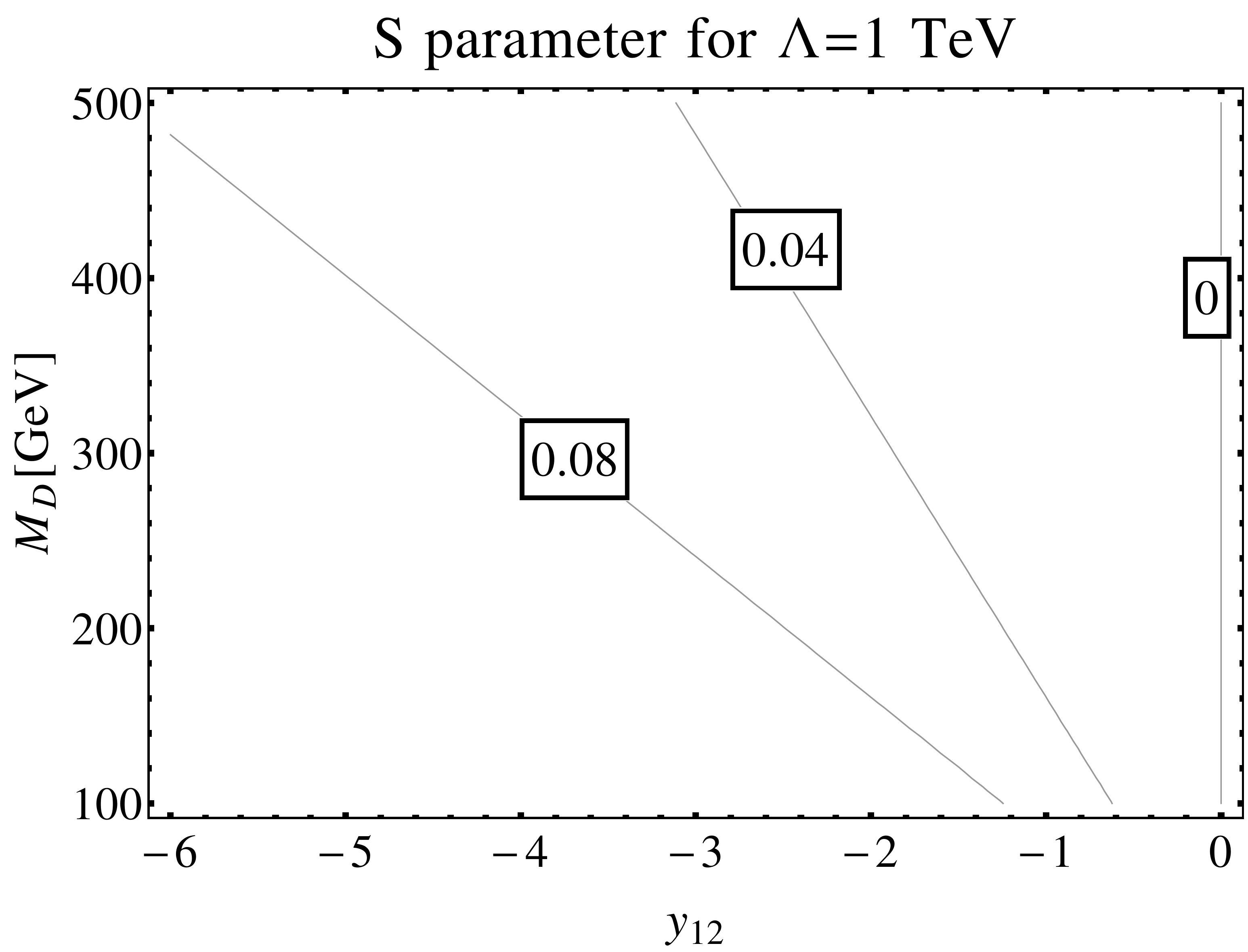}
  \caption{\em Contours of the  $S$-parameter on $y_{12} - M_D$ plane  for $\Lambda = 1 \;\mathrm{TeV}$.}
     \label{STU-plot}
  \end{figure}
%%%%%%%%%%%%%%%%%%%%%%%%%%%%%%%%%%%
%
Experimentally, $S,T$ and $U$-parameters fit the electroweak data  for $U=0$ with values~\cite{Agashe:2014kda}:
%%%%%%%%%%%%%%%%%%%%%%%%
\begin{align}
S = 0.00 \pm 0.08 \;,  \qquad
T = 0.05 \pm 0.07 
\; .
\label{eq:STU-exp}
\end{align}
%%%%%%%%%%%%%%%%%%%%%%%%%%%%%
%The numerical result  for the $S$-parameter is displayed   in Fig.~\ref{STU-plot}. 
In Fig.~\ref{STU-plot}, we present a contour plot for   the $S$-parameter  obtained  from \eqref{STU} as a function of $y_{12}$ and $M_{D}$ 
for $\Lambda=1$ TeV. As expected, stronger (1$\sigma$) bounds from \eq{eq:STU-exp} 
are obtained in the region $M_{D} \approx 100$ GeV,  where  it must be $|y_{12}| \lesssim 1$.
On the other hand, relaxed bounds on $|y_{12}|$ are obtained for higher values of $M_{D}$ 
and/or $\Lambda$.

Apparently  the result of \eq{eq:S}, 
does  not interfere with the bounds discussed before  for  the cases $(b)$ and $(c)$, 
since the allowed values of $y_{12}$,  obtained from \eqref{eq:STU-exp},   are  equivalent to  
those  obtained by the combination of the DM direct searches,   the   $ h\to \gamma \gamma$ decay and LEP $\chi^\pm$ bounds.
 On the contrary, in case (a) where $y=y_{12}$,  the bounds on $y$ arise only from the $S-$parameter.
%since the other bounds do not depend on $y$,
%and depends on $M_D$ as shown in the Fig. above.

%%%%%%%%%%%%%%%%%%%%%%%%%%%
\section{Cosmological and astrophysical constraints}
\setcounter{equation}{0}
\label{sec:astro}
%%%%%%%%%%%%%%%%%%%%%%%%%%

%As discussed in section  \ref{Spectrum-section}, there is in this model a particle, $\chi_1^0$,   which is neutral and stable. 
%It may be a component  of  DM that is  currently subject to the indirect DM experimental bounds.
% 
In the context  of this model,  it is essential to calculate  
 the DM relic density $\Omega h^2$ of the dark fermion  $\chi_1^0$,  in order to impose the 
 cosmological constraint related to 
the Planck satellite measurements~\cite{Agashe:2014kda}, as expressed  in  \eq{Planck}. 
Assuming that $\chi_1^0$ constitutes  the   DM of  the universe,
 we are able to set severe constraints on the parameters of eq.~\eqref{params}, in conjunction 
 to those found previously in section~\ref{Earth constraints}. 
From  now on we focus on benchmark cases (b) and (c) mainly because there is more freedom 
move around the parameter space as compared to cases (a) and (d).

In this section we describe briefly the freeze-out mechanism and discuss the solution of the Boltzmann equation.
Afterwords,  we present general, analytical,  predictions for  $\Omega h^2$, aiming  to understand its  dependences,
and then  numerical solutions are discussed.
Additionally, we  study the  constraints imposed by the  gamma  fluxes  produced 
by DM annihilations in the galactic center (GC)~\cite{Ackermann:2013uma,Ackermann:2015lka} and in
various   dwarf spheroidal  satellite galaxies (dSph)~\cite{Ackermann:2015zua}. Finally, at the end of this section,
we briefly discuss neutrino fluxes from the Sun, which are constrained from IceCube experiment~\cite{Aartsen:2016exj, Aartsen:2012kia}.

\subsection{Dark Matter relic abundance}\label{relic-intro}

The  conventional  way to produce non-relativistic (cold) DM relic particle abundance,  
is the so called freeze-out mechanism~\cite{Lee:1977ua,Hut:1977zn}. 
Although this mechanism is 
well reviewed in the literature\cite{Hooper:2009zm, Jungman:1995df,Kolb:1990vq,Dodelson:2003ft,Weinberg:2008zzc,Lisanti:2016jxe}, 
it would be helpful to outline  the  main steps here.

In the early universe, when the temperature was much higher than $M_D$, the would-be DM particles were in equilibrium, which means that it was equally 
possible to create and destroy pairs  of them due to  the $Z_2$-symmetry. 
As temperature of the universe was  dropping, the thermal 
production of DM pairs  became inefficient. 
Thus, $\chi_{1}^{0}$ pairs started to annihilate into lighter SM particles. As the number of these would-be DM 
particles was dropping, it became increasingly rare for them to interact with each other and annihilate. This yielded   an almost  constant number density 
 of $\chi_{1}^{0}$ particles,  which corresponds to  DM relic density  observed today.

Assuming  that $\chi_{1}^{0}$  is the lighter  particle of  the dark sector, 
one can evaluate 
 the relic density accurately\footnote{Extensive discussion on the solution
of the Boltzmann equation including coannihilation effects can be found  in \cite{Edsjo:1997bg}.} 
by solving  the corresponding   Boltzmann equation: 
\begin{equation}
 \frac{d n_{\chi_{1}^{0}}}{dt}+3H n_{\chi_{1}^{0}}=-\vev{\sigma v_{rel}}\left(n_{\chi_{1}^{0}}^2 -n_{\chi_{1}^{0}}^{(eq)2} \right) \label{Boltz-eq},
\end{equation}
where $H$ is the Hubble parameter defined as
\begin{equation}
H\equiv\dfrac{\dot{\alpha}(t)}{\alpha(t)}\;,
\end{equation}
and $\alpha(t)$ is the cosmic scale factor. Also $n_{\chi_{1}^{0}}$ is the WIMP number density and $n_{\chi_{1}^{0}}^{(eq)}$ is the corresponding quantity in  equilibrium
\begin{equation}
 n_{\chi_{1}^{0}}^{(eq)} \equiv g \left( \frac{m_{\chi_{1}^{0}} T}{2 \pi}\right)^{3/2} e^{-x}\;, \qquad x \equiv\frac{m_{\chi_{1}^{0}}}{T} , \label{neq}
\end{equation}
%%%
where $g$ is the number of the internal degrees of freedom of a particle, 
 $\vev{\sigma v_{rel}}$ is the thermal average of the total annihilation cross-section
of the WIMP to all allowed particles $(k,l)$, multiplied by the relative velocity of the incoming particles, which is usually expanded as 
%%% 
\begin{equation}
\vev{\sigma v_{rel}}=\displaystyle\sum_{k,l} \, \vev{\sigma_{\chi_{1}^{0}\chi_{1}^{0} \to k,l} v_{rel}} \ = \ a +b \, \vev{v_{rel}^{2}} \ + \  ... 
\label{csxv-non_rel_exp}
\end{equation}
%%%
It should be noted, that the second term on  the r.h.s. of  \eq{Boltz-eq} is responsible for creating  $\chi_{1}^{0}$-pairs, while
the first term   for   annihilating them.
According to our description above, at high temperatures, much higher than $m_{\chi_1^0}$,  the r.h.s  of \eq{Boltz-eq} vanishes. This 
results to a constant particle number  density since 
\begin{equation}
 \frac{d n_{\chi_{1}^{0}}}{dt}+3H n_{\chi_{1}^{0}} \ = \ \frac{1}{\alpha^{3}}\, \dfrac{d(\alpha^{3} n_{\chi_{1}^{0}})}{dt}=0 \;. 
\end{equation}
For lower  temperatures than  $m_{\chi_1^0}$, the term $\vev{\sigma v_{rel}}n_{\chi_{1}^{0}}^{(eq)2}$ in \eq{Boltz-eq} should vanish, since the 
WIMP pairs are not produced effectively [see \eq{neq}]. Then the Boltzmann equation  can be   approximated  as
\begin{equation}
  \frac{d n_{\chi_{1}^{0}}}{dt} \approx -\left( \vev{\sigma v_{rel}}\, n_{\chi_{1}^{0}} +3H \right) n_{\chi_{1}^{0}}  \;. \label{Boltz-eq2}
\end{equation}
The freeze-out temperature is defined as this 
where the annihilation rate becomes comparable to the expansion rate of the universe
\begin{equation}
 \vev{\sigma v_{rel}}\, n_{\chi_{1}^{0}} \approx H \;.
\end{equation}
The freeze-out temperature $T_f$
can be  evaluated   iteratively, through 
\begin{equation}
 x_{f}=\log \left[ c(c + 2) \sqrt{\dfrac{45}{8}} \frac{m_{ \chi_{1}^{0} }M_{P} \left(a+6b/x_{f}\right)}{g^{1/2}_{\star} x_{f}^{1/2}}  \right], \label{x_fo}
\end{equation}
%%%%%
where $x_{f}\equiv m_{\chi_1^0}/T_f$. The parameter $c$  is usually  chosen  $c \sim 0.5$,  to get into 
agreement with precise numerical solutions of the Boltzmann equation. Furthermore, $M_{P} \approx 2.435 \times 10^{18} \; \GeV$ is the Planck scale,  and
$g_{\star}$ counts  the relativistic degrees of freedom of the Standard Model at $T_{f}= {m_{ \chi_{1}^{0} }}/{x_{f}}$. 
It turns out that   $x_{f} \simeq 25$.
Calculating  the freeze-out temperature, one can solve the Boltzmann equation 
and find the  present WIMP relic density  
%%%%%
\begin{equation}
 \Omega h^2 \ \approx \ \frac{1.04 \times 10^9 \; \GeV ^{-1}}{M_{P}}\; \frac{x_{f}}{g^{1/2}_{\star}(a+3\,  b \, x_{f}^{-1} )} \;.
 \label{relic-approx}
\end{equation}
%%%%%
For a WIMP mass at the electroweak scale, 
this formula becomes approximately  $\Omega h^2 \approx 0.1 \frac{10^{-8} \; \mathrm{GeV}^{-2}}{a+3 \, b \, x_{f}^{-1}}$. 
From \eq{Planck} we get  $\Omega h^2 \sim 0.1$, so the required cross-section is of order 
$ 10^{-8} \; \mathrm{GeV}^{-2}$ for $a=\mathcal{O}(  \mathrm{GeV}^{-2})$,
which is a typical EW cross section.
%, $\sim \alpha^2/(100 \; \mathrm{GeV})^2 \sim 10^{-8} \; \mathrm{GeV}^{-2}$  with $\alpha$ being the fine structure constant . 

If  other particles  are almost degenerate with
 WIMP, then there could be extra contributions (coannihilation effects) to the total annihilation cross-section due  to them. 
Thus,  the annihilation cross-section modified  in order to incorporate  these  coannihilation effects~\cite{Griest:1990kh}.
Following \cite{Griest:1990kh,Hooper:2009zm}, this change is
\begin{equation}
\displaystyle\sum_{k,l}\sigma_{\chi_{1}^{0}\chi_{1}^{0} \to k,l} \rightarrow \sigma_{eff}=\displaystyle\sum_{k,l}  \; \displaystyle\sum_{i,j}\sigma_{i,j \to k,l} \; 
\dfrac{g_{i} g_{j}}{g^{2}_{eff}(x)}(1+\Delta_{i})^{3/2} \;  (1+\Delta_{j})^{3/2}  \; e^{-x (\Delta_{i}+\Delta_{j})} \;,
\end{equation}
where indices $i,j$ run over all the co-annihilating particles with  $\Delta_{i}=\dfrac{m_{i}-m_{\chi_{1}^{0}}}{m_{\chi_{1}^{0}}} \lesssim 0.1$
and $g_{eff}(x)$ is defined as
%%%%%
\begin{equation}
g_{eff}(x) \equiv  \displaystyle\sum_{i}g_{i} \, (1+\Delta_{i})^{3/2}  \; e^{-x \Delta_{i}} \;.
\end{equation}
%%%%%
Such coannihilation effects,  and other possible  contributions  to the relic abundance~\cite{Griest:1990kh},  have been included
in our numerical analysis  described  in the following.

\subsection{A close look at the relic density}\label{Close_Look-Section}
Before  discussing  the   bounds imposed by the data on $ \Omega h^2$, it would be helpful to study the 
numerical values of the annihilation cross-section that are used to calculate  the relic abundance.  
As discussed in section \ref{sect:htogg}, if   $M_D \gtrsim 90$ GeV, then  the coupling to the Higgs boson is approximately  zero.
Therefore,  the most important annihilation channels,  assuming for the time being  that coannihilation effects  are irrelevant, 
are $\chi_{1}^{0} \chi_{1}^{0} \to W^{+}W^{-}$, $ZZ$, $ \gamma Z $ and $ \gamma \gamma$.
There are no   final states with fermions, since their corresponding interaction  vertices  are absent.
 There are no $\chi_{1}^{0} \chi_{1}^{0}Z / \gamma$
terms in the Lagrangian of \eqs{Zxx}{diag_dipole-3point},
 or they are restricted because of bounds by  direct detection experiments $Y^{h\chi_{1}^{0} \chi_{1}^{0}} \approx 0$.

Keeping  only the first term in  the expansion of  \eq{csxv-non_rel_exp} we obtain
\begin{equation}
  a_{VV} = \frac{  \beta^{3/2}_{V}\, m_{\chi_{1}^{0}}^2  }{32\pi \, S_V\,  v^2 }\, \frac{\left [g^2 v^4- 4\, g \, v^2  \omega \, K_{V} \left(m_{\chi_{1}^{0}} + m_{\chi}  \right) 
  +4\, K_{V}^2\, \omega^2 \left( 2m_{\chi_{1}^{0}}  m_{\chi} +M^2_V \right) \right ]^2 }{v^6 \left( m_{\chi_{1}^{0}}^2 + m_{\chi}^2 -M^2_V \right)^2  }, \label{avv}
\end{equation}
where $V$  denotes $W$ and $Z$ gauge bosons  in the final states for the processes $\chi_{1}^{0} \chi_{1}^{0} \to W^{+}W^{-}$ or 
$\chi_{1}^{0} \chi_{1}^{0} \to ZZ $. Also, we abbreviate, $\beta_V \equiv 1- {M_V^2}/{m_{\chi_{1}^{0}}^2}$, $K_{W} \equiv d_W$, 
$S_W \equiv 1$, $K_{Z}\equiv c_{W} \, (c_{W} d_W+s_{W} d_{\gamma})$ and $S_Z\equiv 2\,c_{W}^{4}$. 
 The mass $m_{\chi}$ denotes  $m_{\chi^{\pm}}$ for $V=W$ and
$m_{\chi_{2}^{0}}$ for $V=Z$.

For the   channels $\gamma Z$ and $\gamma \gamma$, we find
\begin{subequations}
 \begin{align}
  a_{\gamma Z} & = \frac{\beta_{\gamma Z}^3\, m_{\chi_{1}^{0}}^2 }{2 \pi\, c_{W}^2\,  v^2} 
  \frac{C_{\gamma}^2 \, \omega^2\, \left [ g\,v^2\left(m_{\chi_{1}^{0}} + m_{\chi_{2}^{0}}  \right) 
      - \omega\,K_{Z} \left( 4m_{\chi_{1}^{0}}  m_{\chi_{2}^{0}} +M^2_Z \right)  
        \right ]^2}{v^6 \left[ 2\left(m_{\chi_{1}^{0}}^2 + m_{\chi_{2}^{0}}^2\right) -M^2_Z \right]^2 } \, , \label{agammaz}\\
  a_{\gamma \gamma} & =\frac{m_{\chi_{1}^{0}}^4\, m_{\chi_{2}^{0}}^2\, \omega^4\, C_{\gamma}^4}{\pi \,(m_{\chi_{1}^{0}}^2 + m_{\chi_{2}^{0}}^2)^2\, v^8} \,  ,\label{agammagamma}
  \end{align}
\end{subequations}
%%%
with $\beta_{\gamma Z} \equiv 1- {M_Z^2}/{4m_{\chi_{1}^{0}}^2}$ and $C_{\gamma}\equiv (c_{W} d_{\gamma}-s_{W} d_W)$. 
These  channels $\gamma\gamma$ and  $\gamma Z$,   contribute  to   the monochromatic  gamma fluxes  
from  the GC. 
Thus,   in conjunction to the corresponding  bounds  from Fermi-LAT experiment,  one gets  severe constraints for the coupling $C_{\gamma}$.  
Due to absence of $\chi_{1}^{0}$ couplings  to $Z$  and   $\gamma$ and the nearly vanishing  Higgs mediated  $\hat{s}$-channel, 
all the above processes arise  from  $\hat{t}$ and $\hat{u}$  channels.

\Eqsss{avv}{agammaz}{agammagamma},  contain one or more solutions with respect to $d_W$.
This means that $d_W$ could act as a regulator that minimizes the total annihilation cross-section as the 
(required) low mass $M_D$ tends to amplify it 
(generally the cross section scales as $M_{D}^{-2}$ if we ignore magnetic dipole interactions). This minimization,  will be  proved  essential 
when trying to obtain  cosmologically acceptable relic abundance  at the electroweak scale.

Qualitatively, concerning the minimum of the {\it total} annihilation cross-section as a function of the dipole couplings
one anticipates that each cross-section should be minimized for almost the 
same value of $d_W$, in order for the {\it total} annihilation cross-section to be at its minimum.
In addition, $d_{\gamma}\approx \frac{s_{W}}{c_{W}}  d_W$  so that $C_{\gamma}$ is quite small.
This keeps $d_{\gamma}$ from obtaining large negative values, because $a_{WW}$ can be minimized only for $d_W>0$.

%%%%%%%%%%%%%%%%%%%%%
 \begin{figure}[t!]
  \centering
 %    \hspace{-1.2 cm}
    \includegraphics[width=0.78\linewidth]{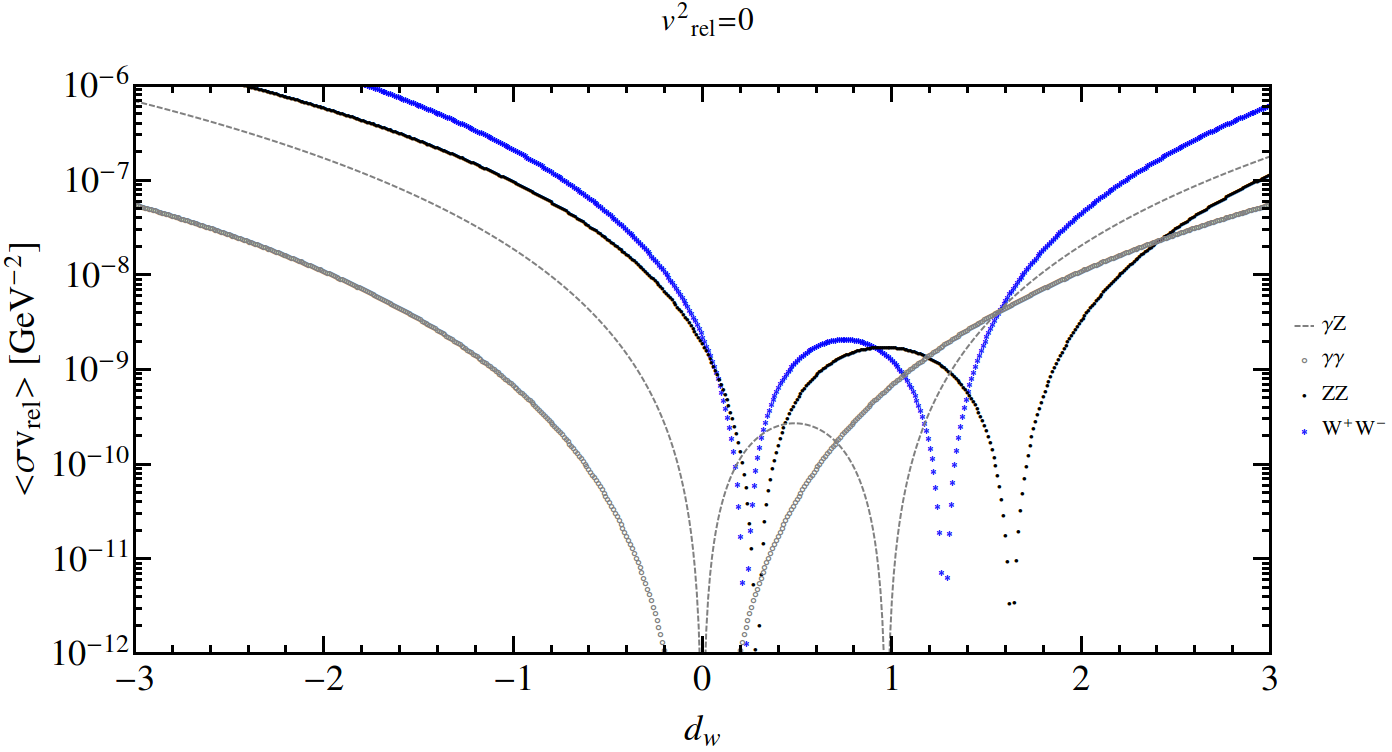}
  \caption{\em The dependence of different annihilation channels on $d_W$ for $M_D =400  \GeV$, $\Lambda =1  \TeV$, 
  $y=-y_{12}=-\frac{\xi_{12}}{2}=-0.8$ and $d_{\gamma}=0$.
  Notice that, in a certain range of $d_W$ values, there is at least one dip  for each channel cross section.}
     \label{cross_sections_Vs_dw-a}
  \end{figure}
  %%%%%%%%%%%%%%%%%%%%%%%%%%%%%

%%%%%%%%%%%%%%%%%%%%%%%%%%%
\begin{figure}[H]
\hspace*{-0.0cm} \begin{subfigure}[b]{0.52\textwidth}
        \includegraphics[width=\textwidth]{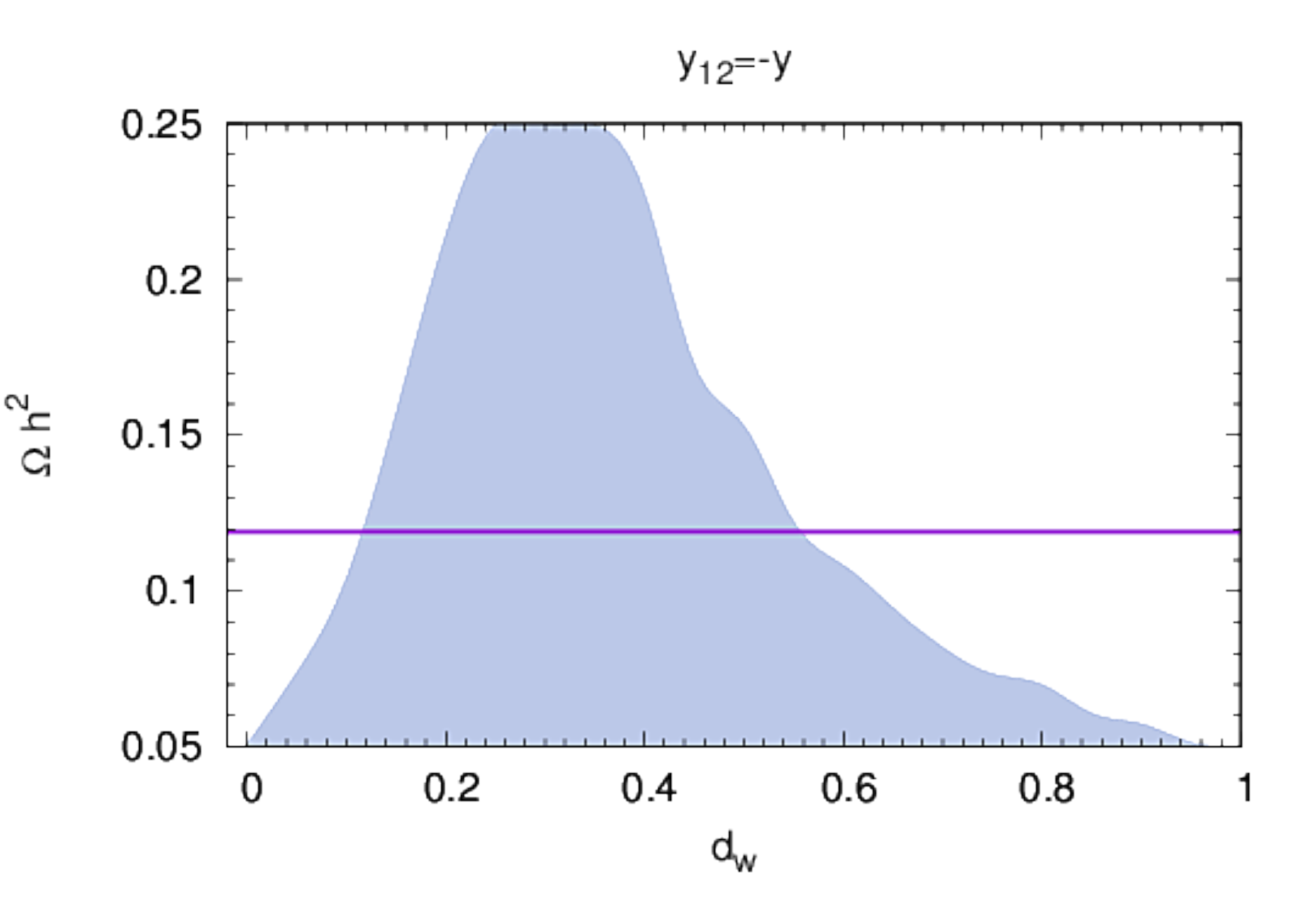}
            \caption{}
        \label{dwvsrelic}
    \end{subfigure}
 \begin{subfigure}[b]{0.52\textwidth}
        \includegraphics[width=\textwidth]{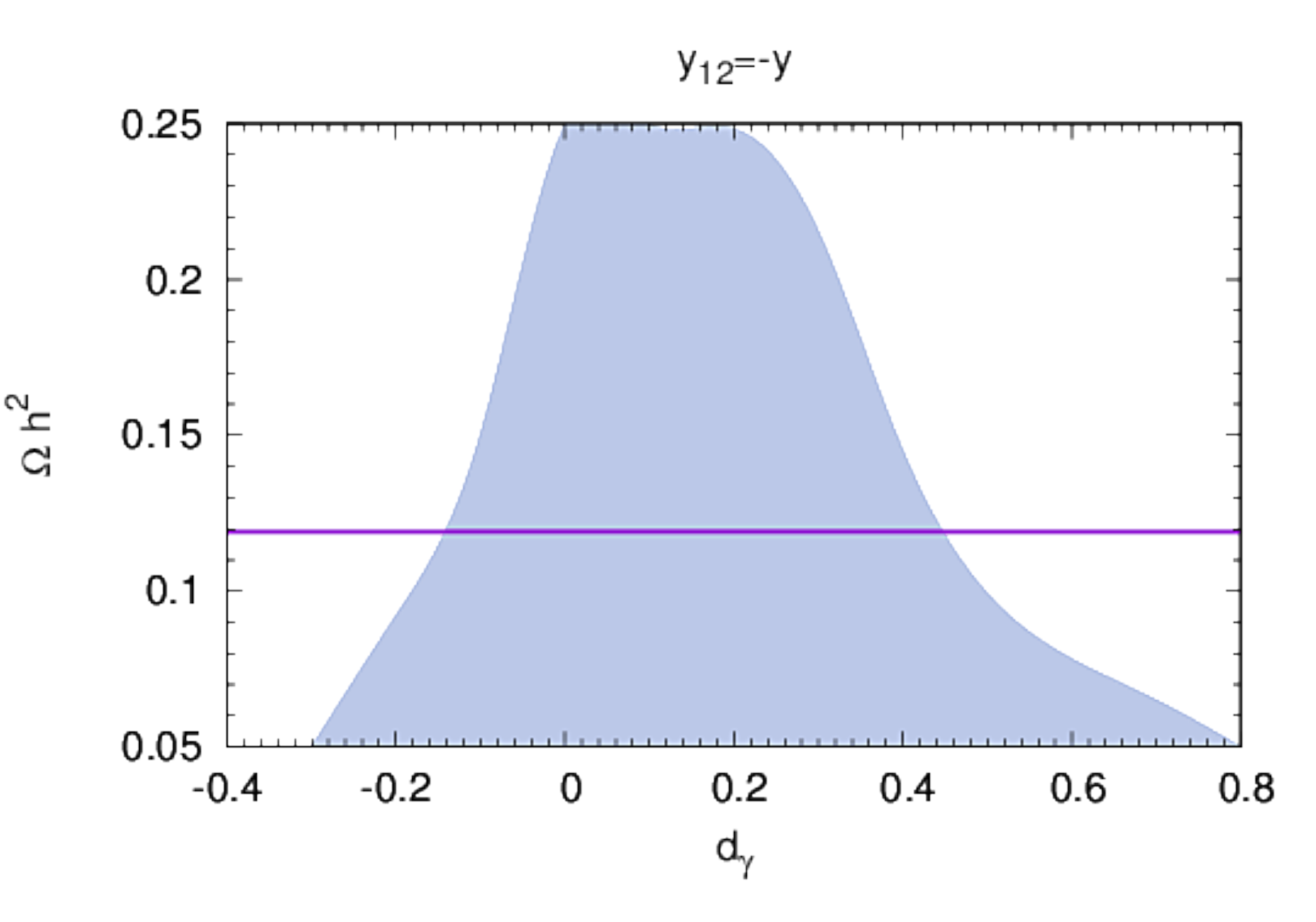}
            \caption{}
         \label{dgvsrelic}
    \end{subfigure} 
\vspace*{-1.5cm}     \begin{center}
 \begin{subfigure}[b]{0.52\textwidth}
        \includegraphics[width=\textwidth]{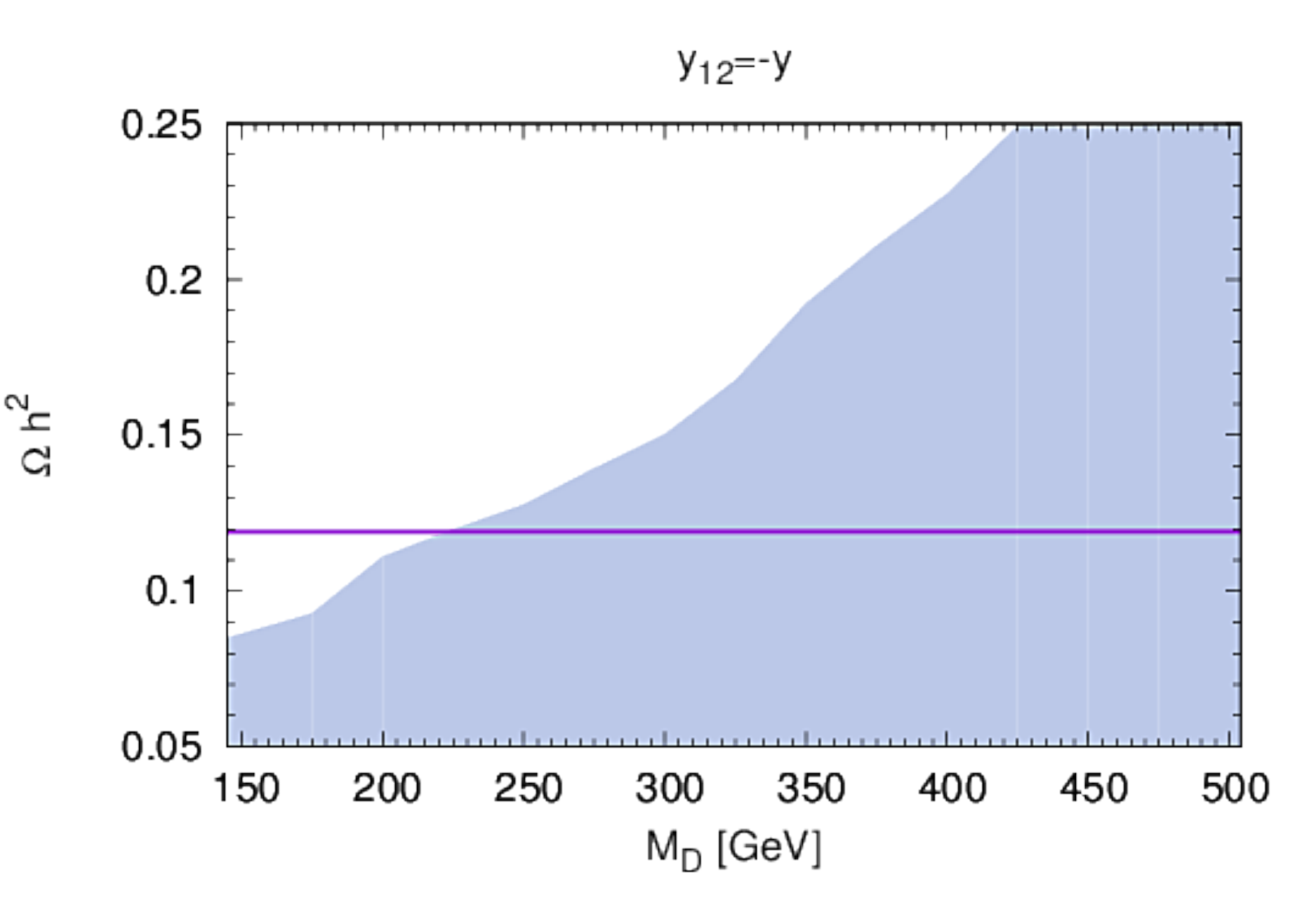}
            \caption{}
         \label{MDvsrelic}
    \end{subfigure}  \end{center}
%  \begin{subfigure}[b]{0.6\textwidth}
%         \includegraphics[width=\textwidth]{y12=-y_y1vsrelic.png}
%             \caption{}
%          \label{yvsrelic_y12=-y}
%     \end{subfigure}    
\caption{\em Relic abundance dependence  on the parameters
 $(a) \, d_W$, $(b) \, d_{\gamma}$, $\text {and } (c) \, M_D$,
for $\Lambda=1 \TeV$ and $y_{12}=-y$.
The   cosmologically  allowed (shaded) region  corresponds to the  variation of the other parameters  in \eqref{params} not shown in the plot. 
   The horizontal line stands for $\Omega h^2 =0.12$.}
\label{Omega_dependence}
\end{figure}
%%%%%%%%%%%%%%%%%%%%%%%%%%%%%%

A numerical example is shown in Fig.~\ref{cross_sections_Vs_dw-a}. 
We observe that there are two minima for the annihilation cross-sections to $ZZ$, $W^{+}W^{-}$ 
and $\gamma Z$ and one minimum for $\gamma \gamma$. 
The first minimum of $a_{ZZ}$ and $a_{WW}$ coincides with the vanishing point of $C_{\gamma}$, which
gives small cross-sections for $\chi_{1}^{0}\chi_{1}^{0} \to \gamma \gamma \text{ and } \gamma Z$. 
On the other hand, the second minimum of $a_{ZZ}$ and $a_{WW}$ is in a 
region where the annihilation to $\gamma\gamma$ and $\gamma Z$ blows  up.
Furthermore, for negative $d_W$, there are no such minima and, as can be seen from Fig.~\ref{cross_sections_Vs_dw-a}, every cross-section becomes quite  large.  
 
Since \eq{relic-approx} is an approximation
 which could lead to an error up to  $ \sim 10 \%$ (as discussed in \Ref{Griest:1990kh}),   the Boltzmann
equation must be solved  numerically.
To do this we implement the $d=4$ and $d=5$ operators  to the computer program microOMEGAs\cite{Belanger:2014hqa}  via the  LanHEP\cite{Semenov:1998eb} package\footnote{More information 
about these packages can be found in https://lapth.cnrs.fr/micromegas/ and  http://theory.sinp.msu.ru/$\sim$semenov/lanhep.html. } in order to obtain more accurate 
results for the relic abundance.

In Figs.~\ref{dwvsrelic}, \ref{dgvsrelic} and \ref{MDvsrelic}
we  examine  the dependence  of the relic abundance $\Omega h^2$ on the parameters, $d_{W}$, $d_{\gamma}$ and $M_{D}$,  respectively. 
Because all parameters in \eqref{params}, run freely, the corresponding  plots are given as shaded areas in   Fig.~\ref{Omega_dependence}.
We remark that:
{\it a)} The  minimization  effects  on the various cross-sections  discussed before, are evident  in the  numerical results too. 
{\it b)} As expected, when $M_D$ increases,   $\Omega h^2$   increases too.
{\it c)}  For acceptable $\Omega h^2 $ and  $M_D$ of a few hundred $\GeV$,  $d_W$ must lie  in the region $0.1 \lesssim d_W \lesssim 0.5$, which does not include the zero node.
The   dipole moment to photon $d_\gamma$ should be in the region $-0.2 \lesssim d_{\gamma} \lesssim 0.5$, which includes the zero node.     
{\it d)} The minimization of the total annihilation cross-section, is not enough to produce the observed DM density for $M_D \lesssim 200 \, \mathrm{GeV}$. 
%
%%%%%%%%%%%%%%%%%%%%%
 \begin{figure}[H]
  \centering
%  \hspace{-1.2 cm}
    \includegraphics[width=0.70\linewidth]{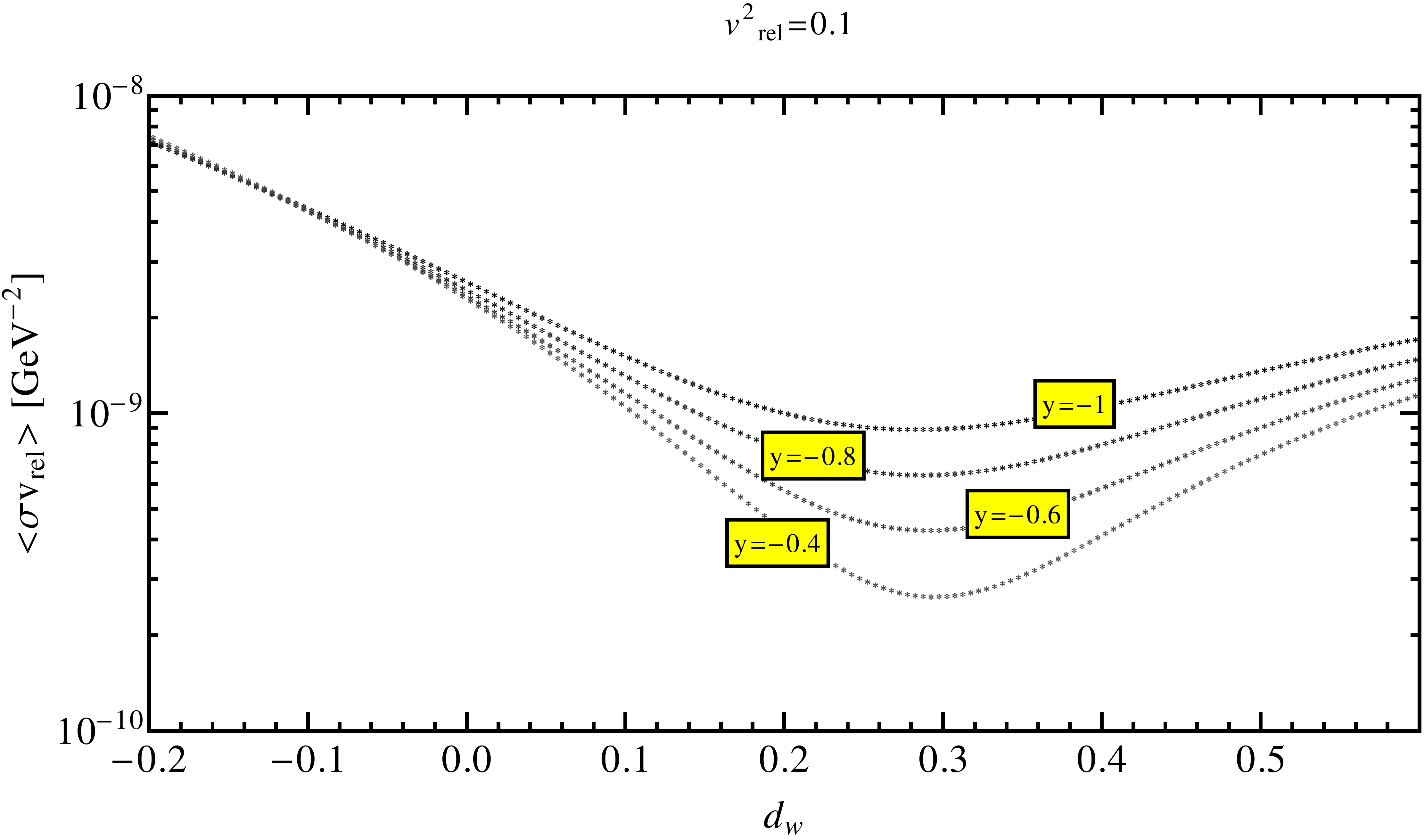}
\caption{\em $ZZ$ annihilation cross section dependence on $d_W$ for different values of $y=-y_{12}=-\frac{\xi_{12}}{2}$, 
$\Lambda =1  \TeV$, $M_D =400  \GeV$, $d_{\gamma}=0$ and $v_{\mathrm{rel}}^2 =0.1$. 
At the minimum, the values of the cross-section  decreases as we lower the values of $|y|$. 
The behaviour of  the $W^{+}W^{-}$ annihilation channel is similar. 
}
     \label{csxvab-dw-MD=400_y}
  \end{figure}
  %%%%%%%%%%%%%%%%%%%%%%%%%%%%%
%
%
%%%%%%%%%%%%%%%%%%
\begin{figure}[H]
\hspace*{-0.3cm} \begin{subfigure}[b]{0.52\textwidth}
        \includegraphics[width=\textwidth]{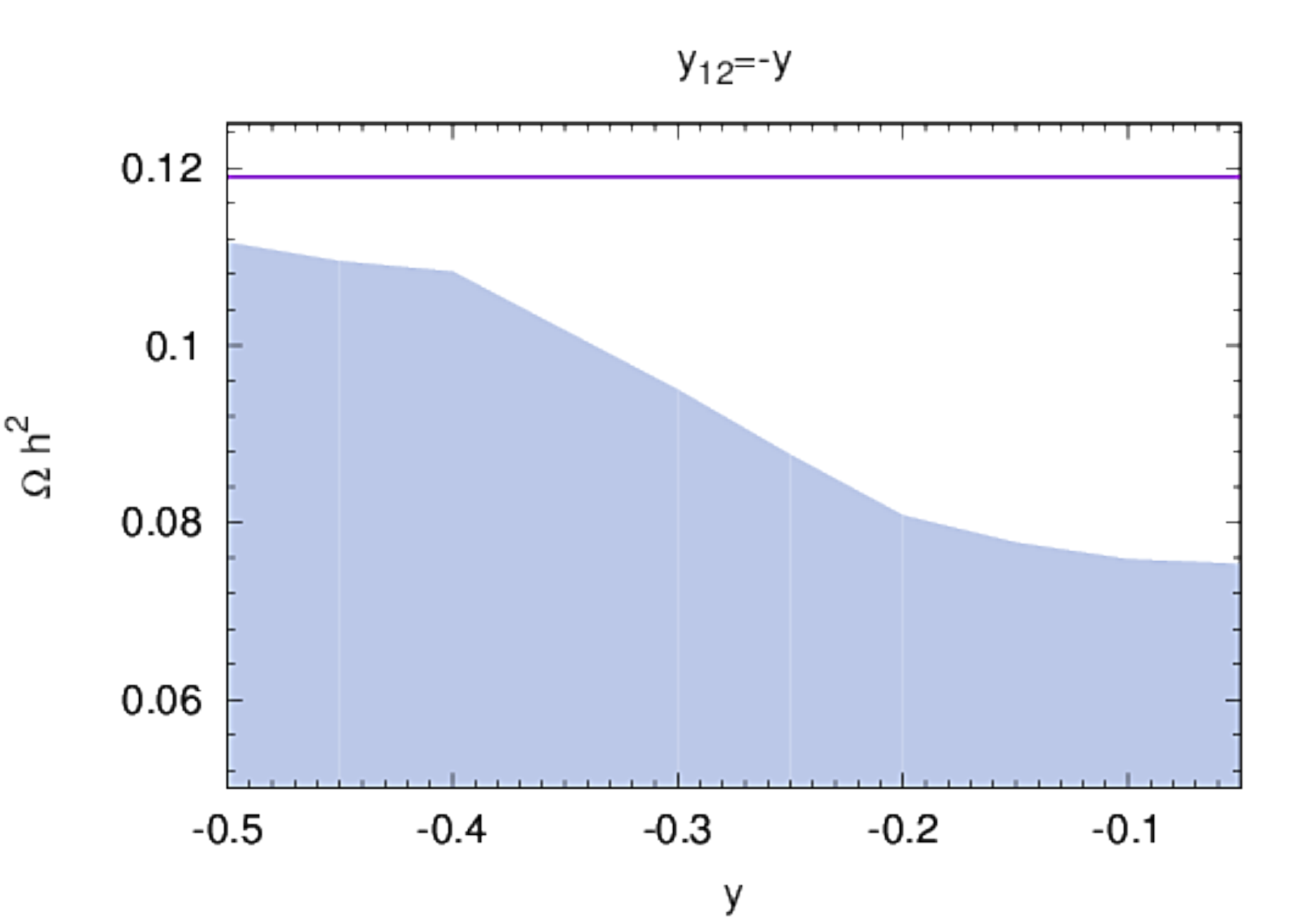}
            \caption{}
        \label{yvsrelic_MD=200GeV}
    \end{subfigure}
 \begin{subfigure}[b]{0.52\textwidth}
        \includegraphics[width=\textwidth]{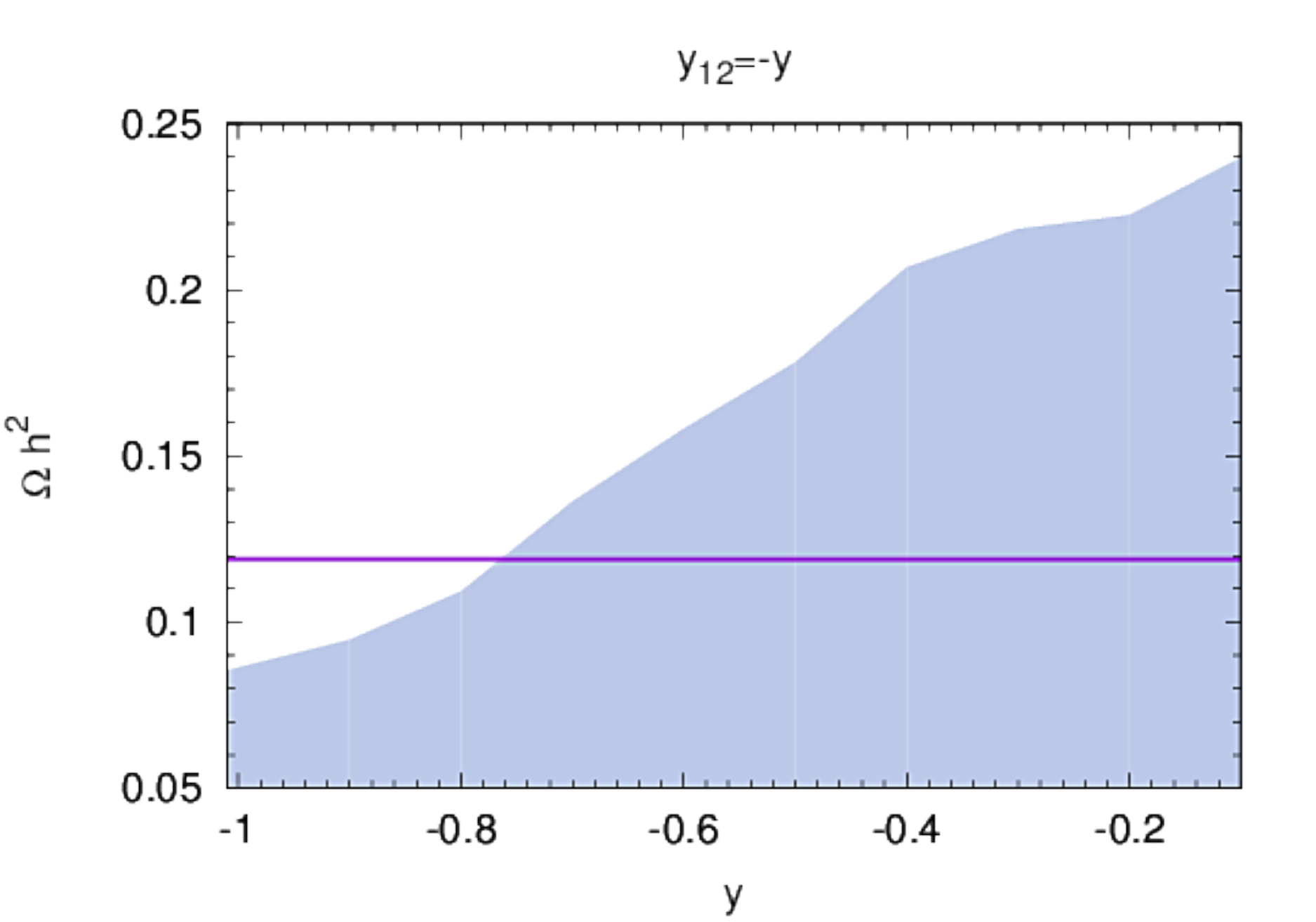}
            \caption{}
         \label{yvsrelic_MD=400GeV}
    \end{subfigure}   
\caption{\em $y$  vs. $\Omega h^2$, for $\Lambda =1 \; \TeV$ and  $(a) \, M_D =  200 \; \GeV$ and $  (b) \, M_D = 400 \; \GeV$.
 Other  parameters    from the list  \eqref{params}  vary  in the range  constrained from ``Earth" constraints and for  $y_{12}=-y$.
The dependence of the relic density on $y$ changes for different values of $M_D$.}
\label{Omega_dependence_y_MD=200|400}
\end{figure}
%%%%%%%%%%%%%%%%%%%%%%

The dependence of the relic density on the parameter $y$ is complicated due to  the following  competing  effects: 
The coannihilation channels,  increase the total annihilation cross-section as $|y|$ tends to zero, since  the
mass differences of the initial particles involved become smaller and smaller.
But, as shown in Fig.~\ref{csxvab-dw-MD=400_y}, 
the $b-$term in the expansion of \eq{csxv-non_rel_exp}, tends to decrease the value of the cross-sections (around the minimum), 
at least for the annihilation to $ZZ$ and $W^{+}W^{-}$.

Moreover, in Fig.~\ref{Omega_dependence_y_MD=200|400}  we study   the dependence of $\Omega h^2$ on $y$,  for various values of the mass $M_D$. 
In the region $M_D\lesssim 260 \;\GeV$,  the relic abundance becomes smaller for smaller $y$ (an example for $M_D =200 \; \GeV$ is shown in
Fig.\ref{yvsrelic_MD=200GeV}), which means that the coannihilation effects   dominate  over  the $b-$term, and vice-versa  for larger values of $M_D$ (Fig.\ref{yvsrelic_MD=400GeV}). 
%
%%%%%%%%%%%%%%%%%%%%%%%%%%%%%%%%%%%5
\begin{figure}[t]
\hspace*{-0.3cm} \begin{subfigure}[b]{0.52\textwidth}
        \includegraphics[width=\textwidth]{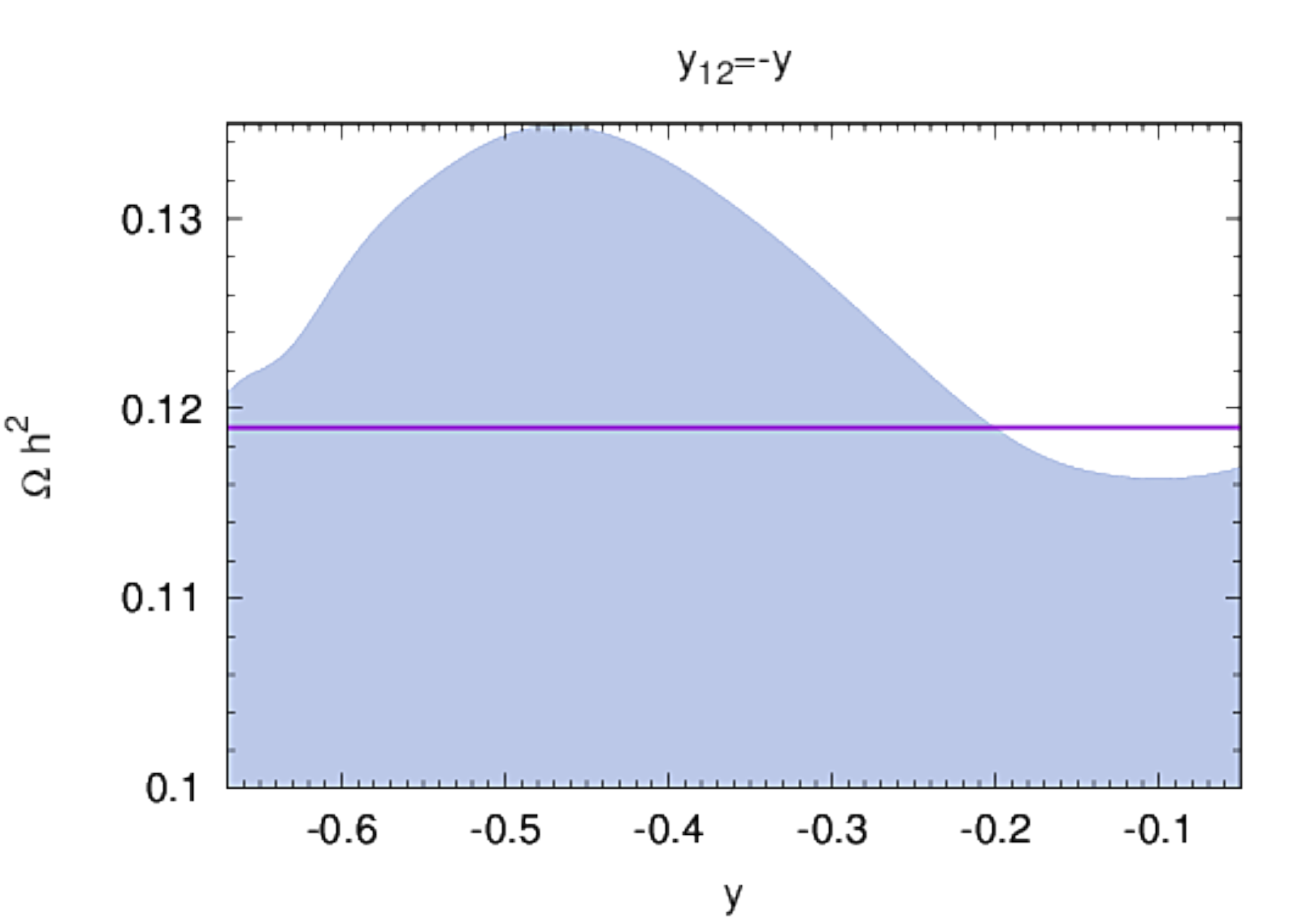}
            \caption{}
        \label{turning_point_y}
    \end{subfigure}
 \begin{subfigure}[b]{0.52\textwidth}
        \includegraphics[width=\textwidth]{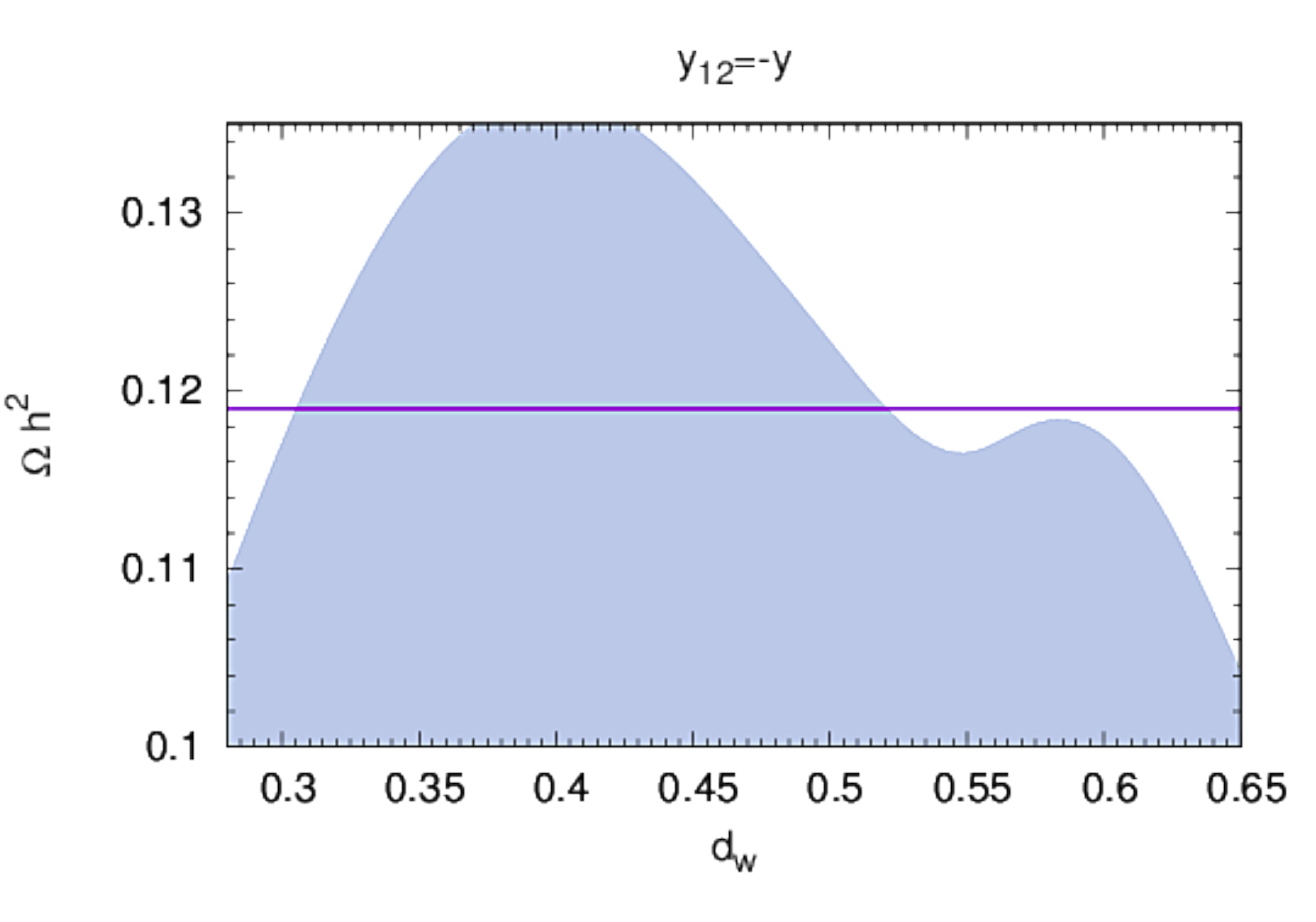}
            \caption{}
         \label{turning_point_dw}
    \end{subfigure}   
\caption{\em  Turning point as can be seen in (a) $y$ versus $\Omega h^2$ and (b) $d_W$ versus $\Omega h^2$, for $y_{12}=-y$, $\Lambda =1 \TeV$ and $M_D = 260 \; \GeV$. 
The shaded area and the curves are as in Fig.~\ref{Omega_dependence}.}
\label{turning_point}
\end{figure}
%%%%%%%%%%%%%%%%%%%%%%%%%%%%%%%%%%%%%
%%%%%%%%%%%%%%%%%%%%%%%%%%%%%%%%%%%%%
 \begin{figure}[t]
 \centering
    \includegraphics[width=.52\linewidth]{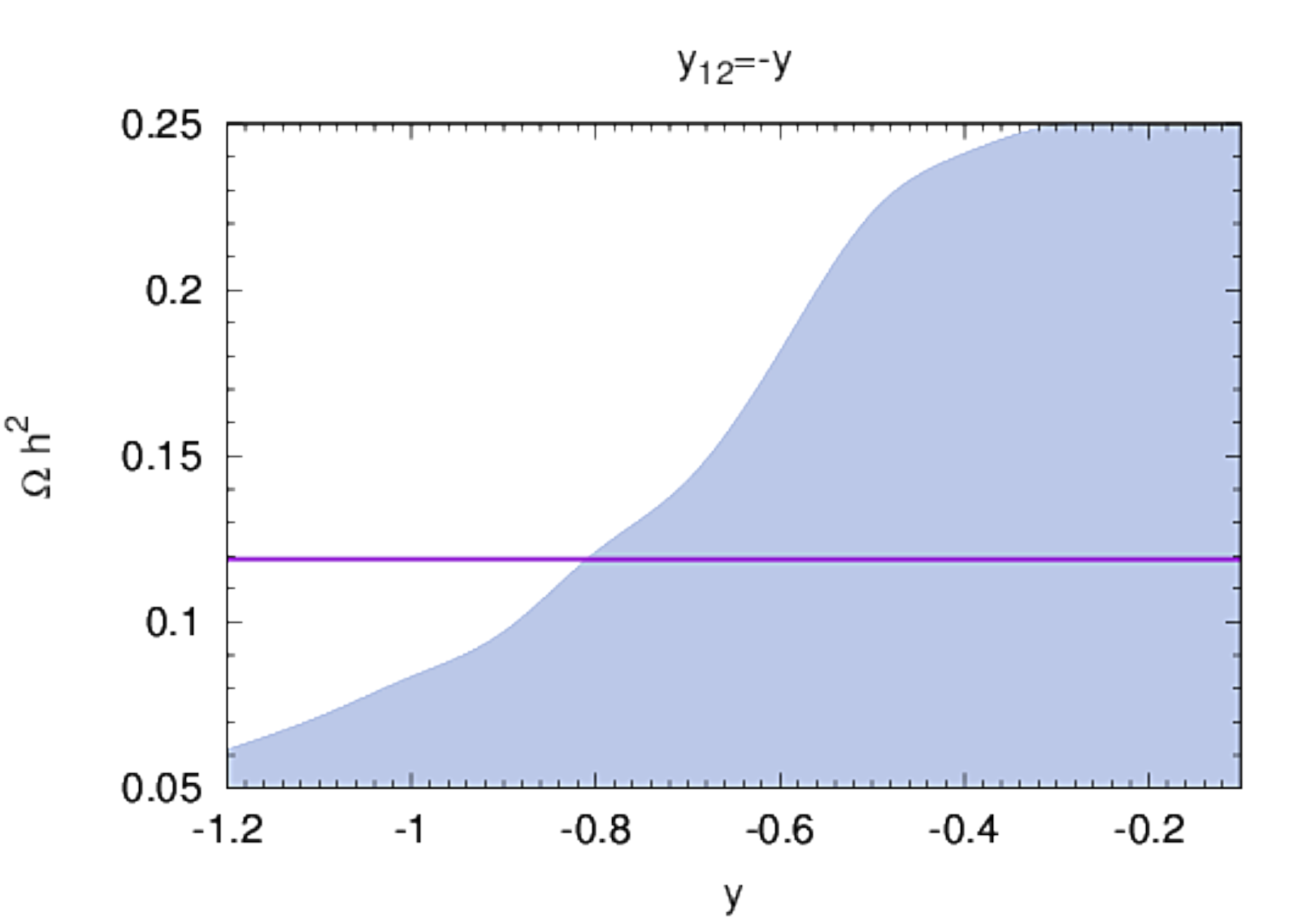}
  \caption{\em $\Omega h^2$ versus $y$, for $y_{12}=-y$, $\Lambda =1 \; \TeV$ and $M_D \leq 500 \; \GeV$. 
  The shaded area and the curves are as in Fig.~\ref{Omega_dependence}.}
     \label{Omega_dependence_y}
  \end{figure}
  %%%%%%%%%%%%%%%%%%%%%%%%%%%%%%%%%%%%%

There is a small region at $M_D \approx 260 \; \GeV$ where this dependence is mixed. We call this value of $M_D$ 
{\it ``turning point''}. 
An example of this behavior is shown in Fig.\ref{turning_point_y}. As we can see, the relic abundance rises until $y \sim -0.4$
and then decreases, but for $y\sim -0.06$ it starts to increase again. 
Also, as shown in Fig.\ref{turning_point_dw}, 
we obtain  two maxima for $\Omega h^2$ with respect to $d_W$,  as a result of this effect, since the value of
$d_W$ which minimizes the annihilation cross-section depends on $y$.

Although $y$ has no definite effect on  $\Omega h^2$,
% {\sak what then we discussed do far???}{\dmtr (by this we mean that the effect $y$ has on $\Omega h^2$ depends also on $M_D$)}
 the relic density  increases  as $M_D$ increases.  
Therefore, if we calculate the relic density in the  allowed parameter space, the 
dependence of the relic on $y$ would be dominated by its dependence for larger $M_D$.
In Fig.~\ref{Omega_dependence_y}, we show the dependence of $\Omega h^2$ on $y$. The relic density
decreases as $|y|$ becomes larger and for $|y| \gtrsim 0.9$ the DM becomes under-abundant.

Finally,   the case where $y_{12}=0$  yields similar  results with the case   $y_{12}=-y$ just discussed, as can be deduced from  
Figs.~\ref{Omega_dependence} and \ref{Omega_dependence_y}. 
Also, for other values of the cut-off scale $\Lambda$, the parameters $d_{\gamma , \, W}$ and $y$ should 
be rescaled in order for the ratios ${d_W}/{\Lambda}$, ${d_{\gamma}}/{\Lambda}$ and ${y}/{\Lambda}$ to remain unchanged.

\subsection{Cosmological constraints due to  relic density}

Having studied  the constraints from $LEP$, $R_{h\to\gamma\gamma}$, the direct detection DM experiments
as well as the Planck bound on  the relic density for this effective theory, we are able delineate the cosmologically acceptable regions 
of the parameter space.
For this reason,  we perform a combined scan in the  so far allowed parameter space 
which is also cosmologically preferred, for the cases $y_{12}=-y$ and $y_{12}=0$ at $\Lambda =1 \TeV$.
%%%%%%%%%%%%%%%%%%%%%%%%%%%%%%%%% 
 \begin{figure}[t!]
    \centering\vspace{-.3cm}
    \includegraphics[width=0.52\linewidth]{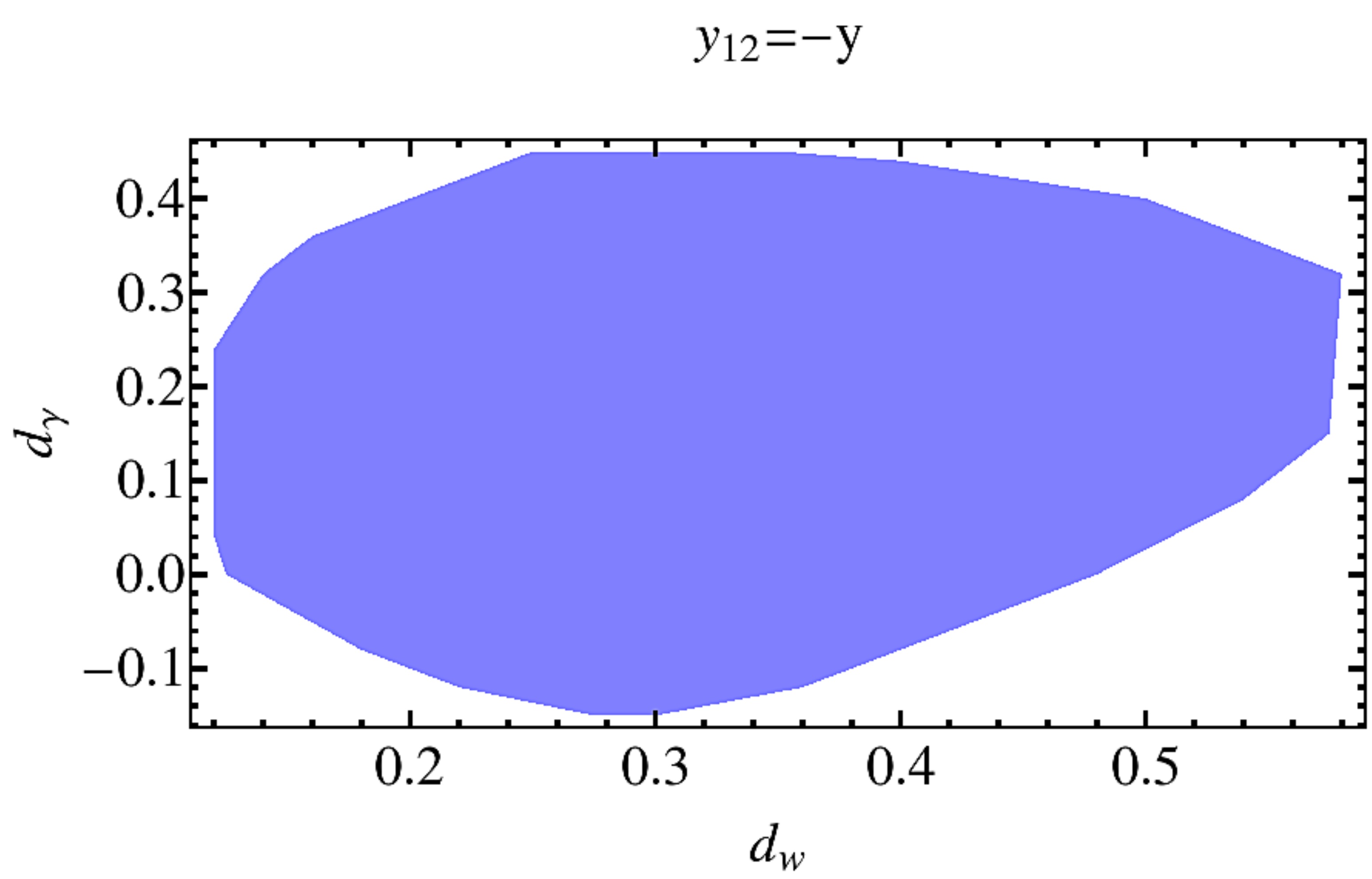}
           \caption{\em The plane $d_W - d_{\gamma}$ of the parameter space that gives
            the observable relic abundance, for $\Lambda =1 \TeV$ and $y_{12}=-y$. 
           The same region holds also for $y_{12}=0$. 
           We  allow variation of other  parameters in \eqref{params} consistently with observational data.}
    \label{dw-dgamma}
  \end{figure}
%%%%%%%%%%%%%%%%%%%%%%%%%%%%%%%%%%%
First, for $\Lambda =1\,\TeV$,  in Fig.~\ref{dw-dgamma} we display  the part of the  $d_{\gamma} - d_{W}$ plane,  
that is compatible to the DM relic density,  varying  all the other parameters, but 
 keeping $M_D \lesssim 500  \GeV$.  Apparently  the parameter $d_{W}$ is bounded to be
 positive in order to explain the DM relic abundance for a
 WIMP mass at electroweak scale. 
Also, the region where $d_{\gamma}$ is positive, is larger than the region where it is negative, 
a situation  explained in the preceding analysis. 
A similar  region is also found for  $y_{12}=-y$ and $y_{12}=0$. 
%%%%%%%%%%%%%%%%%%%%%%%%%%%%%%%%%%%%%%
 \begin{figure}[t]
% \centering
\hspace*{-0.3cm} \begin{subfigure}[b]{0.52\textwidth}
        \includegraphics[width=\textwidth]{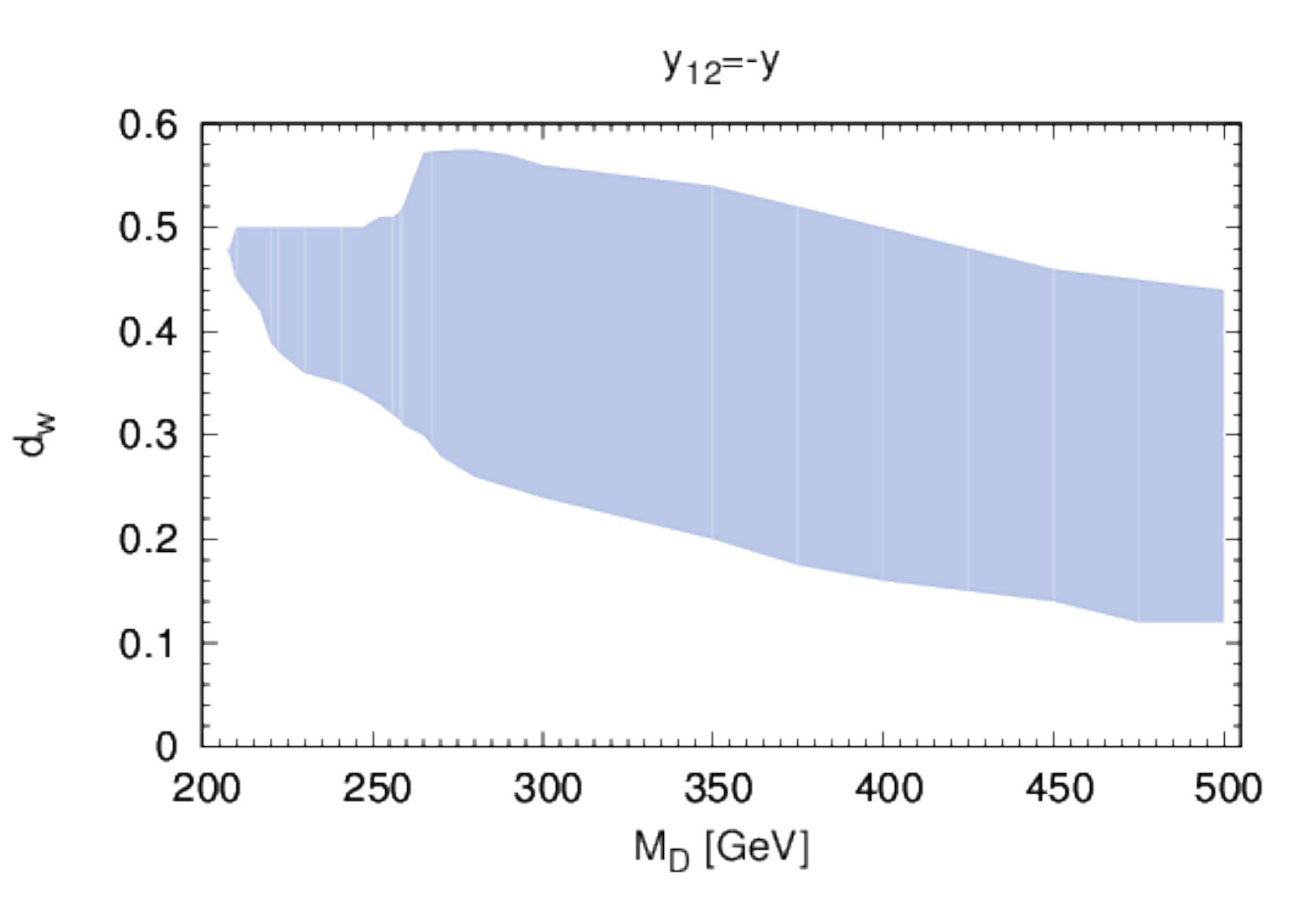}
           \caption{$M_D$ vs $d_W$, for $\Lambda =1 \; \TeV$ and $y_{12}=-y$.}
        \label{y12=-y_MDvsdw_1TeV}
    \end{subfigure}
\begin{subfigure}[b]{0.52\textwidth}
        \includegraphics[width=\textwidth]{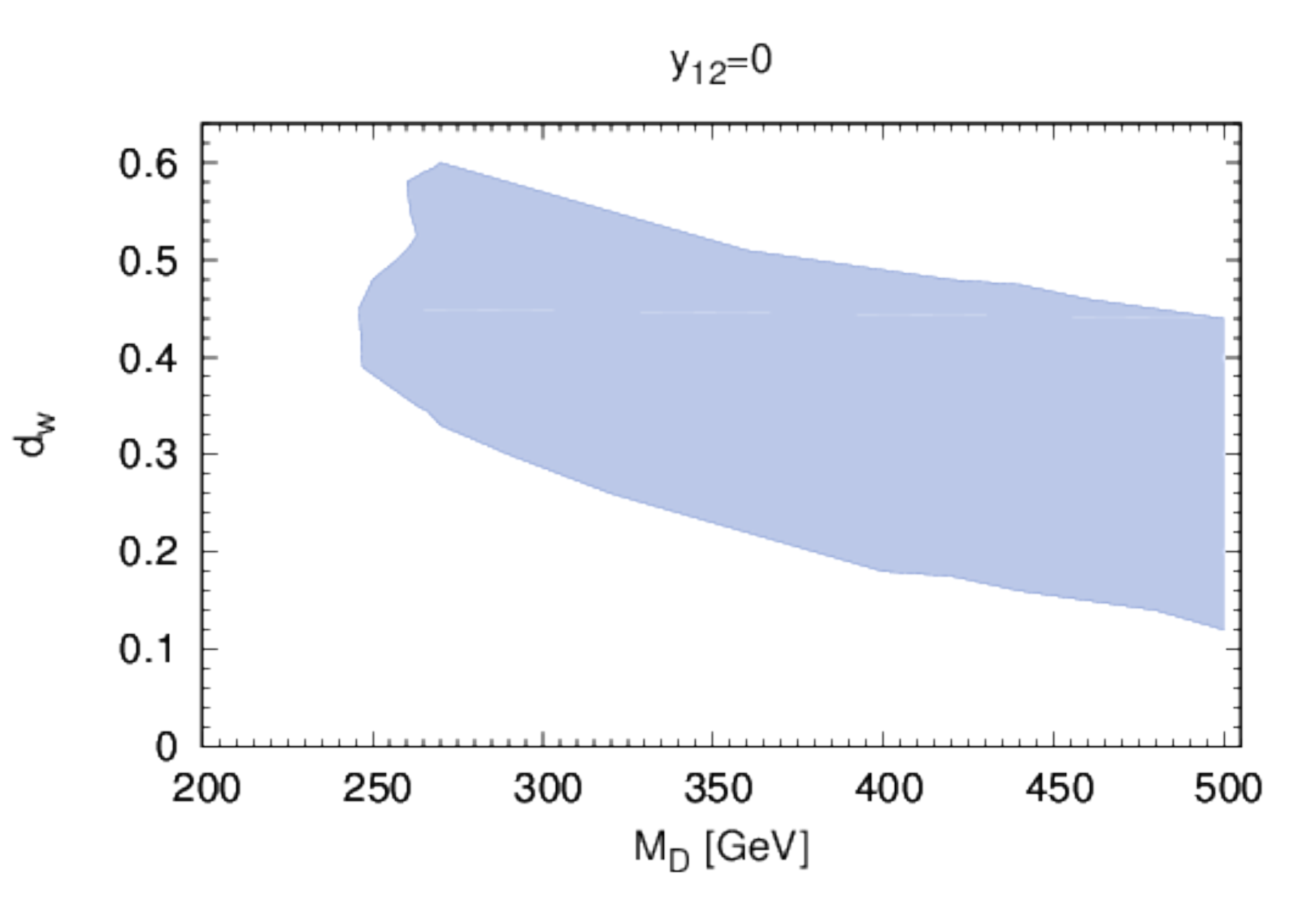}
           \caption{$M_D$ vs $d_W$, for $\Lambda =1 \; \TeV$ and $y_{12}=0$.}
        \label{y12=0_MDvsdw_1TeV}
    \end{subfigure}
\caption{\em As in Fig.~\ref{dw-dgamma} but for acceptable values on the plane $M_D - d_W$.
%We  allow variation of the no-shown  parameters that are cosmologically acceptable, for $\Lambda =1 \; \TeV$.
}
        \label{MDvsdw}
\end{figure}
%%%%%%%%%%%%%%%%%%%%%%%%
 \begin{figure}[H]
\hspace*{-0.3cm} \begin{subfigure}[b]{0.52\textwidth}
        \includegraphics[width=\textwidth]{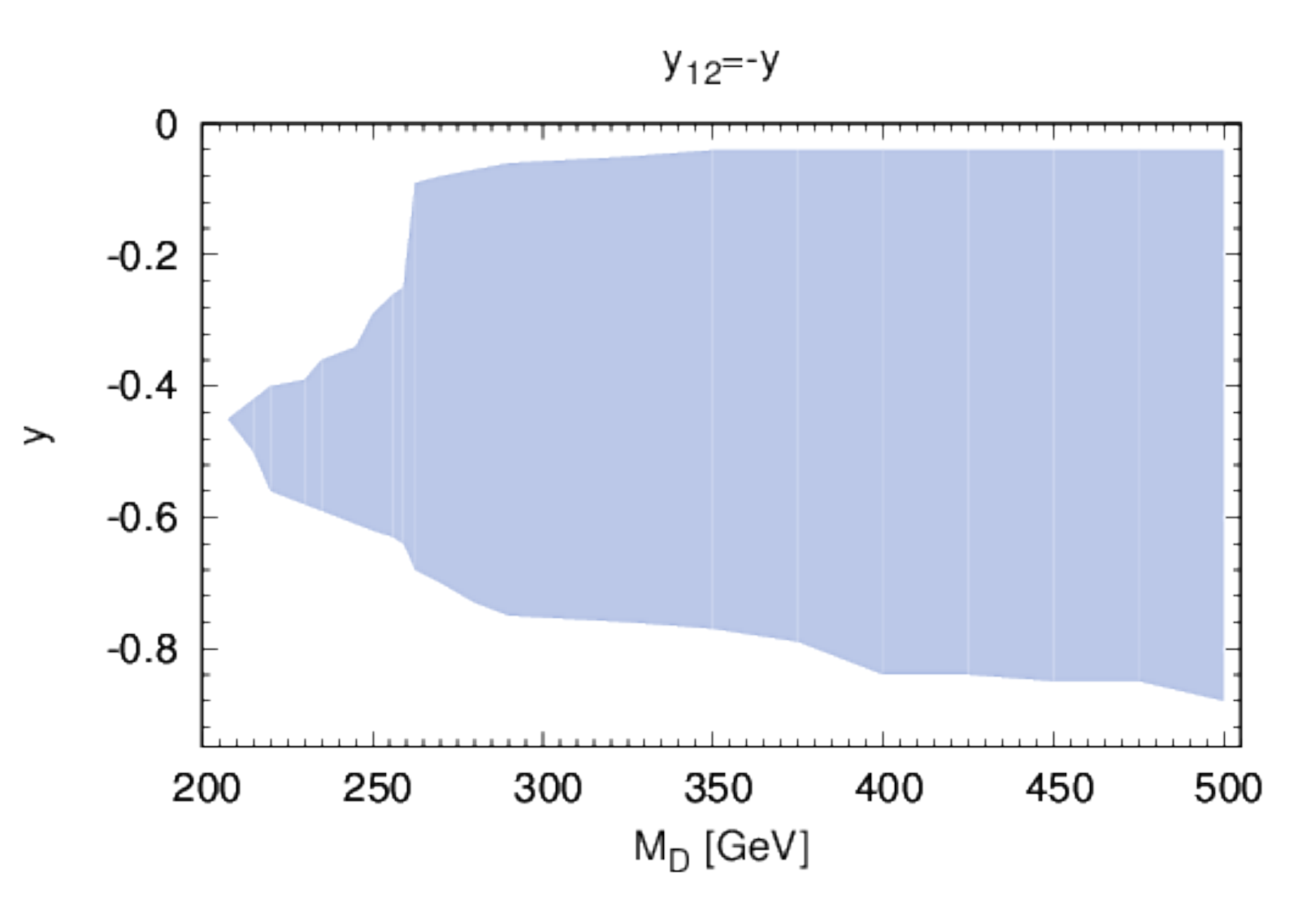}
           \caption{$M_D$ vs $y$, for $\Lambda =1 \; TeV$ and $y_{12}=-y$.}
        \label{y12=-y_MDvsy1_1TeV}
    \end{subfigure}
 \begin{subfigure}[b]{0.52\textwidth}
        \includegraphics[width=\textwidth]{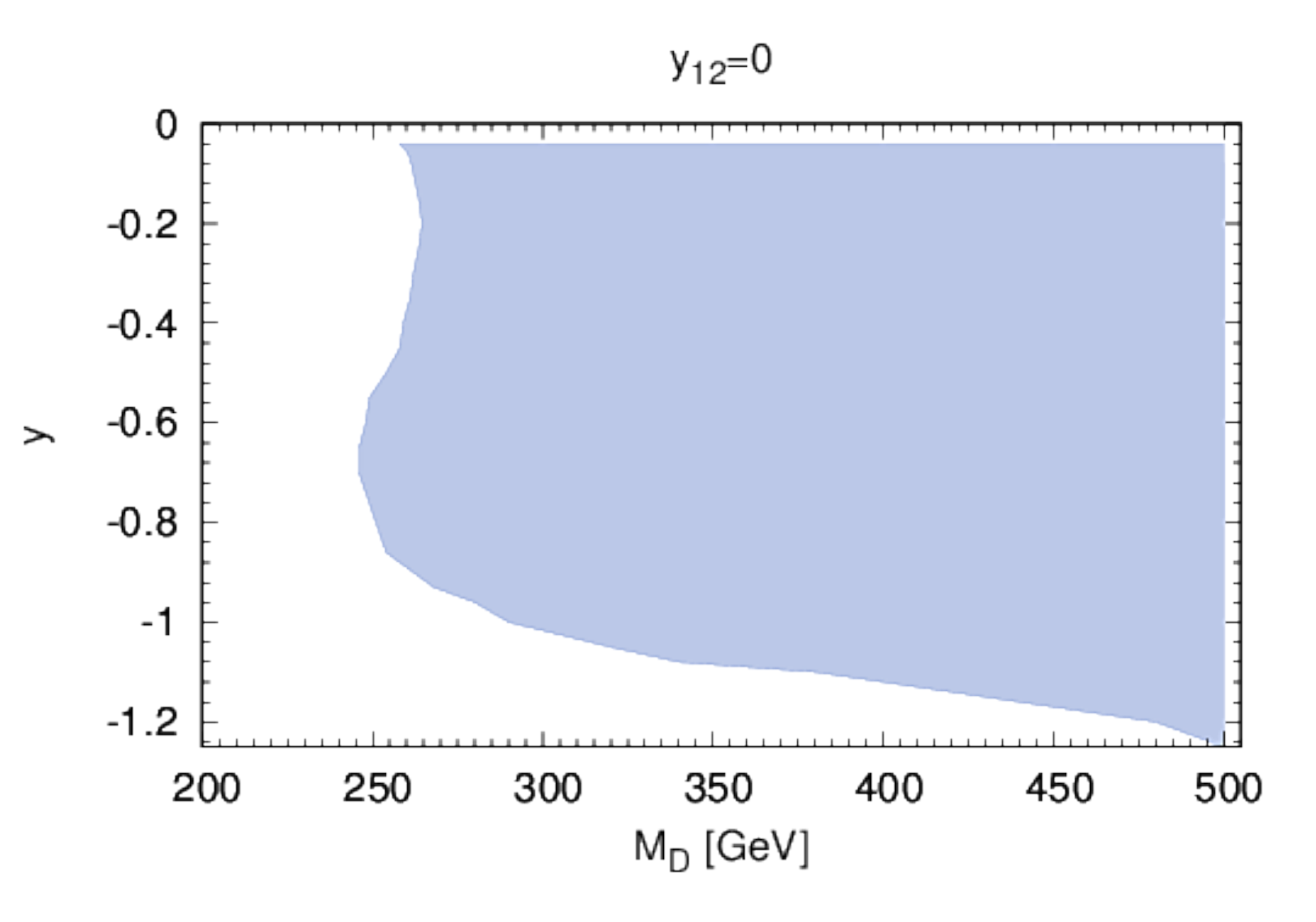}
           \caption{$M_D$ vs $y$, for $\Lambda =1 \; \TeV$ and $y_{12}=0$.}
        \label{y12=0_MDvsy1_1TeV}
    \end{subfigure}
\caption{\em Values on  $M_D - y$ plane that provide acceptable  DM relic abundance.  }
 \label{MDvsy1}
\end{figure}
%%%%%%%%%%%%%%%%%%%%%%%%%%%%%5
In Fig.~\ref{MDvsdw} we observe that  $M_D$ vastly  affects the  allowed values for $d_W$ that provide the correct relic abundance. 
This is due to  the fact that the minimum of the total annihilation cross-section 
depends on the mass $M_D$, as can be seen from  \eqss{avv}{agammaz}{agammagamma}  and also from the fact that 
the maximum   of $\Omega h^2$ varies as $M_D$ changes,  see  also Fig.~\ref{MDvsrelic}. Moreover, as $M_D$ 
becomes larger, the minimization of the cross-section becomes less necessary. Note that, for $y_{12}=0$ there is a gap for $d_W$ at $M_D \approx 260 \GeV$, 
a result of the ``turning point'' discussed at the end of the previous paragraph (see Fig.~\ref{turning_point_dw}). For $y_{12}=-y$, 
this ``turning point'' is ineffective.
%%%%%%%%%%%%%%%%%%%%%%%%%%%%%%%%%%%5
 \begin{figure}[t]
% \centering
\hspace*{-0.3cm} \begin{subfigure}[b]{0.52\textwidth}
        \includegraphics[width=\textwidth]{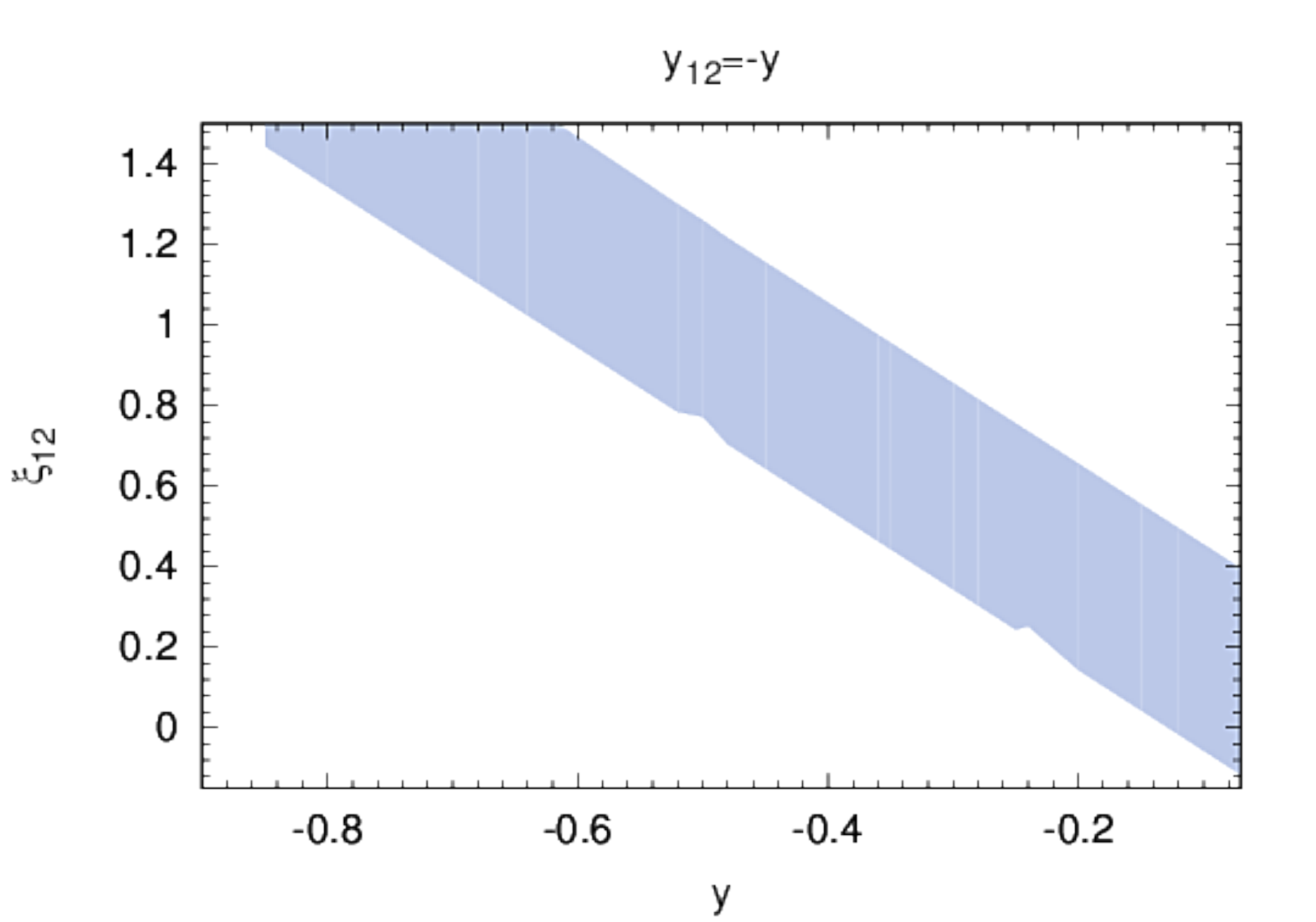}
           \caption{$y$ vs $\xi_{12}$, for $\Lambda =1 \; \TeV$ and $y_{12}=-y$.}
        \label{y12=-y_y1vsxi12_1TeV}
    \end{subfigure}
\begin{subfigure}[b]{0.52\textwidth}
        \includegraphics[width=\textwidth]{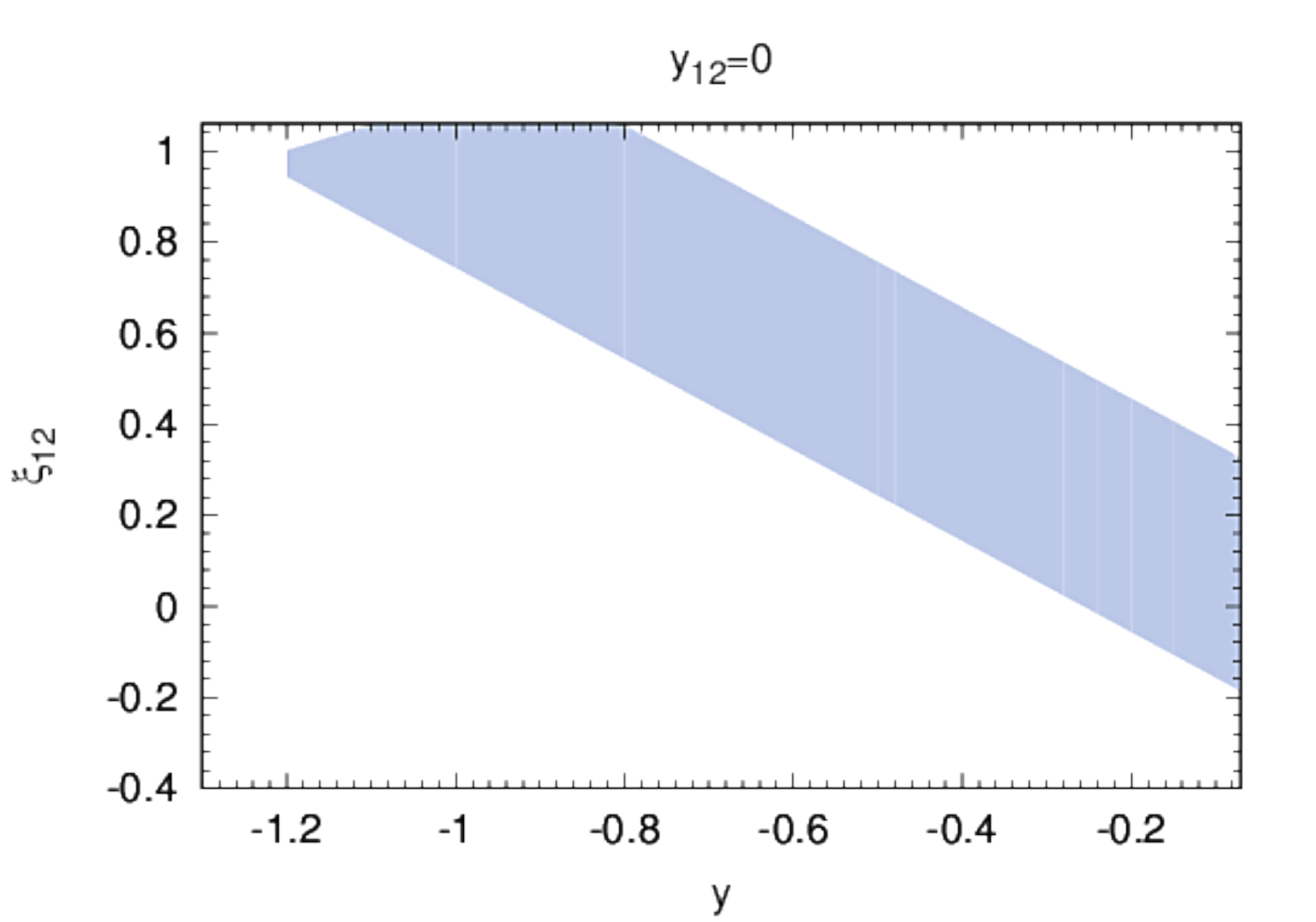}
           \caption{$y$ vs $\xi_{12}$, for $\Lambda =1 \; \TeV$ and $y_{12}=0$.}
        \label{y12=0_y1vsxi12_1TeV}
    \end{subfigure}
     \caption{\em As in Fig.~\ref{MDvsy1}, but for the Yukawa parameters 
     $y-\xi_{12}$.
%     We  allow variation of the no-shown  parameters that are cosmologically acceptable.
     }
        \label{yvsxi12}
    \end{figure}

In Fig.~\ref{MDvsy1} one  can see the dependence of  $M_D$ on  $y$,   
in the region where the DM density complies  the current cosmological bound. We observe that for large values of $M_D$, for 
$|y|<~0.85 (1.25)$ for the case $y_{12}=-y \ (y_{12}=0)$ we obtain the desired $\Omega h^2$. 
On the contrary, when $M_D \lesssim 300 \GeV$  in both cases for $y_{12}$, $|y|$ seems to be strongly dependent on $M_D$.
This happens because the bound on $|y|$ from Earth-based experiments becomes stronger than the one from the 
relic abundance for smaller masses. 
In addition to that, since $\Omega h^2$ tends to decrease as 
$|y|$ becomes smaller for $M_D \lesssim 260 \; \GeV$, $|y|$ is also bounded from below.
Furthermore, due to the ``oscillation'' of the relic abundance (Fig.~\ref{turning_point_y}), at $M_D \sim 260 \; \GeV$ there 
is a ``gap'' on the allowed values of $y$ (similar to $d_W$). 
Additionally, in Fig.~\ref{yvsxi12}, we see that $\xi_{12}$ follows $y$, a remaining  result from
the direct detection bound (similar to Fig.~\ref{DirDet-region}).

 \begin{figure}[t]
% \centering
\hspace*{-0.3cm} \begin{subfigure}[b]{0.51\textwidth}
        \includegraphics[width=\textwidth]{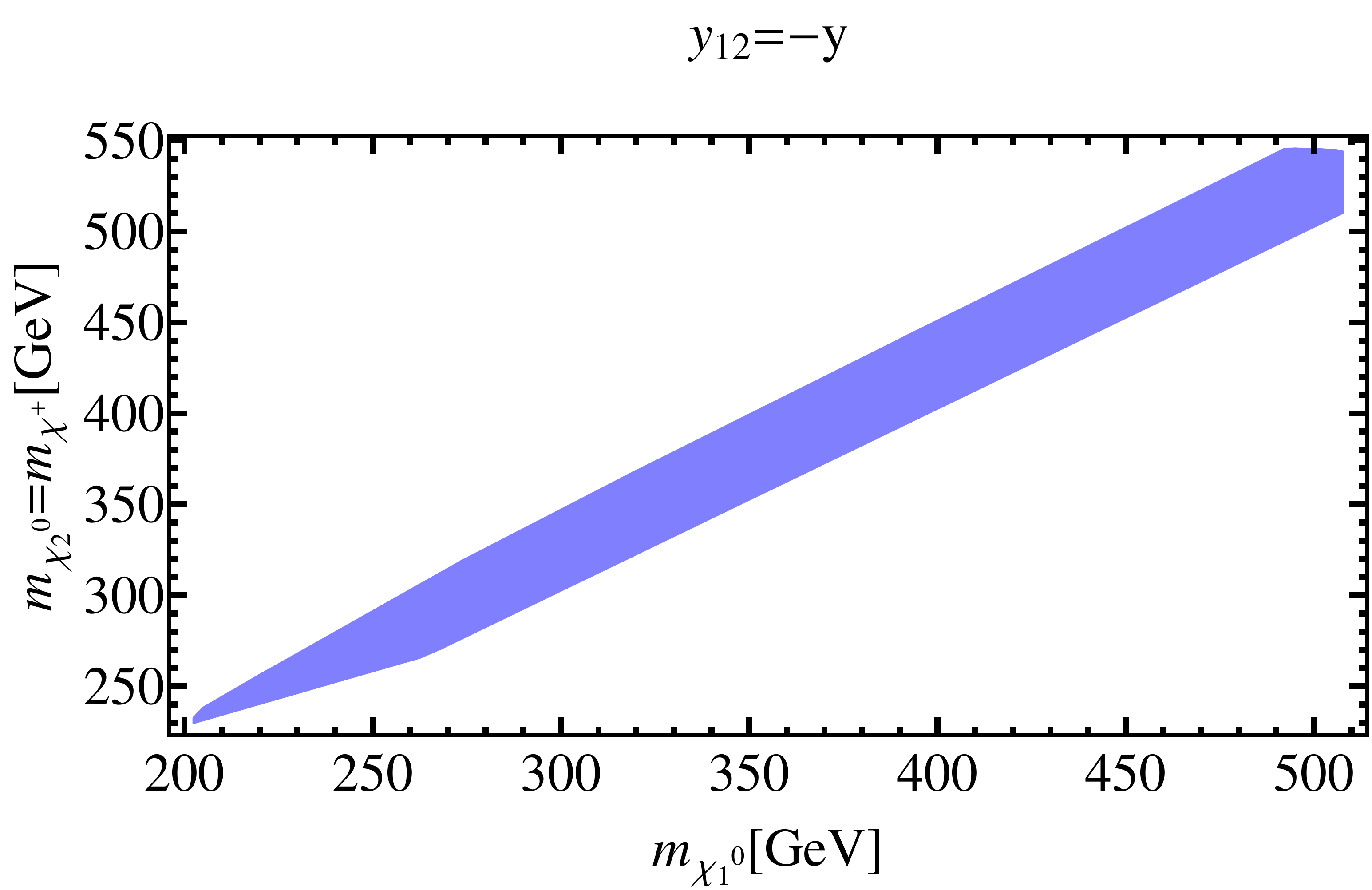}
           \caption{}
        \label{}
    \end{subfigure}
\begin{subfigure}[b]{0.50\textwidth}
        \includegraphics[width=\textwidth]{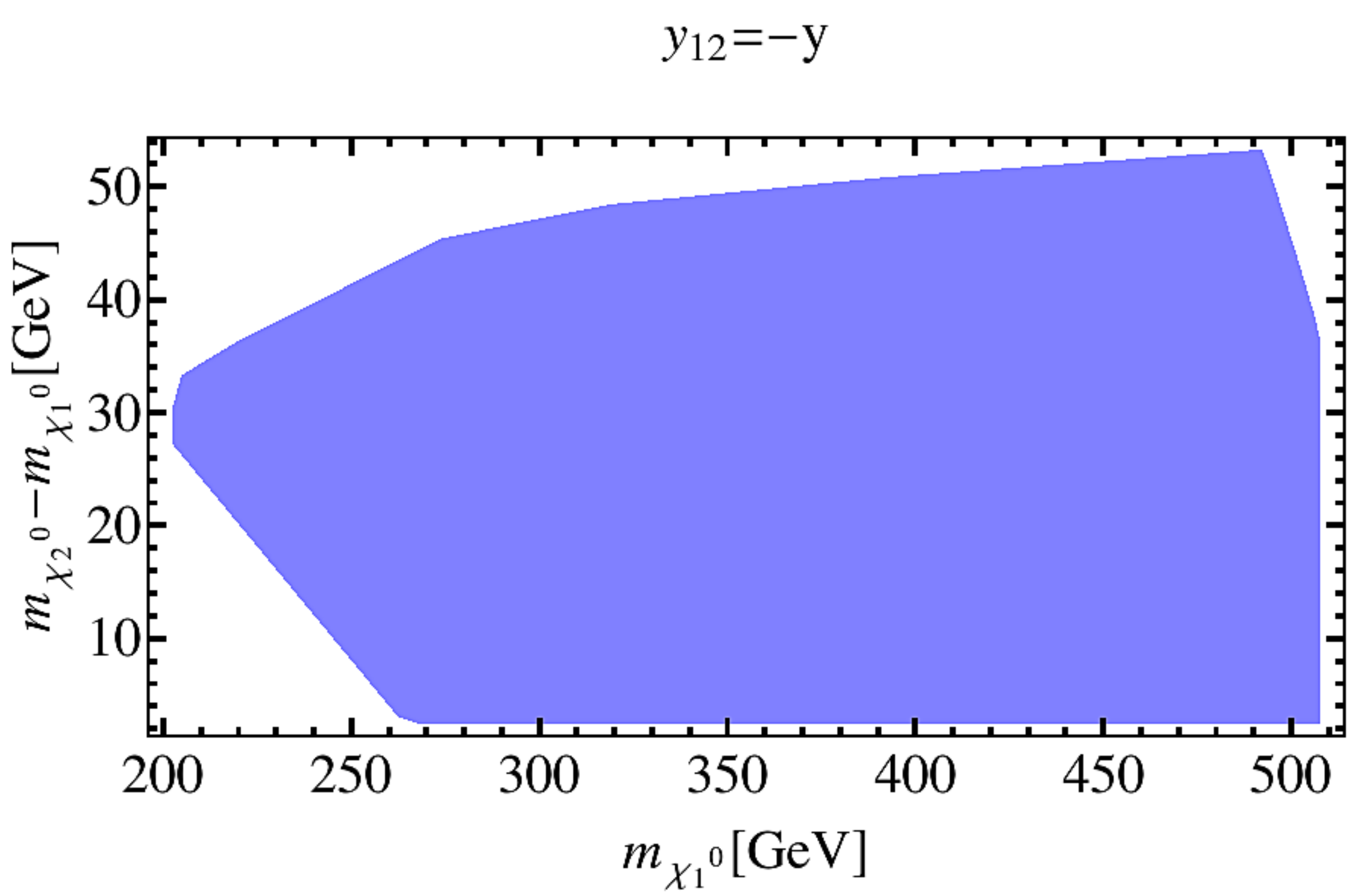}
           \caption{}
        \label{}
    \end{subfigure}
     \caption{\em (a) The cosmologically allowed mass of the WIMP versus the mass of the heavy fermions  and (b) 
     their mass  difference  for $y_{12}=-y$. Similar regions can be obtained  for $y_{12}=0$.}
        \label{mass-region}
    \end{figure}

The Yukawa couplings  and the mass parameter $M_D$ displayed, fix the masses and their differences.
For the sake of completeness, the masses and their difference from $m_{\chi_{1}^{0}}$ 
are shown in Fig.~\ref{mass-region} for $y_{12}=-y$  (similar region holds also for $y_{12}=0$). 
We observe that $m_{\chi_{1}^{0}} \gtrsim 200 \; \GeV$, for $y_{12}=-y$, which is also what one should expect from Fig.~\ref{MDvsy1}.   
In addition to this, the mass difference $m_{\chi_{2}^{0}}-m_{\chi_{1}^{0}}$ is in the region $\sim 2 - 50 \; \GeV$. 
Finally, we note that this mass difference takes slightly larger values ($\sim 2-70 \; \GeV$ ) for the other case
of the  symmetric limit for $y_{12}$, while $m_{\chi^{\pm}}-m_{\chi_{1}^{0}}$ is always half that [see \eq{eq:spec}]. 
Accordingly, the smallest possible 
mass of the WIMP in this case is $\sim 250 \; \GeV$ (which again can be seen also from Fig.~\ref{MDvsy1}).

\subsection{Gamma-rays}\label{Gamma-rays-sec}
Having delineated  the cosmologically acceptable regions concerning the  DM abundance,    we 
will  proceed calculating   other astrophysical observables, like  
 the  gamma-ray  fluxes (monochromatic and continuous) originating  from  the Milky Way GC and dSphs.

\subsubsection{Continuous Gamma spectrum}\label{Gamma-rays-Con}
In our model  the DM pair annihilation cross-sections  have been  studied in section \ref{Close_Look-Section}. 
In particular, the  relevant relations can be found in 
 \eq{avv}. From \Refs{Ackermann:2015zua, Hooper:2012sr} we observe that the bounds on the cross 
sections $a_{ZZ}$ and  $a_{WW}$ are above  
the required $\sim 3 \times 10^{-26} \; \mathrm{cm}^{3} \mathrm{s}^{-1}$ (for masses above $200 \; \GeV$) which generally
gives the desired relic abundance. More precisely, for $m_{\chi_{1}^{0}}\gtrsim 200 \; \GeV$, the bound from 
dSphs is below $\sim 5 \times 10^{-26} \; \mathrm{cm}^{3} \mathrm{s}^{-1}$
for the annihilation $\chi_{1}^{0} \chi_{1}^{0} \to W^{+}W^{+}$ (assuming that the branching ratio is $100 \%$). 
The same bound holds the annihilation to a pair of $Z$-bosons, since their gamma spectra are quite similar. 
When applied to our model, which generally gives smaller branching ratios, these bounds should be even weaker.
%%%%%%%%%%%%%%%%%%%
%
%%%%%%%%%%%%%%%%%%%
\begin{figure}[t]
% \centering
\hspace*{-0.5cm} \begin{subfigure}[b]{0.53\textwidth}
        \includegraphics[width=\textwidth]{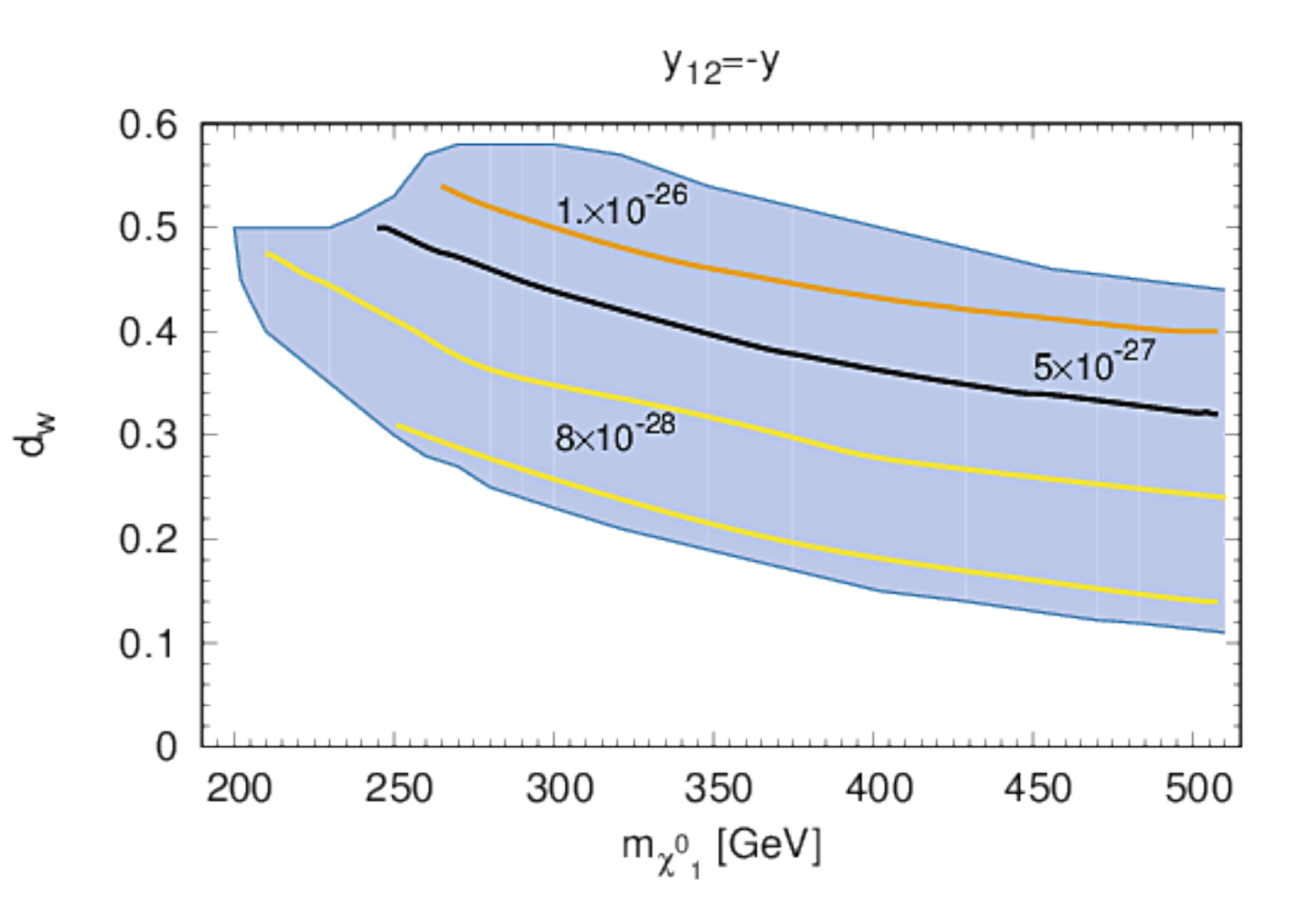}
           \caption{ }
        \label{}
    \end{subfigure}
\hspace*{.0cm}    
\begin{subfigure}[b]{0.53\textwidth}
        \includegraphics[width=\textwidth]{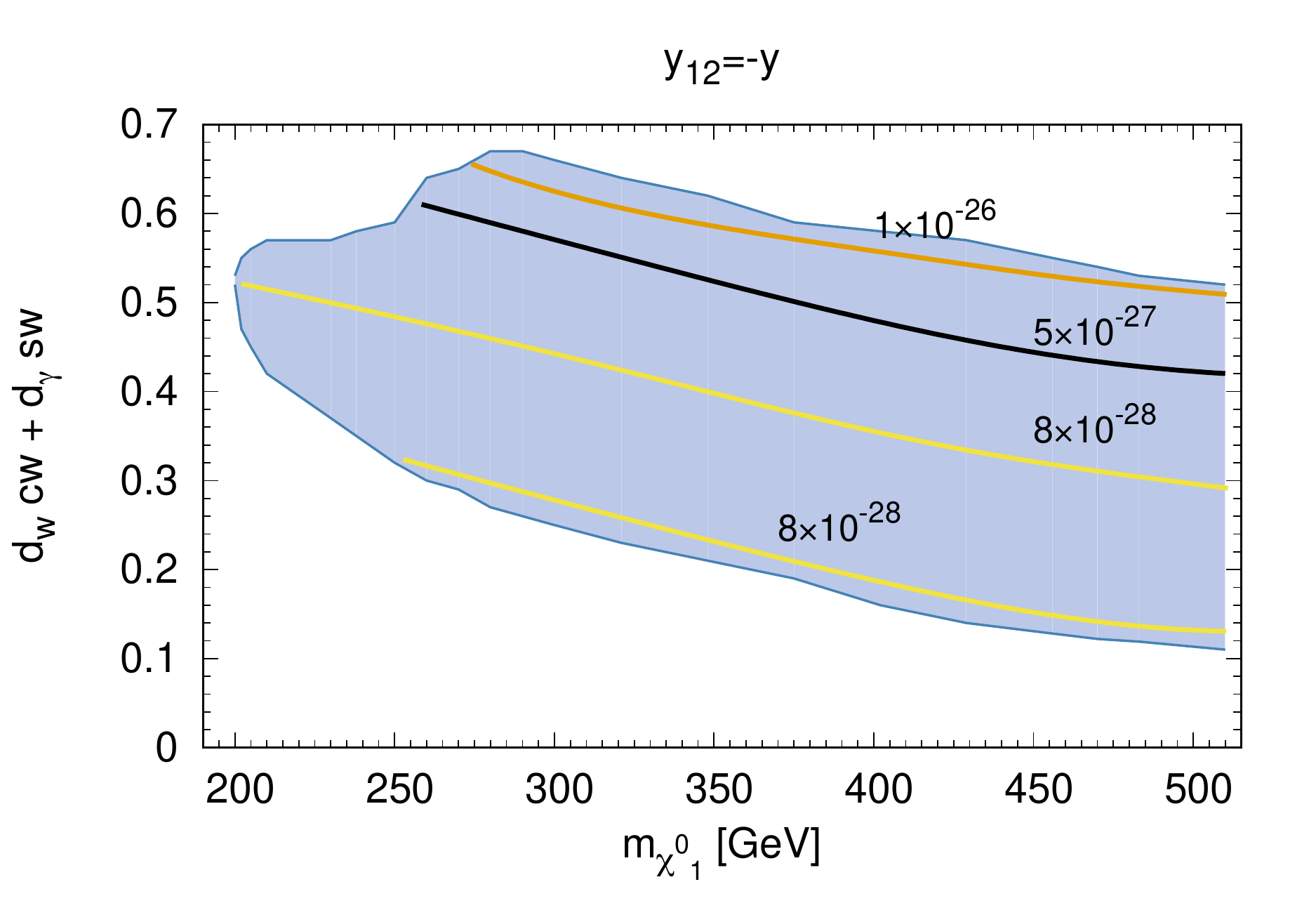}
           \caption{}
        \label{}
    \end{subfigure}   
        \caption{ \em Allowed region in the parameter space from  collider, DM direct detection  
        and relic density constraints discussed in sections 4 and 5.3, respectively, as a function 
         of the WIMP mass and the couplings $d_W$ and ${K_{Z}}/{c_{W}}$. 
         The contours show the values of the thermally averaged  cross-sections (a) for  $a_{WW}$  and (b) for 
          $a_{ZZ}$ in $\mathrm{cm}^{3}\mathrm{s}^{-1}$ for  $y_{12}=-y$. Similar for  $y_{12}=0$.
  We take $\Lambda =1\;\TeV$.}
        \label{WW-ZZ-gamma_rays_cont}
\end{figure}
%%%%%%%%%%%%%%%%%%%%%%%%%%
%
As it is shown  in Fig.~\ref{WW-ZZ-gamma_rays_cont},
 the relevant to continuous emission of photons cross-sections, $\sigma_{\chi_{1}^{0}\chi_{1}^{0} \to W^{+}W^{-}, \, ZZ }$
are safe with experimental bounds from continuous gamma ray spectrum discussed in this paragraph.

%%%%%%%%%%%%%%%%%%%%%%%%%%%%%%%%%%%%%%%%%%%%%%%%%%%%%
\subsubsection{Constraints from Gamma-ray monochromatic spectrum}\label{Gamma-rays-lines}
%%%%%%%%%%%%%%%%%%%%%%%%%%%%%%%%%%%%%%%%%%%%%%%%%%%%%

As we have seen, this effective theory relies on the various WIMPs magnetic  
dipole moment operators in order to give us the observed relic abundance. This could result to annihilations
of pairs of WIMPs into photons which could be detectable from observations of gamma ray monochromatic spectrum 
originated  from the GC.   
In this paragraph, we will calculate the cross-sections for processes that could give such gamma rays (\eqs{agammaz}{agammagamma}). 
As input, we use the  parameter space 
that evade all the other, previously examined, bounds and use the results from Fermi-LAT~\cite{Ackermann:2013uma,Ackermann:2015lka} 
to set additional bounds  to the parameters of this model.
%%%%%%%%%%%%%%%%%%
%
%%%%%%%%%%%%%%%%%
\begin{figure}[H]
% \centering
\hspace*{-0.5cm} \begin{subfigure}[b]{0.53\textwidth}
        \includegraphics[width=\textwidth]{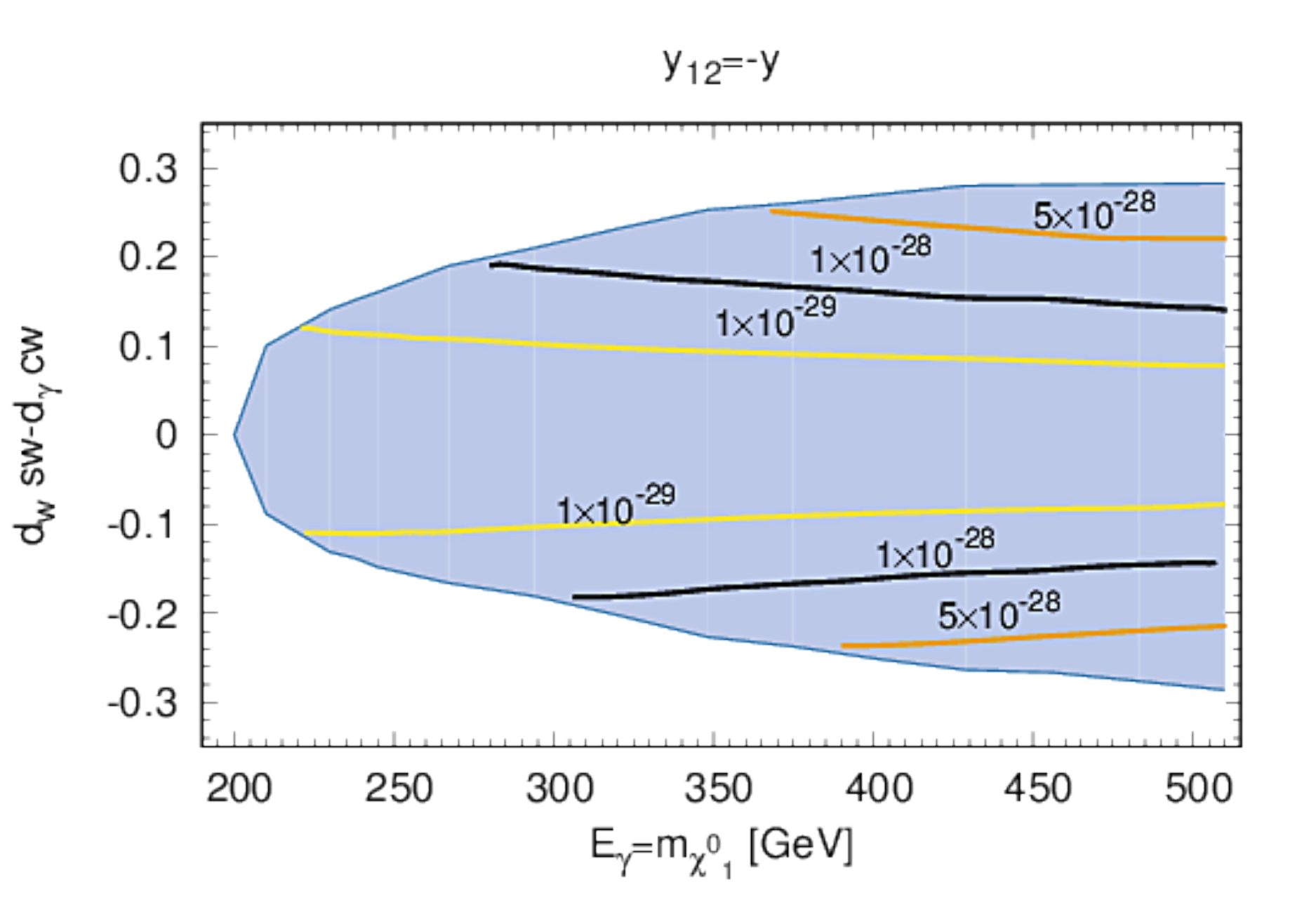}
           \caption{}
%         \label{}
    \end{subfigure}
\hspace*{0.0cm} \begin{subfigure}[b]{0.53\textwidth}
        \includegraphics[width=\textwidth]{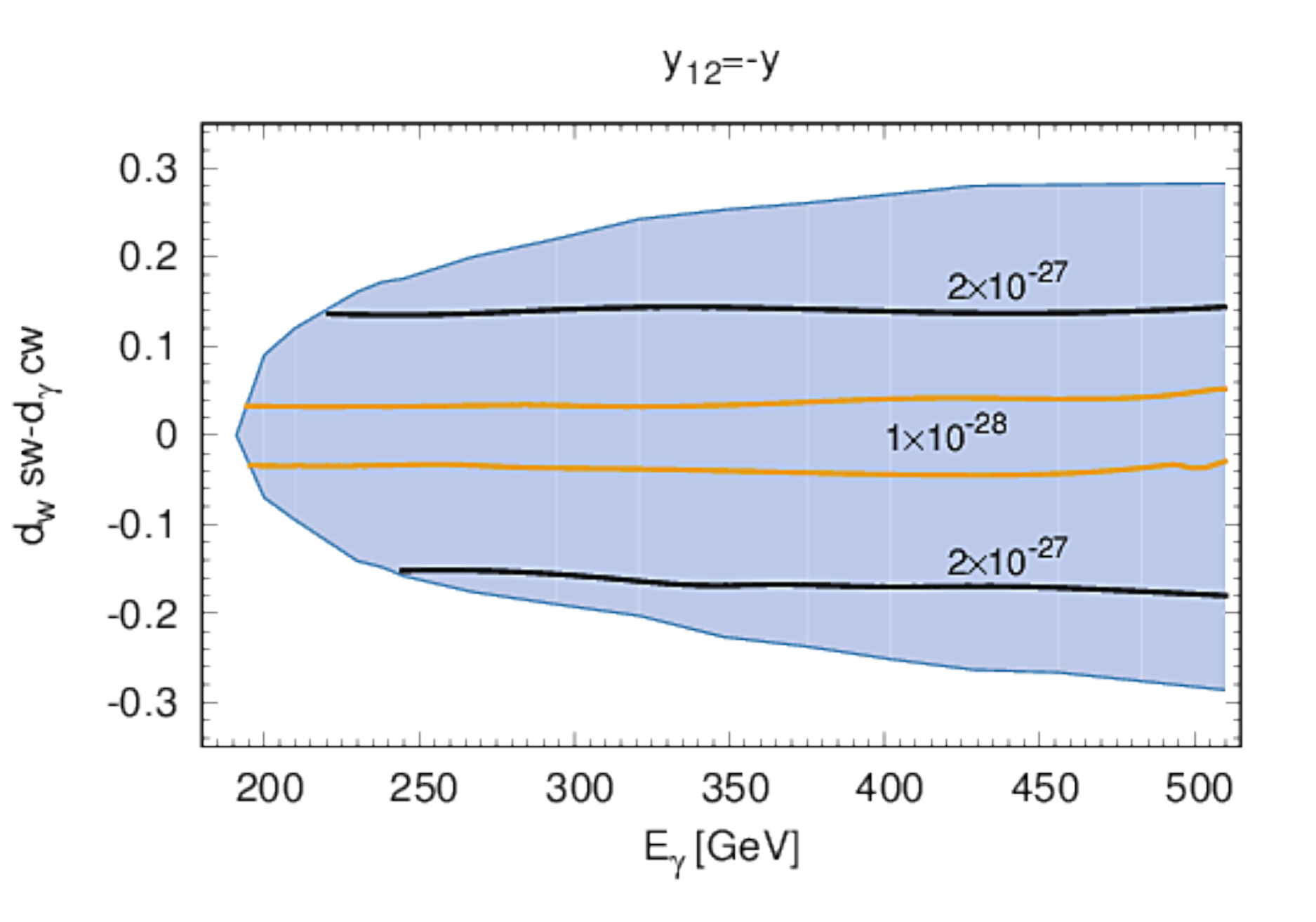}
           \caption{}
%         \label{}
    \end{subfigure}
\caption{\em The allowed, as in Fig.~\ref{WW-ZZ-gamma_rays_cont},  region of the parameter space, in terms of the photon 
energy and the coupling $C_{\gamma}$. The contours show the values of the thermally averaged cross-sections 
		    $a_{\gamma \gamma}$ (a) and $a_{\gamma Z}$ (b) in $cm^{3}s^{-1}$
		     for $y_{12}=-y$. Again $y_{12}=0$ results in  an almost identical plot.}
        \label{gammagamma-gammaZ-gamma_rays_lines}
\end{figure}
%%%%%%%%%%%%%%%%%%
%
These bounds  depend strongly on the DM halo 
profile\footnote{The bounds have up to a factor of $15$ difference for different 
profiles and regions.} (and the region of interest) that one follows. Thus, 
we study  the profile which gives the strongest bound. 
This comes from the $R3$ region which is optimized for the 
Navarro-Frenk-White NFWc($\gamma=1.3$) profile~\cite{Navarro:1995iw} (the relevant discussion on these regions
of interest is found in \cite{Ackermann:2013uma}). So, the annihilation cross-section for 
$\chi_{1}^{0}\chi_{1}^{0}\to \gamma \gamma$ for this region of interest  is
bounded to be smaller than
 $\sim 10^{-28} \; cm^{3}s^{-1}$ for photon energy ($E_{\gamma}=m_{\chi_{1}^{0}}$) at 
 $200 \; \GeV$ up to $\sim 3.5 \times 10^{-28}$ for $E_{\gamma}\sim 450 \;\GeV$ (and if we 
extrapolate up to $\sim 5 \times 10^{-28}$ for $E_{\gamma}\sim 500 \;\GeV$). 
For the process $\chi_{1}^{0}\chi_{1}^{0}\to \gamma Z$, we need to rescale this bound by a factor of two, since there is one photon
in the final state instead of two. This process results to different value of
 $E_{\gamma}=m_{\chi_{1}^{0}}\, (1-  {m_{Z}^2}/{4m_{\chi_{1}^{0}}^2} )$.

Fig.~\ref{gammagamma-gammaZ-gamma_rays_lines}
illustrates that   the annihilation to $\gamma Z$ (and less to $\gamma \gamma$), violates the Fermi-LAT bound, mainly for larger values of $E_{\gamma}$.
Thus, the values of $d_{W}$ and $d_{\gamma}$ are constrained so that $C_{\gamma}$ is even smaller than 
the cosmologically acceptable values.    
%%%%%%%%%%%%%%%%%
\begin{figure}[t]
% \centering
\hspace*{-0.5cm} \begin{subfigure}[b]{0.53\textwidth}
        \includegraphics[width=\textwidth]{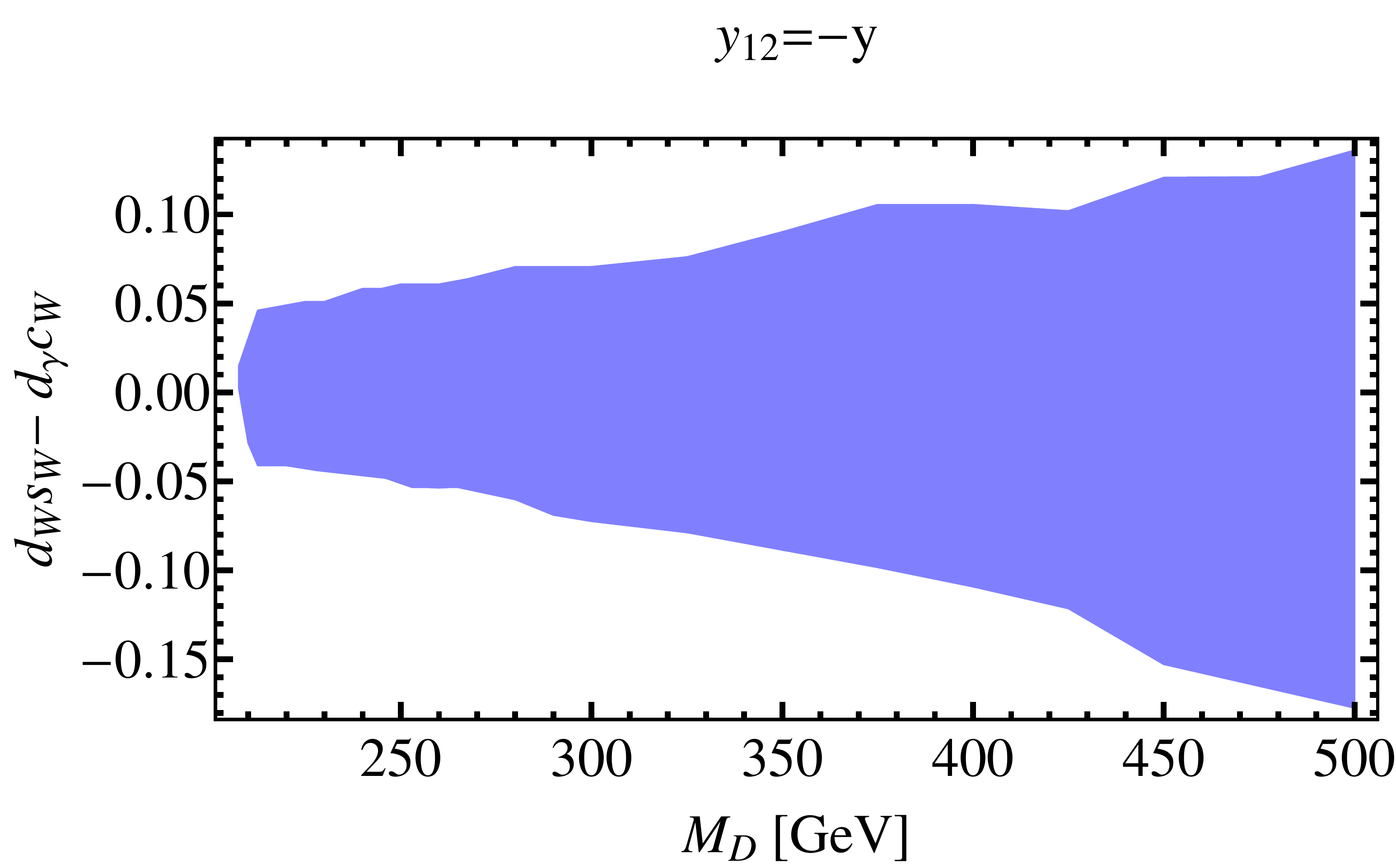}
           \caption{}
         \label{MDvsCgamma}
    \end{subfigure}
\hspace*{0.0cm} \begin{subfigure}[b]{0.50\textwidth}
        \includegraphics[width=\textwidth]{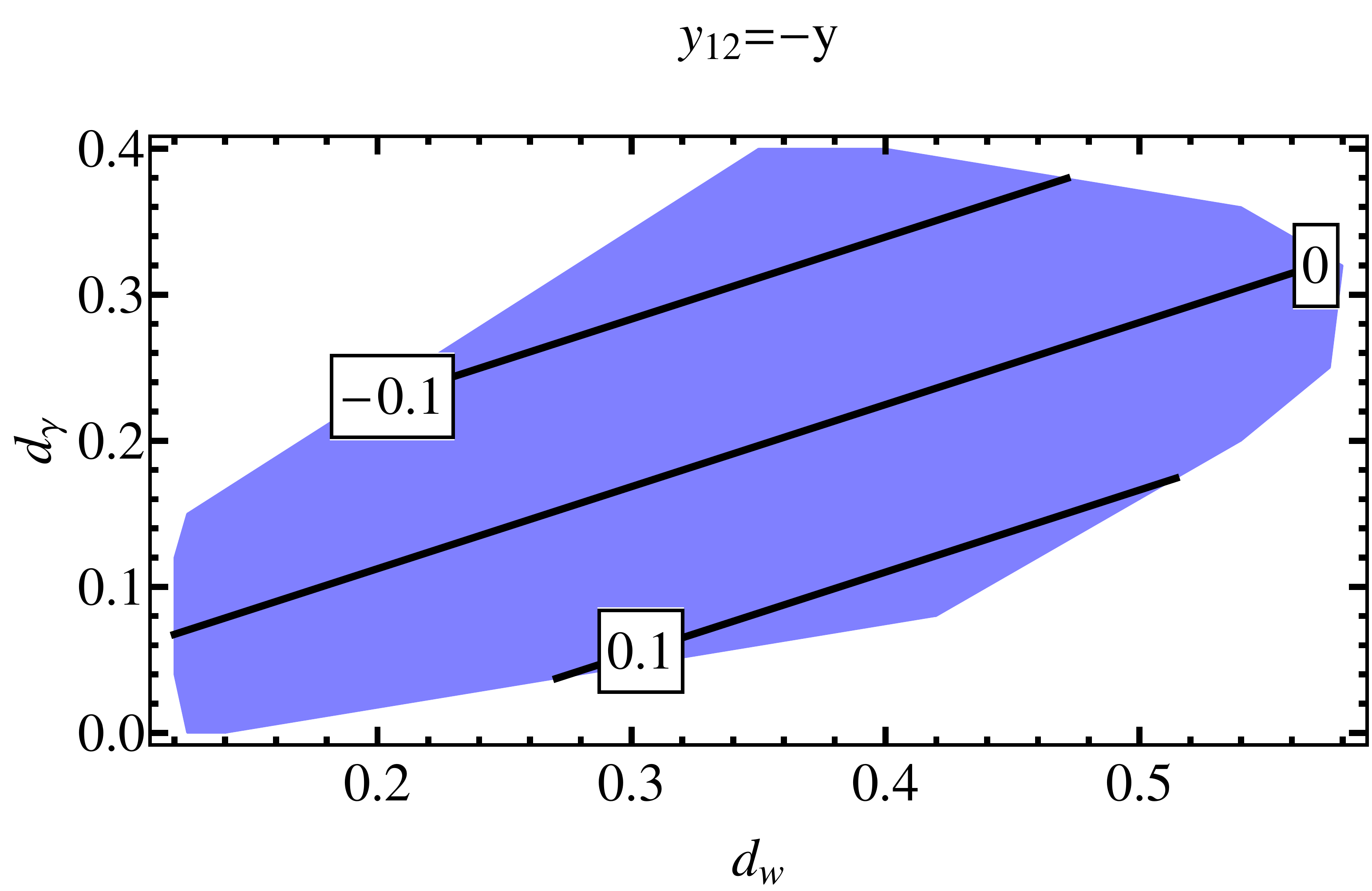}
           \caption{}
         \label{dwvsdgamma_update}
    \end{subfigure}
\caption{\em  Allowed  regions on  a) $M_D - C_{\gamma}$ plane and 
b) $d_W - d_{\gamma}$ plane  for $y_{12}=-y$, consistent with ``Earth" constraints, the observed  relic abundance and 
 the bounds from gamma-ray monochromatic spectrum,  discussed sections 4, 5.3 and 5.4.2, respectively, in the text. 
Almost identical regions  are allowed  for $y_{12}=0$. 
The contour lines in (b) show the value of the $\chi_{1}^{0}\chi_{2}^{0}$-photon coupling 
$C_{\gamma}$.}
        \label{fig:gammagamma}
\end{figure}
%%%%%%%%%%%%%%%%%%
It is evident from Fig.~\ref{MDvsCgamma},
 that in order this model   to  deceive  the current  monochromatic  gamma ray  bounds from GC,
 we should limit the dipole couplings so they satisfy the relation 
 $|d_W s_{W}-c_{W} d_{\gamma}| \lesssim 0.05$ ($\Lambda =1 \; \TeV$) for $M_D = 200 \; \GeV$ 
up to $|d_W s_{W}-c_{W} d_{\gamma}| \lesssim 0.15$ for $M_D = 500 \; \GeV$. 
Therefore, one can delineate accordingly  
 the parameter space on   the $d_W - d_{{\gamma}}$ 
plane,  that evades all bounds and yields  the correct relic density, which is shown in Fig.~\ref{dwvsdgamma_update}.
 It should be noted, that the other parameters remain unchanged as in the previous section, 
since they do not affect  WIMP pair  annihilation rates   to two photons or to  a photon and a $Z$ boson.
Other values of $y_{12}$ result to almost identical regions to these in Fig.~\ref{fig:gammagamma}.

Concluding  this paragraph, we note that the Fermi-LAT data  set  upper bounds  to the annihilation 
cross-section of two WIMPs into one or two photons,
relating strongly  the two dipole couplings, resulting  to positive values for $d_{\gamma}$.
 Therefore,  the two neutral particles of the model have an almost  zero coupling to photon
  ($C_{\gamma}\approx 0$), while the other parameters are intact.

%% Sommerfeld Enhancement 
It is worth pointing out that there is a non-relativistic non-perturbative
effect, known as ``Sommerfeld enhancement"~\cite{ANDP:ANDP19314030302},
that can boost the annihilation cross-section, sometimes even, by orders of magnitude. 
For the bi-doublet case here, it has been
calculated in the literature and the results are shown in
\Refs{Hisano:2003ec,Hisano:2004ds,Cirelli:2007xd}. As it turns out, for the masses
we are considering here, this effect is non-important. It becomes only sizeable for
WIMP, ``higgsino-like" masses greater than about $1~\mathrm{TeV}$ or so. 
%%

%%%%%%%%%%%%%%%%%%%%%%
\subsection{Neutrino flux from the Sun}\label{Neutrinos}
%%%%%%%%%%%%%%%%%%%%%%
Another interesting indirect signal could come from solar neutrino flux.
The cross-section for neutrino production from WIMP annihilations in the Sun, 
can be decomposed to the spin-dependent and spin-independent WIMP-nucleon 
cross-sections. Therefore, such experiments compete with direct detection ones. 
Recent results from IceCube~\cite{Aartsen:2012kia}, show that the 
spin-independent cross-section bound is relaxed as compared to the one obtained from 
direct detection experiments~\cite{Akerib:2015rjg}. On the other hand, the 
latest spin-dependent cross-section bound from solar neutrino 
flux~\cite{Aartsen:2016exj}, is much stronger than the one derived from 
LUX~\cite{Akerib:2016lao} for $m_{\chi_{1}^{0}} \gtrsim 200 \GeV$. 
In our  study, the spin-independent bound from IceCube is evaded, since the 
constraints from LUX have been introduced from the beginning of this analysis.
In addition, due  to the c.c. symmetry, the spin-dependent cross-section vanishes, 
since $\chi_{1}^{0 \dagger}\bar{\sigma}^{\mu}\chi_{1}^{0}Z_{\mu}$ is 
odd under the transformation introduced in section~\ref{sec:eft}.
Thus, these bounds, leave the allowed parameter space unaffected.

%%%%%%%%%%%%%%%%%%%%%
\section{LHC searches}
\setcounter{equation}{0}
\label{sec:LHC}
%%%%%%%%%%%%%%%%%%%%%

Having found that there is a viable area in the parameter space, which produces the observed DM relic 
abundance of the universe while avoiding all the other experimental and observational constraints, we move on
to find out whether this theory can provide us with  observational effects at the LHC. First, we calculate the cross-sections 
for some channels at $\sqrt{\hat{s}}=8 \; \TeV$ and compare them to the current bounds from LHC (Run I) and then we 
 do the same at RunII with  $\sqrt{\hat{s}}=13 \; \TeV$.

In this section we are looking
at the mono-$Z$ channel for which the experimental analysis is performed by ATLAS~\cite{Aad:2014vka}, 
the mono-$W$ channel where we use the results from ATLAS~\cite{ATLAS:2014wra} (a weaker bound is obtained from the analysis of
CMS~\cite{Khachatryan:2014tva}), the hadronically decaying $W/Z$ boson channel searched for by ATLAS~\cite{Aad:2013oja}.
%%%%%%%%%%%%%%%
DM interacting with vector bosons can be probed by dijet searches through vector boson 
fusion as discussed in \Refs{Delannoy:2013ata,Berlin:2015aba}. The analysis has been performed 
by ATLAS~\cite{Aad:2015txa}\footnote{The fermions considered here, do not contribute to the invisible decays of the Higgs boson, but  bounds
from \Ref{Aad:2015txa} still apply for a  dijet + $\slashed{E}_{T}$ final state.} (which gives a somewhat stronger bound
than CMS~\cite{Chatrchyan:2014tja}). Furthermore, there are mono-jet searches from CMS~\cite{Khachatryan:2014rra}.
Finally, there is also the mono-photon channel searches~\cite{Aad:2014tda,Aad:2013oja}, 
but in our case it is not very important  due to the Fermi-LAT bound discussed previously in section~\ref{Gamma-rays-lines}.
 
We note that, for these processes, an extensive study has been performed in ref.~\cite{Crivellin:2015wva} with singlet Dirac DM particle and
for  operators with dimension $d=7$.
However, in the analysis we perform here  there are differences:  {\it a)} The set of operators is different, since we consider Yukawa, dipole and renormalizable operators. These operators produce mass splittings between the Dark-sector fermions. In addition to this, the interactions with the gauge bosons come from both 3- and 4-point terms in the Lagrangian with different Lorentz structure than the $d=7$ ones. 
{\it b)} The parameter space in which we calculate the cross-sections for these processes, respects other experimental and observational constraints.       
%c) The number of Dark-sector fermions, where the mass splittings may be imortant (e.g. for the mono-jet channel).      
In addition, for the dijet channel and at $\sqrt{\hat{s}}=13 \TeV$, another dedicated study has been performed in \Ref{Brooke:2016vlw}. Again our case is different because 
of the inclusion of  $d=4 \text{ and } 5$ operators in the calculations of the LHC cross-sections, while at the same time 
the parameter space is also constrained by all the other bounds discussed in sections~\ref{sec:earth} and \ref{sec:astro}.

\subsection{LHC constraints at 8 TeV}
\label{sec:LHC_8TeV}

In this paragraph we calculate the cross-sections for the relevant channels at $8 \; \TeV$ and compare them to the current bounds from LHC. 
The bounds we use throughout this analysis are:
\begin{itemize}
  \item {\it Mono-Z}: $pp \to   \chi^{0}_{1}  \chi^{0}_{1} \,+\, (Z\to l^{+} l^{-})$, $l=e, \, \mu $, with cross-section $\lesssim 0.27 \, fb$~\cite{Aad:2014vka}.
  \item {\it Mono-W}: $pp \to   \chi^{0}_{1}  \chi^{0}_{1} \,+\, (W \to \mu \nu_{\mu} )$, with  cross-section  
  $\lesssim 0.54 \;fb$~\cite{ATLAS:2014wra}.
  \item {\it Hadronically decaying} $Z/W$: $pp \to   \chi^{0}_{1}  \chi^{0}_{1} \,+\, (W/Z \to hadrons)$, with  $\sigma_{\slashed{E}_T+hadrons} \lesssim 2.2 \;fb$~\cite{Aad:2013oja}.
  \item {\it Dijet}: $pp \to   \chi^{0}_{1}  \chi^{0}_{1} \,+\, 2 \, jets$, with $\lesssim 4.8 \;fb$~\cite{Aad:2015txa}.
  \item {\it Mono-jet}: $pp \to   \chi^{0}_{1} \, (\chi^{0}_{2} \to \chi^{0}_{1} \,+\, \nu \bar{\nu}) \,+\, jet$ with  
  $\sigma_{\slashed{E}_T+jet} \lesssim 6.1 \,fb$~\cite{Khachatryan:2014rra}.
\end{itemize}
%
%%%%%%%%%%%%%%%%%%%%%%%%%%%%%%%%
\begin{figure}[t]
% \centering
\hspace*{-0.5cm} \begin{subfigure}[b]{0.52\textwidth}
        \includegraphics[width=\textwidth]{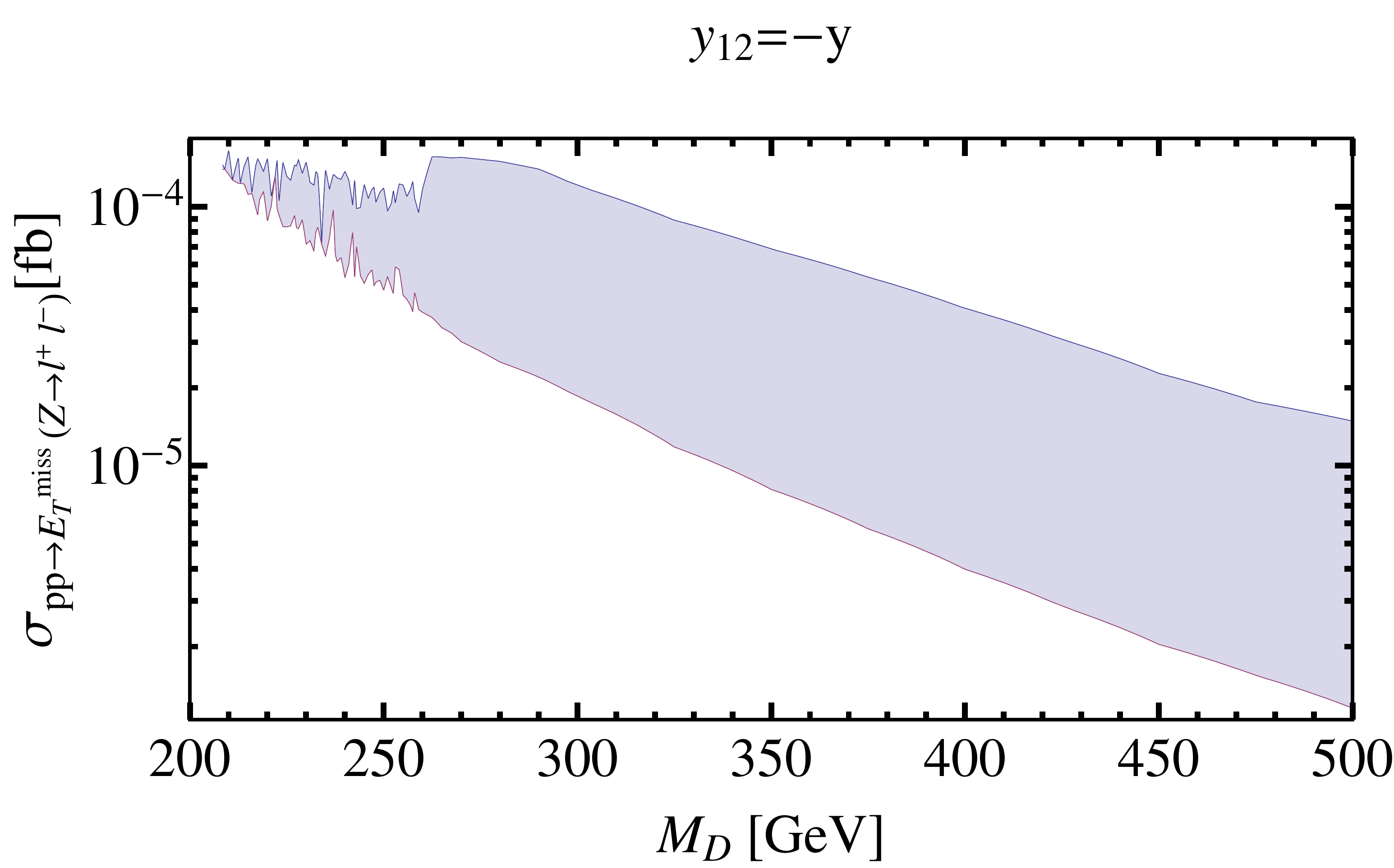}
           \caption{}
%         \label{}
    \end{subfigure}
\hspace*{0.0cm} \begin{subfigure}[b]{0.52\textwidth}
        \includegraphics[width=\textwidth]{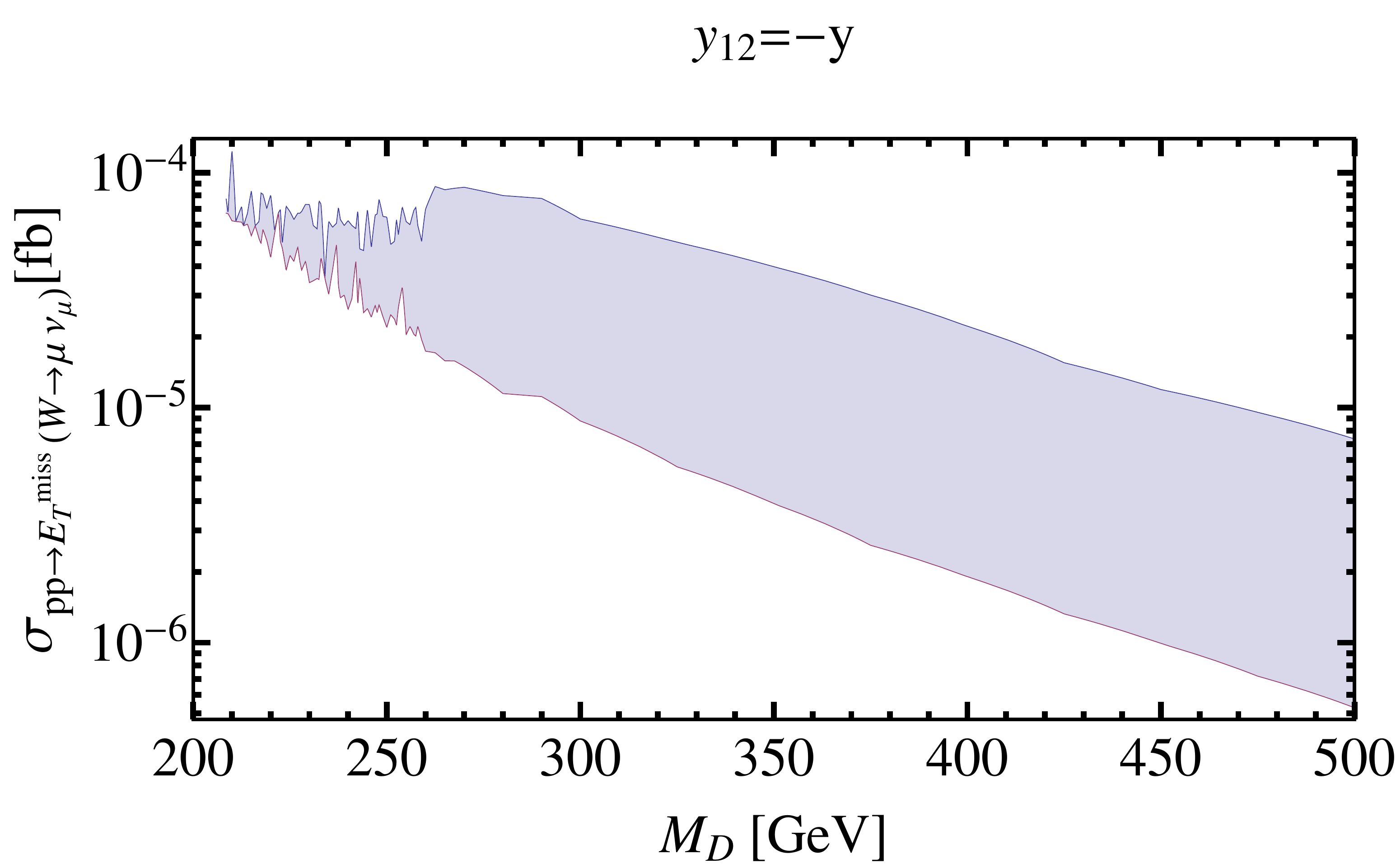}
           \caption{}
%         \label{}
    \end{subfigure} \\[2mm]
\hspace*{-0.5cm} \begin{subfigure}[b]{0.52\textwidth}
        \includegraphics[width=\textwidth]{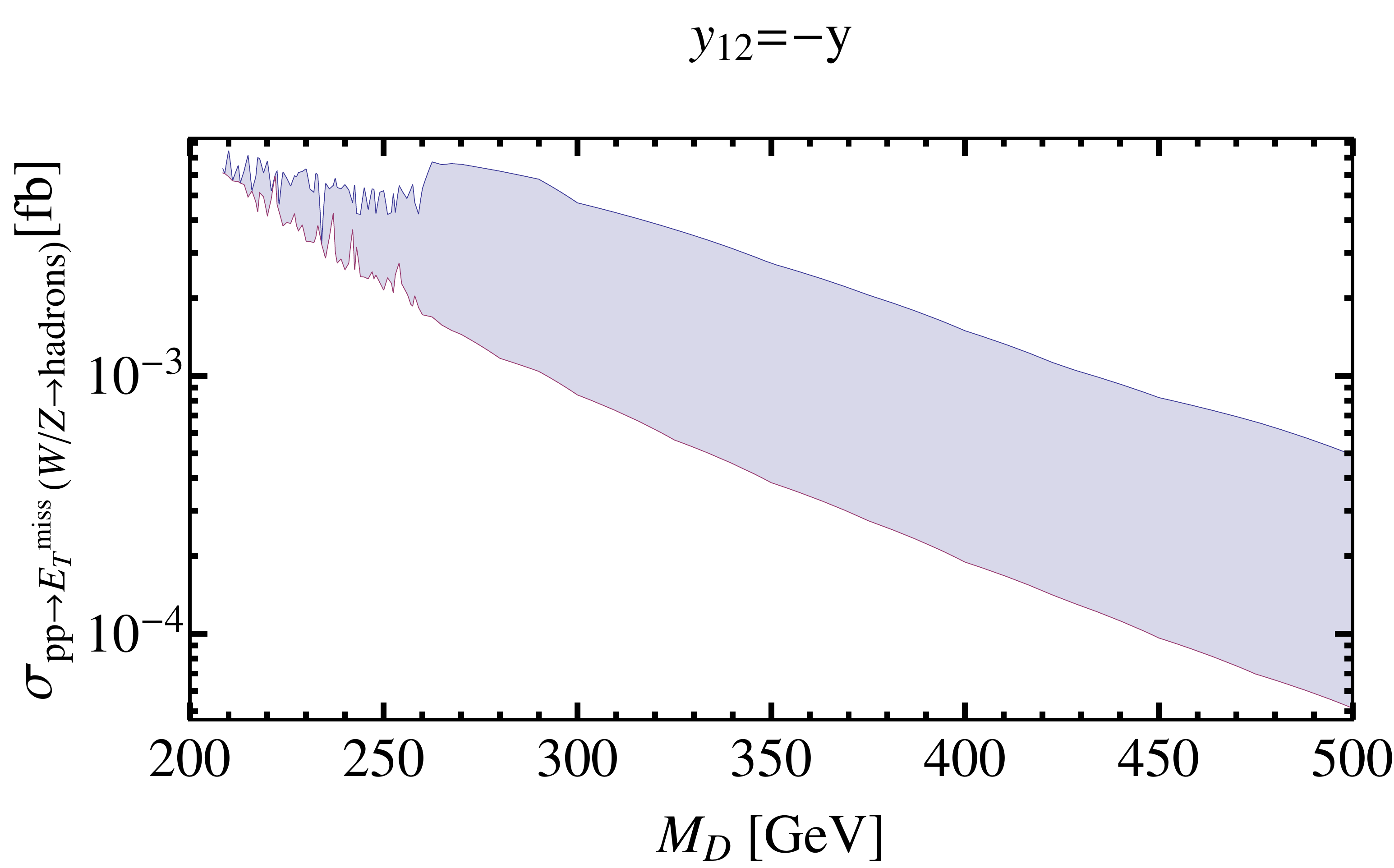}
           \caption{}
%         \label{}
    \end{subfigure}
\begin{subfigure}[b]{0.52\textwidth}
        \includegraphics[width=\textwidth]{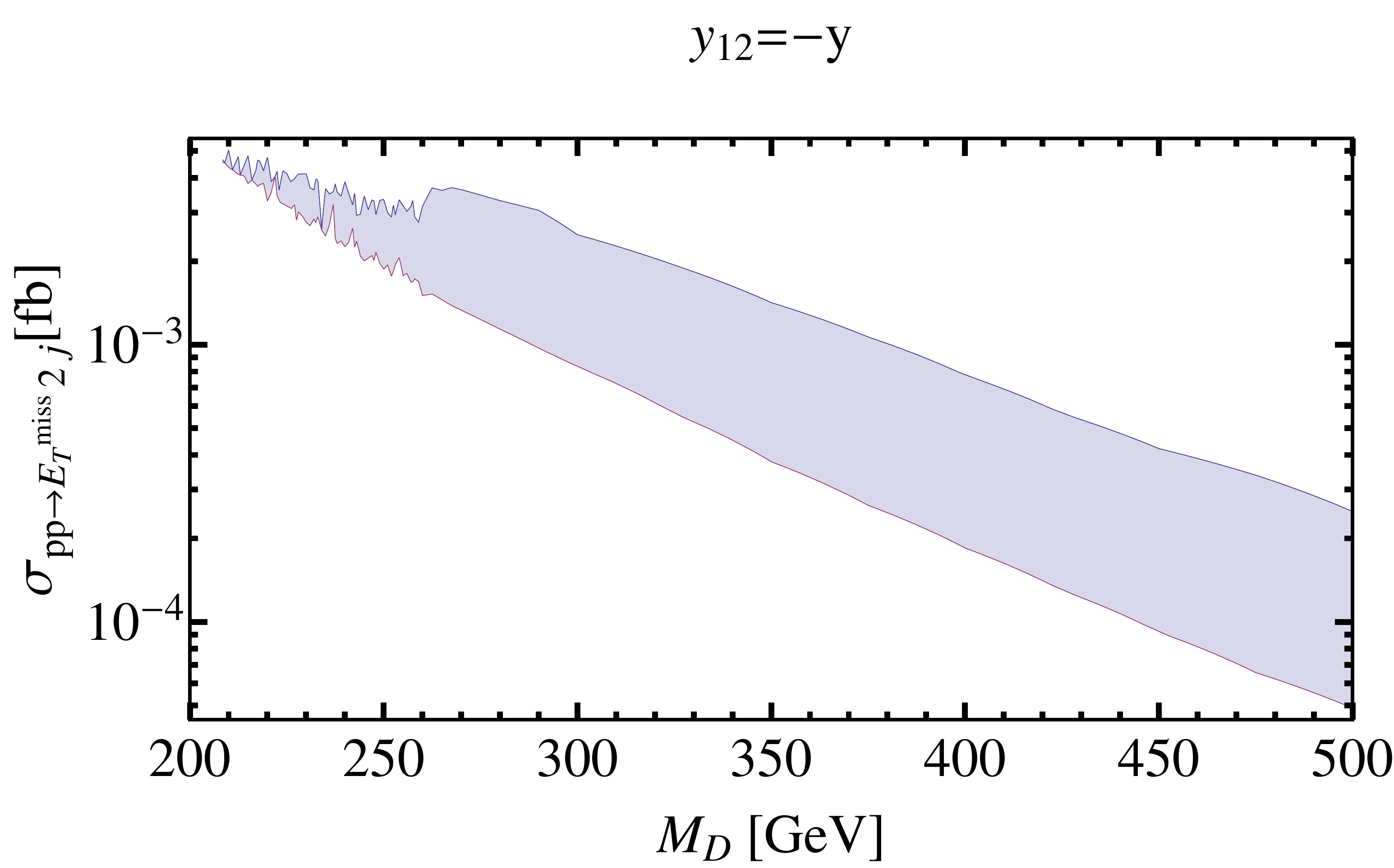}
           \caption{}
%         \label{}
    \end{subfigure}
\caption{\em The cross-sections for the (a) mono-Z, (b) mono-W, (c) hadronically decaying 
$Z/W$ and (d) the dijet processes at $\sqrt{\hat{s}}=8 \TeV$ versus the doublet-mass parameter $M_D$, while all the 
        other parameters run freely and for $y_{12}=-y$. 
        The other case with $y_{12}=0$ gives almost identical results.  
        The ``spikes'' appeared   is a result of varying a random selection of parameters.}
        \label{small-LHC}
\end{figure}
%%%%%%%%%%%%%%%%%%%%%%%%%%%%%
The cross-sections for the first four channels in the allowed parameter space are shown in Fig.~\ref{small-LHC}.
It is apparent that the current bounds of LHC for these processes cannot put any further restrictions to the allowed parameter space. On top of that, as $M_D$ becomes larger, the cross-sections decrease. There are two reasons for this. First, as $M_D$ increases, the masses increase,  and, second, 
the dipole moments $d_{W}$ and $ d_{\gamma}$ relevant for the observed relic abundance, move to smaller values as $M_D$ becomes larger (see Fig.~\ref{MDvsdw}), which reduces the interaction strength of the WIMP to  the gauge bosons.

We should point out that we only calculate the cross-sections of the hard processes (before showering, 
jet reconstruction, etc.).\footnote{For the calculation we use the program CalcHEP v.3.6 of \Ref{Belyaev:2012qa}.}
This means that in general, the actual cross-sections should be smaller than the ones we present here, 
since the cuts we are able to use for the hard processes are weaker than the cuts used in the experimental analyses.

%%%%%%%%%%%%%%%%%%%%%%%%%%%%%%%%%%%%%%%
\begin{figure}[H]
% \centering
\hspace*{-0.5cm} \begin{subfigure}[b]{0.52\textwidth}
        \includegraphics[width=  \textwidth]{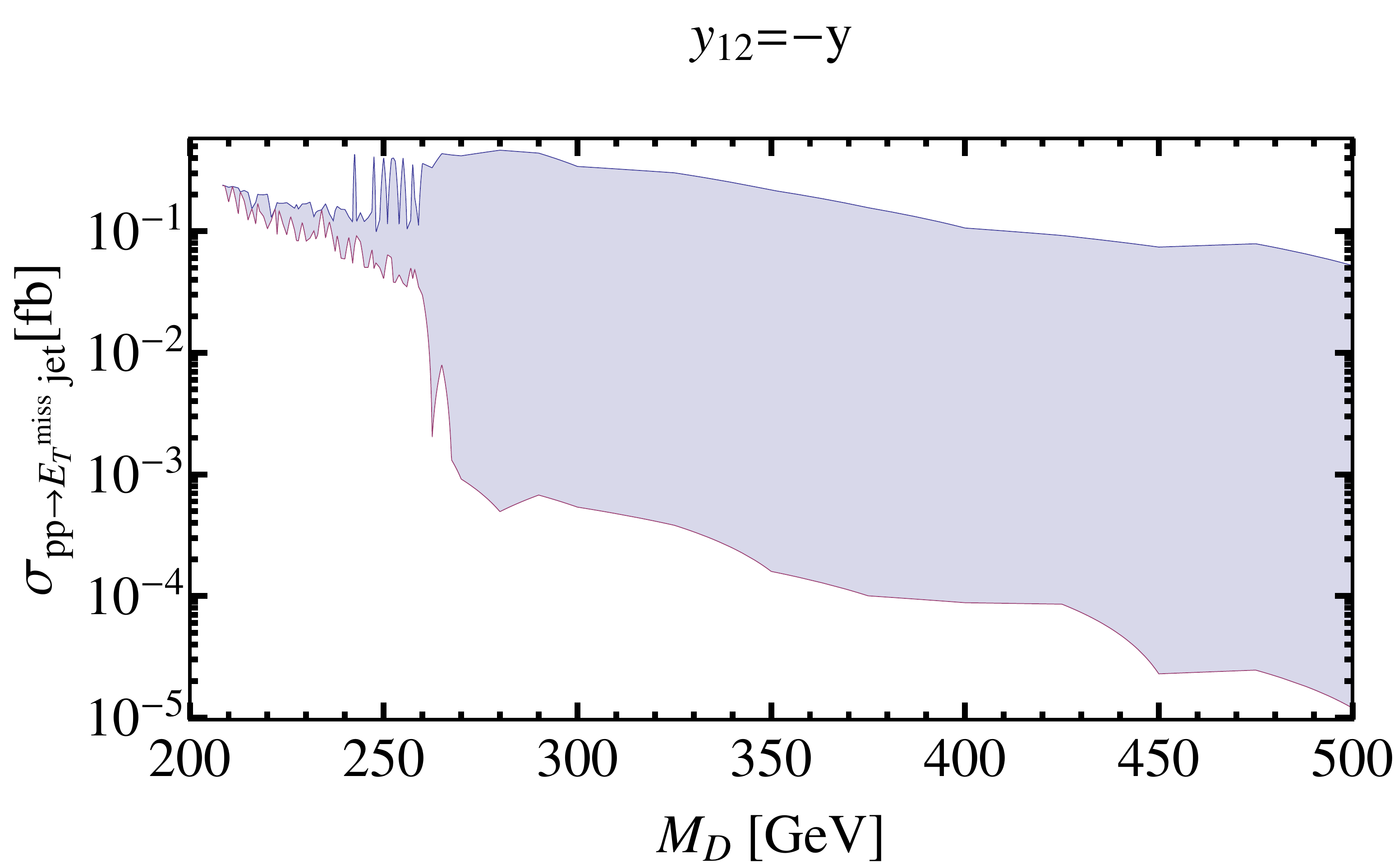}
           \caption{}
         \label{mono-jet_8TeV}
    \end{subfigure}
\hspace*{-0.0cm} \begin{subfigure}[b]{0.52\textwidth}
        \includegraphics[width=  \textwidth]{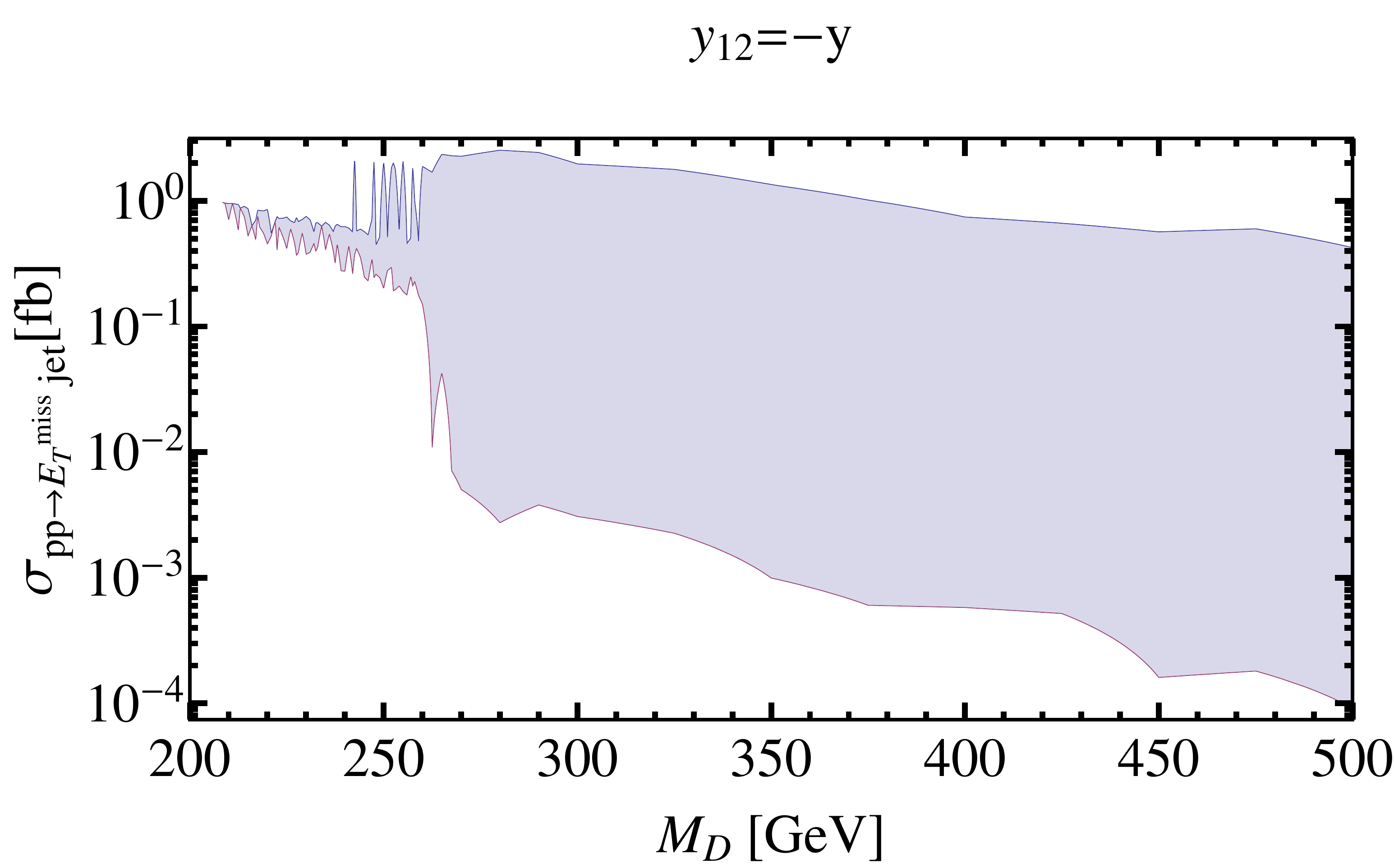}
           \caption{}
         \label{mono-jet_13TeV}
    \end{subfigure}
\caption{\em  As in Fig.~\ref{small-LHC} for the mono-jet channel with (a) $\sqrt{\hat{s}}=8 \TeV$ and (b)  $\sqrt{\hat{s}}=13  \TeV$. }
        \label{mono-jets}
\end{figure}
%%%%%%%%%%%%%%%%%%%
The cross-section for the mono-jet channel\footnote{We approximate this cross-section by 
$\sigma_{pp \to \chi_{1}^{0} \chi_{2}^{0} } \times\displaystyle\sum_{i=e, \mu , \tau}  BR_{ \chi_{2}^{0} \to  \chi_{1}^{0} \bar{\nu}_i \nu_i}$.} 
is shown in Fig.~\ref{mono-jet_8TeV}.
Again, as it can be seen, the cross-section is significantly smaller than the current bound from LHC. Additionally, similar to the other channels discussed here, as $M_D$ increases, the cross-section tends to decrease. But, since this cross-section depends strongly on both the dipole moments, $d_\gamma$ and $d_W$,
 and  the Yukawa coupling $y$ (through the branching ratio of $\chi_{2}^{0} \to \chi_{1}^{0} +\nu \bar{\nu}$), 
the shaded area is larger than the areas in Fig.~\ref{small-LHC}, because the available values of $y$ do not depend strongly on $M_D$ (see Fig.~\ref{MDvsy1}).
Also, it is apparent from Figs.~\ref{small-LHC} and \ref{mono-jets}(a), that for future DM searches at the LHC (for the model we study here), the mono-jet channel seems to be the most promising, 
since it could result to the largest number of events compared to other channels discussed here.

\subsection{Mono-jet searches at 13 TeV}

For LHC (RunII) with  $\sqrt{\hat{s}}=13\;\TeV$, the mono-jet channel provides the biggest number of events when compared to other channels.
From Fig.~\ref{mono-jet_13TeV}, we observe that the production 
of a jet accompanied with missing $E_T$, can reach cross sections  up to $\sim 2.5 \, fb$ for both cases $y_{12}=0$ and $y_{12}=-y$
and for $M_D\approx 300 \GeV$. 
This means that the number of events that can, in principle, 
be observed~\footnote{Very recently, a mono-jet+photon  search has been
proposed  in \Ref{Ismail:2016zby}. Emphasised for higgsinos, this final state can often be
as competitive as the  monojet channel.}
 is around $250 \, (750) $ for LHC  expected luminosity 
reach of $100 \, (300) \; fb^{-1}$.
%%%%%%%%%%%%%%%%%%
%
%
%  Furthermore, there are  other proposed  searches that could also  give rise to missing energy signals. Such a process is the 
% mono-jet+{\em soft} photon considered in~\Ref{Ismail:2016zby}. It is shown there, that even though the number of events
% may be smaller for this channel than the mono-jet one, the signal-to-background ratio is increased. {\sak \bf This is incomplete:
% if it is similar to monojets it has to be taken seriously. Dimitris mentioned it is supressed in our case; why this is not written here?}

%
 \begin{figure}[t]
    \centering
    \includegraphics[width=0.3\linewidth]{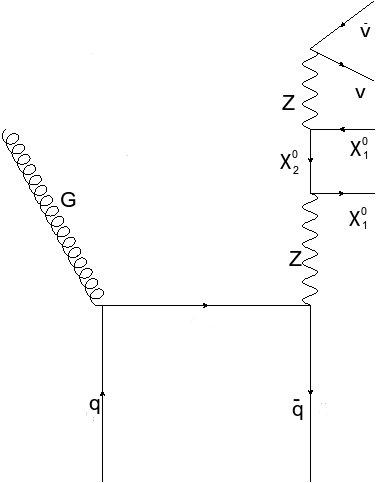}
  \caption{\em A Feynman diagram for the mono-jet process. The time ``runs'' upwards.}
     \label{mono-jet-diagram}
  \end{figure}

Before closing this section, we should remark issues about the validity of our calculations at such high center-of-mass  energy.
The validity of calculations for such theories at the LHC depends on the cut off energy and the couplings.
The energy for which the calculation of an observable  becomes invalid is $\sim {\Lambda  }/{C}$,\footnote{This holds under 
the assumption that the couplings of the {\it UV complete} model are $\sim 1$.} where $C$ is the Wilson coefficient for the relevant operator.
In our case, and for the mono-jet searches, the relevant $d=5$ term   (a Feynman diagram is shown in Fig.\ref{mono-jet-diagram}) is
$C\: \chi_{1}^{0} \: \sigma_{\mu\nu} \: \chi_{2}^{0} \: F_{Z}^{\mu\nu}$ with $C \sim c_{W} d_W + s_{W} d_{\gamma}$. 
Thus, if  the pair of $\chi_1^0\, \chi_2^0$ particles %($\chi_{1}^{0} \chi_{2}^{0}$, since we assume that $\chi_{2}^{0}$ is produced on-shell and then decays)
are produced with energy larger than $\sim {\Lambda }{(c_{W} d_W + s_{W} d_{\gamma})^{-1}}$, the calculation is considered to be inaccurate.
%
%%%%%%%%%%%%%%%%%%%%%%%%%%%%%%%%%%%%%%%%%%%%%%%%%%%
 \begin{figure}[t]
    \centering
    \includegraphics[width=0.6\linewidth]{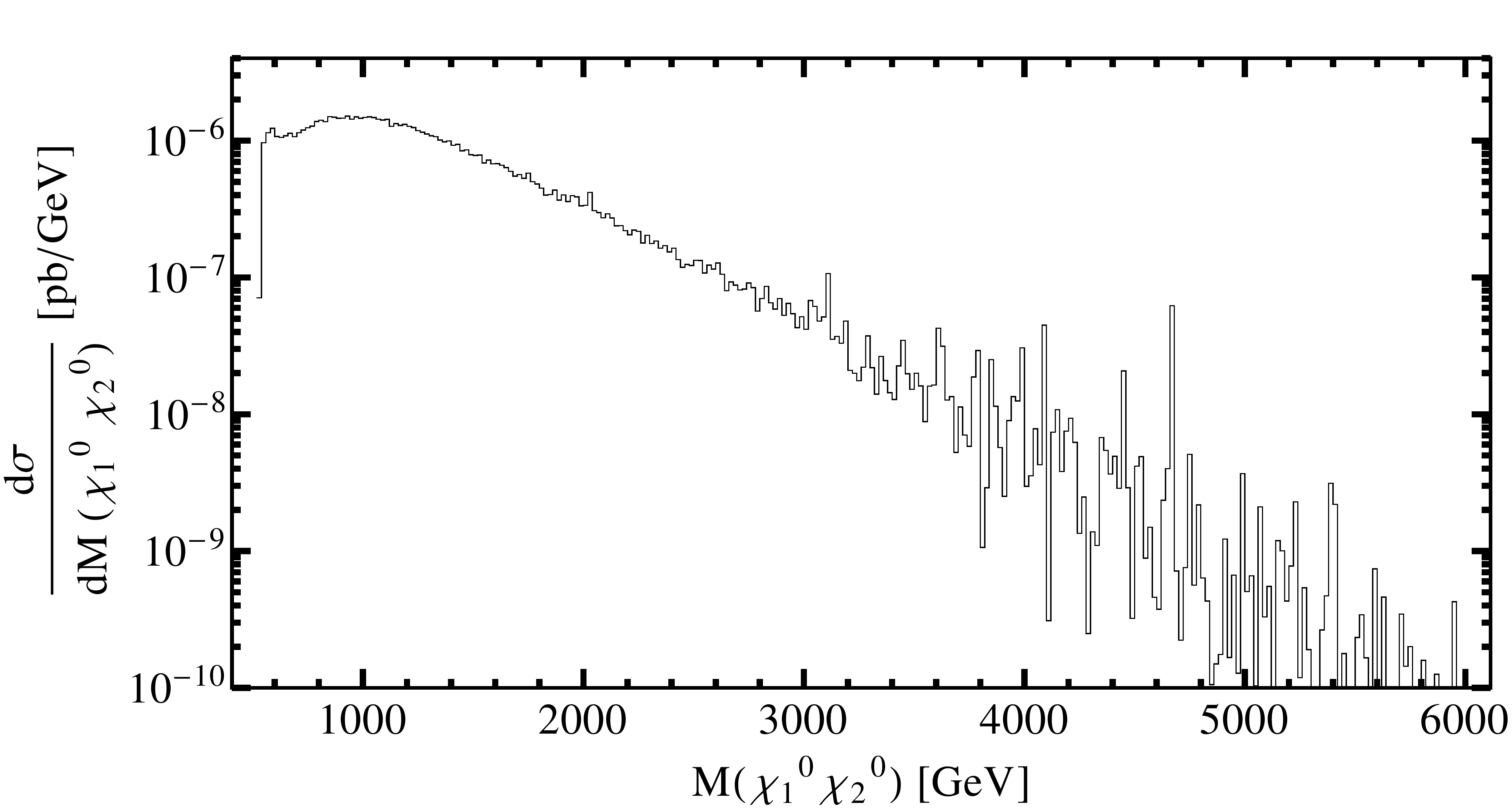}
  \caption{\em The dependence of the mono-jet differential cross-section, 
  ${d\sigma}/{dM(\chi_{1}^{0} \chi_{2}^{0})}$, on the invariant mass for the 
  pair of the neutral fermions, for $\sqrt{\hat{s}}=13 \TeV$,  $M_D=250 \GeV$, $d_W=0.45$ and $d_{\gamma}=0.25$.}
     \label{mono-jet-inv_mass}
  \end{figure}
%%%%%%%%%%%%%%%%%%%%%%%%%%%%%%%%%%%%%%%%%%%%%%%%%%%                           
In order to understand this, a numerical example is given in Fig.~\ref{mono-jet-inv_mass}, where the dependence 
of the differential cross-section on the invariant mass of the dark sector particles (which measures the energy that 
would be transferred by the integrated out particle) is 
shown for $M_D=250 \; \GeV$, $d_W=0.45$ and $d_{\gamma}=0.25$. 
We observe that above $\sim 2 \, \TeV$, the mono-jet differential 
cross-section falls rapidly, and, the main contribution to the inclusive cross-section, around $85\%$  for this particular example,  arises
 for invariant masses with  $M(\chi_{1}^{0} \chi_{2}^{0})\lesssim 2 \TeV$. 
 In addition, since  $C \approx 0.5$ and $\Lambda=1 \TeV$, the energy scale where this calculation 
 is inaccurate is $ \frac{\Lambda}{C} \sim 2 \TeV$,
 and therefore  this calculation is, in principle, reliable.

Furthermore,  the limit discussed above could be different, since the expansion of the {\it $UV$ complete} model is generally written 
in powers of $\frac{\lambda}{M}$, where $\lambda$ is a generic coupling (or a function of couplings) 
of this model and $M$ is the mass of the particle which is integrated out. The convergence of this
 expansion depends on the value of $\frac{\lambda}{M}$ which is, in principle,  different from $\frac{C}{\Lambda}$. 
An extensive discussion on the limitations of  effective theories at the LHC can be found in \Refs{Busoni:2013lha,Busoni:2014sya}.
Finally, as shown in \Ref{Buchmueller:2013dya}, there are cases where the decay width of the  particle that is integrated out
vastly affects the cross-section. There are also UV independent bounds coming from unitarity, 
discussed in \Refs{Endo:2014mja,Bell:2016obu}.  A detailed study of these effects is beyond the scope of this paper.

%%%%%%%%%%%%%%%%%%%%%%
\section{Conclusions}
\setcounter{equation}{0}
\label{sec:concl}
%%%%%%%%%%%%%%%%%%%%%

We have introduced in the SM particle spectrum a fermionic bi-doublet: a pair of Weyl fermion 
$SU(2)_L$-doublets, $\mathbf{D}_1$ and $\mathbf{D}_2$, with opposite hypercharges. 
In addition, we assume a discrete $Z_2$-symmetry that distinguishes $\mathbf{D}_1$ and $\mathbf{D}_2$ from the SM 
fields. This anomaly free set of fermions, together with the $Z_2$-symmetry are quite common features in 
non-supersymmetric SO(10) GUT constructions for light dark matter. Light $SU(2)_L$ doublets, whose components are parts
of the WIMP have been also considered countless of times in ``UV-complete'' non-supersymmetric 
or supersymmetric models (\ie higgsino dark matter).
Our work is related to these UV models when all other particles but the doublets have been integrated out in their low energy spectrum.

At the renormalizable level the mass spectrum consists of a electromagnetically neutral, and a charged Dirac, fermions. 
Under the presence of $d=5$ operators, the neutral Dirac fermion is split into two Majorana states, 
the WIMP, $\chi_1^0$, and its excited state,  $\chi_2^0$. 
Moreover, the $d=5$ operators include magnetic and electric dipole transitions which 
are, in principle, generated by a UV-complete theory, possibly at the TeV scale.
We ask here the question whether the dark matter particle $\chi_1^0$, with mass $(m_{\chi_1^0})$, 
{\emph {around the EW scale}}, is compatible  to various collider,  astrophysical and cosmological data.

In order to reduce fine tuning and extensive scans of the parameter space, 
in section~\ref{Spectrum-section} we adopted four scenarios, a,b,c and d,
based on well motivated symmetry limits of the theory such as a charge conjugation  or a custodial 
symmetry that act on $\mathbf{D}$'s and Higgs field $H$. 
 These low energy symmetries simplify enough the analytical expressions 
of the interactions and  possibly help to construct UV-completions of the model. After collecting all relevant
$d=5$, and  $d=6$  (though the latter 
not used in the analysis), operators in the \ref{sec:ope}, we went on to investigate their
implications into collider and astrophysical processes. 

In section~\ref{sec:earth},  we performed a constraint analysis based {\it (i)} on scattering 
WIMP-nucleus recoiling experiments, such as LUX, {\it (ii)} on LEP searches for
new fermions, as well as {\it (iii)}  on LHC searches for the decay $h\to \gamma\gamma$. 
Bounds on the model parameters \eqref{params} are
collected in Fig.~\ref{Combined}. Only in cases (b) and (c) there is still enough freedom to carry on. 
In the same section, we also studied contributions from the new fermion interactions into oblique electroweak $S,T$ and $U$
parameters. Only the $S$ parameter is affected, and, as a consequence, only case (a) is further constrained.

 Focused on   the more interesting cases (b) and (c), in section~\ref{sec:astro}  we calculated the relic density
 $\Omega h^2$ for $\chi_1^0$.  In the presence of $d=5$ dipole operators there are destructive  interference effects
 in the (dominant) amplitudes for WIMP annihilations (or co-annihilations) into SM vector bosons. The minima
 in the cross sections correspond to certain, usually non-zero, values for  the coefficients of the dipole operators
 $d_W$ and $d_\gamma$ [see \eqs{agammaz}{agammagamma}]. Nearby these minima the relic density is found
 to be consistent with observation [\eq{Planck}] for $m_{\chi_1^0} \gtrsim 200 \GeV$. Although continuous gamma
 ray spectrum constraints are harmless, constraints from monochromatic gamma ray spectrum are serious
 for the photon dipole coupling as it is shown in Figs.~\ref{gammagamma-gammaZ-gamma_rays_lines} and \ref{fig:gammagamma}. 
 The coefficient $d_W$ has to be more than $10\%$ a value which is non-negligible
 for UV models with Dark matter at the EW scale. $d_\gamma$ on the other hand can be tuned to zero 
 without a problem.
 
 Apart from possible aesthetics, the main reason in insisting for EW dark matter mass, $m_{\chi_1^0} \approx M_Z$,
 has to do with enhancing the possibility of observing the dark sector at the LHC (or, in any case, to be as close as
 visible in the RunII phase). In section~\ref{sec:LHC} we estimated the cross section for producing $\chi_1^0$ 
 at LHC with center of mass energy $\sqrt{\hat{s}} = 8, 13$ TeV and in  association with a jet (monojet) or 2 jets or
 a $W$ or a $Z$. We found that  the monojet process is the most promising   with a 
 few hundred of events at  $\sqrt{\hat{s}} = 13 \TeV$ and with $m_{\chi_1^0} \simeq 200 - 350 \GeV$ (see 
 Fig.~\ref{mono-jets}). 
 
Searching for dark matter and/or related particles at LHC consists in a major effort from 
physicists in high energy physics and astrophysics. An effective field theory for an electroweak
dark matter described in this article may guide us closing that goal.  

\vspace*{3cm}
%%%%%%%%%%%
\section*{Acknowledgements}

AD would like to thank CERN Theory Division for the  kind hospitality and Apostolos Pilaftsis for
useful discussions. DK would like to thank Alexander Pukhov for his helpful advices in setting up CalcHEP.

This research has been co-financed by the
European Union (European Social Fund - ESF) and Greek national funds
through the Operational Program ``Education and Lifelong Learning" of
the National Strategic Reference Framework (NSRF) - Research Funding
Programs: THALIS and ARISTEIA - Investing in the society of knowledge
through the European Social Fund.

%%%%%%%%%%%%%%%%%%%%%%%%%%%%%%%%%%%%%%%%
% APPENDICES
\newpage
%%%%%%%%%%%%%%%%%%%%%%%%%%%%%%%%%%%%%%%%
\renewcommand{\thesection}{Appendix~\Alph{section}}
\renewcommand{\theequation}{\Alph{section}.\arabic{equation}}

\setcounter{equation}{0}  % reset counter
\setcounter{section}{0}
\bigskip

%%%%%%%%%%%%%%%%%%%%%%%
\section{Non-renormalizable operators}
\setcounter{equation}{0}
\label{sec:ope}

%%%%%%%%%%%%%%%%%
%%%%%%%%%%%%%%%%%%%%%%%%%%%%%%%%%
%\section{Non-renormalizable operators}
%\setcounter{equation}{0}
%\label{sec:ope}
%%%%%%%%%%%%%%%%%%%%%%%%%%%%%%%%%

Apart from the mass term in \eq{eq:md}, and the renormalizable couplings to gauge 
bosons discussed in section~\ref{sec:pheno}, 
the $``D"$-doublets   couple to the  bosons of the theory through non-renormalizable $d=5,6$ interactions.
Gauge numbers, denoted as $(SU(3)_C,SU(2)_L)_{U(1)_X}$, for the  particles here are:  for quarks $\mathbf{Q\sim (3^c,2)_{\frac{1}{3}}}$,
 $\mathbf{\bar{u}\sim (3^c,1)_{-\frac{4}{3}}}$, $\mathbf{\bar{d}\sim (3^c,1)_{+\frac{2}{3}}}$, for leptons: 
 $\mathbf{L\sim (1^c,2)_{-1}}$, $\mathbf{\bar{e}\sim (1^c,1)_{+2}}$,  for the Higgs doublet: $\mathbf{H\sim (1^c,2)_{+1}}$
and  finally for the new bi-doublets:  $\mathbf{D_1\sim (1^c,2)_{-1}}$ and $\mathbf{D_2 \sim (1^c,2)_{+1}}$.
Schematically,  the possible  interactions are: 
$ffHH$, $ff \mathscr{D} H$,  $ff \mathscr{D}\mathscr{D}$,
where $\mathscr{D}$ is the covariant derivative acting in both Weyl fermions $f$
or to the Higgs fields. We  arrange all Weyl fermions $f$ to be left-handed.  

We list below  all relevant possible independent $d=5$ and $d=6$ operators. An analogous list has been constructed in
\Ref{Duch:2014xda} but for the fermionic \emph{singlet} extension of the SM. The complete set of $d=5,6$ Standard Model operators
can be read from \Ref{Grzadkowski:2010es}.

%%%%%%%%%%%%%%%%%%%%%%%%%%%%%%%%%
\subsection{$d=5$ non-renormalizable operators}
%%%%%%%%%%%%%%%%%%%%%%%%%%%%%%%%%

\begin{itemize}
\item $ff HH$ :
The $d=5$ operators alter the DM mass  spectrum and the Higgs-boson interactions 
with the dark sector obtained for $f=D_{1}, D_{2}$, when integrating out heavy particles.
Examples of possible simplified models that result into these operators, are obtained 
by integrating out fermion  neutral singlets ($S_0$) and triplets ($T$), fermion charged  singlets ($S^\pm$) and triplets ($T^\pm$),
or scalar  singlets, ($\Phi_{S_0}, \Phi_{S^\pm})$ and triplets,  ($\Phi_{T^0}, \Phi_{T^\pm}$). In fully $SU(2)_L$-invariant form we have
%
%%\footnote{$\epsilon^{12}=-\epsilon^{21}=1$ is the antisymmetric tensor.}

%%%%%%%%%%%%%%%%%%%%%
\begin{align}
&- \mathscr{L}_{\rm dim=5} \   \supset  \
 + \frac{\lambda_{1}^{2}}{2\, M_{S_{0}}} \: ( \epsilon^{ab} H_{a} D_{1b} )\: (\epsilon^{cd} H_{c} D_{1d} )
\ + \ \frac{\lambda_{2}^{2}}{2 M_{S_{0}}} \: (H^{\dagger a} D_{2a} )\: (H^{\dagger b} D_{2b} )
\nonumber \\[2mm]
& + \frac{\lambda_{12}}{M_{S_{0}}} \:  ( \epsilon^{ab} H_{a} D_{1b} )\: (H^{\dagger c} D_{2c} ) 
\ +\ \frac{\lambda^{\prime}_{12}}{M_{S^{\pm}}} \: 
( \epsilon^{ab} H_{a} D_{2b} )\: (H^{\dagger c} D_{1c} )
\nonumber \\[2mm]
& + \frac{Y_{1}^{2}}{2\, M_{T}} \: [ \epsilon^{ab} H_{a} (\tau^{A})^{c}_{b} D_{1c} ]\: 
                                               [\epsilon^{fg} H_{f} (\tau^{A})^{h}_{g} D_{1h} ]
\ +\ \frac{Y_{2}^{2}}{2\, M_{T}} \: [H^{\dagger a} (\tau^{A})_{a}^{b} D_{2b} ] \:
[H^{\dagger c} (\tau^{A})_{c}^{d} D_{2d} ]
\nonumber \\[2mm]
&+\frac{Y_{12}}{M_{T}} \: [ H^{\dagger a}  (\tau^{A})_{a}^{b} D_{2b} ] \:
                      [ \epsilon^{cd} H_{c}  (\tau^{A})_{d}^{f} D_{1f} ]
                      \ + \ 
                      \frac{Y^{\prime}_{12}}{M_{T^{\pm}}} \: [ H^{\dagger a}  (\tau^{A})_{a}^{b} D_{1b} ] \:
                      [ \epsilon^{cd} H_{c}  (\tau^{A})_{d}^{f} D_{2f} ]
\nonumber \\[2mm]
%& -\frac{Y_{12}^{2}}{M_{T}} \: [ H^{\dagger a}  (\tau^{A})_{a}^{b} D_{2b} ] \:
 %                     [ \epsilon^{cd} H_{c}  (\tau^{A})_{d}^{f} D_{2f} ]              
%\nonumber \\[2mm]
& + \frac{\xi_{12}}{M_{\Phi_{0}}} \: (\epsilon^{ab} D_{1a} D_{2b} ) (H^{\dagger c} H_{c}) 
\nonumber \\[2mm]
& + \frac{k_{1}^{2}}{2\, M_{\Phi_{T}^{\pm}}} \: [ \epsilon^{ab} D_{1a} (\tau^{A})^{c}_{b} D_{1c} ]\: 
                                               [\epsilon^{fg} H_{f} (\tau^{A})^{h}_{g} H_{h} ]
+ \frac{k_{2}^{2}}{2\, M_{\Phi_{T}^{\pm}}} \: [\epsilon^{ab} D_{2a} (\tau^{A})_{b}^{c} D_{2c} ] \:
[H^{\dagger d} (\tau^{A})_{d}^{f} \epsilon_{fg} H^{\dagger g} ]
\nonumber \\[2mm]
&+ \frac{k_{12}}{M_{\Phi_{T_{0}}}}\: [\epsilon^{ab} D_{1a} (\tau^{A})_{b}^{c} D_{2c} ]\:
[H^{\dagger d} (\tau^{A})_{d}^{g} H_{g} ]  \ + \
\frac{k^{\prime}_{12}}{M_{\Phi_{T_{0}}}}\: [\epsilon^{ab} D_{2a} (\tau^{A})_{b}^{c} D_{1c} ]\:
[H^{\dagger d} (\tau^{A})_{d}^{g} H_{g} ] 
\nonumber \\[2mm]
&\ + \ \mathrm{H.c.} \;,
\label{eq:op1}
\end{align}
%%%%%%%%%%%%%%%%%%%%%%%
where the meaning of various mass scales is rather obvious e.g., those suppressed by 
$M_{S_{0}}, M_{S^{\pm}}$ and  
$M_{T,T^{\pm}}$ are derived from integrating out  heavy fermionic neutral and/or 
charged singlets and  triplets $S_{0}, S^{\pm}$ and, $T, T^{\pm}$ respectively,  and so on.

However, not all operators in eq.~(\ref{eq:op1}) are independent;  in fact most of them are not.
Using a standard  identity  for   Pauli matrices,
%%%%%%%%%
$(\tau^{A})_{ab} (\tau^{A})_{cd} = 2 (\delta_{ad} \delta_{bc} -\frac{1}{2} \delta_{ab}\delta_{cd})$, one can arrive at
the most general form of \eqref{eq:op1} written as
%%%%%%%%%%%%%%%%%%%%%
\begin{align}
& -\mathscr{L}_{\rm dim=5} \   \supset  \
 \ + \ \frac{y_{1}}{2\, \Lambda } \: ( \epsilon^{ab} H_{a} D_{1b} )\: (\epsilon^{cd} H_{c} D_{1d} )
\ + \ \frac{y_{2}}{2 \Lambda } \: (H^{\dagger a} D_{2a} )\: (H^{\dagger b} D_{2b} )
\nonumber \\[2mm]
& \ + \ \frac{y_{12}}{\Lambda } \:  ( \epsilon^{ab} H_{a} D_{1b} )\: (H^{\dagger c} D_{2c} ) 
%\ +\ \frac{\kappa_{12}}{\Lambda } \: 
%( \epsilon^{ab} H_{a} D_{2b} )\: (H^{\dagger c} D_{1c} )
%\nonumber \\[2mm]
\ + \ \frac{\xi_{12}}{\Lambda} \: (\epsilon^{ab} D_{1a} D_{2b} ) (H^{\dagger c} H_{c}) 
%\nonumber \\[2mm]
\ + \ \mathrm{H.c.}
\label{eq:op2}
\end{align}
%%%%%%%%%%%%%%%%%%%%%%%
where we use a common mass scale $\Lambda$ at which heavy particles are integrated
out and the complex valued Yukawa couplings $y_{1},y_{2},y_{12},\xi_{12}$.    
 We should also remark that the last 
operator in \eqref{eq:op2}  is somewhat  trivial and it can appear in 
any powers of the Higgs polynomial. At EW vacuum it adds 
 a common mass to $D_{1}$ and $D_{2}$ as in \eq{eq:md} does.
 All operators in \eqref{eq:op2} give masses to neutral components of the WIMPs
 except from the last one that gives  mass also to the charged components.

Furthermore,   in this class  belongs the famous 
Weinberg operator for neutrino masses,  with  $f=L$ being the SM lepton doublet
\begin{eqnarray}
\frac{y_{\nu}}{2\, \Lambda } \: ( \epsilon^{ab} H_{a} L_{b} )\: (\epsilon^{cd} H_{c} L_{d} )
\ + \ \mathrm{H.c.} 
\end{eqnarray}
The  origin of this operator is not necessarily related to the DM sector. 
Note that the first three terms in \eq{eq:op2}, can also be obtained by integrating 
out heavy right-handed  neutrino states, $\mathbf{\bar{\nu}\sim (1^c,1)_0}$, from renormalizable Yukawa couplings,
$H^{\dagger}\: {D}_{2}\: \bar{\nu} + H\: {D}_{1}\: \bar{\nu} + \mathrm{H.c.}$ as in the see-saw model for neutrino masses.

Of course there are additional terms, \eg $L D_{2} H H$, but these  in general, break
the $Z_{2}$-discrete (or lepton number) symmetry that keeps the DM particle stable. 
Interestingly enough,  these terms  are  connecting the DM particle  to  neutrinos, see for instance~\cite{Huang:2014bva}. 
These  independent operators are
%%%%%%%%%%%%
\begin{align}
 & \frac{\eta_{1}}{2\, \Lambda } \: ( \epsilon^{ab} H_{a} D_{1b} )\: (\epsilon^{cd} H_{c} L_{d} )
 \ + \ \frac{\eta_{12}}{\Lambda } \:  ( \epsilon^{ab} H_{a} L_{b} )\: (H^{\dagger c} D_{2c} ) 
\nonumber \\[2mm]
%&\ +\ \frac{\kappa_{12}}{\Lambda } \: 
%( \epsilon^{ab} H_{a} D_{2b} )\: (H^{\dagger c} L_{c} )
%
&\ + \ \frac{\zeta_{12}}{\Lambda} \: (\epsilon^{ab} L_{a} D_{2b} ) (H^{\dagger c} H_{c}) 
%\nonumber \\[2mm]
\ + \ \mathrm{H.c.}
\label{leptonDM}
\end{align}
%%%%%%%%%%%%%%%

\item $ff\mathscr{D}H$ : In this case the fermion bilinear must be a weak doublet
with hypercharge $-1$. The only such combination, 
$D_{2}^{\dagger} \bar{\sigma}^{\mu} \bar{e} \mathscr{D}_{\mu} H^\dagger + \mathrm{H.c.}$, 
is not invariant under  the $Z_{2}$-symmetry.

\item $ff \mathscr{D} \mathscr{D}$ : Under $Z_{2}$-symmetry there are three
possibilities : $D_{1}  \mathscr{D} \mathscr{D} D_{2}, \mathscr{D} \mathscr{D} D_{1} D_{2}$ and
$\mathscr{D} D_{1} \mathscr{D} D_{2}$. After some algebra, and taking the equations of motion into 
account we find that these lead to dipole operators of the form
%%%%%%%%%%%%%%%%
\begin{align}
\frac{d_{\gamma}}{\Lambda}\: \epsilon^{ab} D_{1a} \: \sigma^{\mu\nu}  \: D_{2b}\: 
 B_{\mu\nu} & \ + \
\frac{d_{W}}{\Lambda}\: \epsilon^{ab} D_{1a} \: \sigma^{\mu\nu} \: (\tau^{A})_{b}^{\ c} \: D_{2c}\: 
 W_{\mu\nu}^{A} \ + \ %\mathrm{H.c.}\;, 
 \nonumber \\[2mm]
 \frac{i\, e_{\gamma}}{\Lambda}\: \epsilon^{ab} D_{1a} \: \sigma^{\mu\nu}  \: D_{2b}\: 
 \widetilde{B}_{\mu\nu} & \ + \
\frac{i\, e_{W}}{\Lambda}\: \epsilon^{ab} D_{1a} \: \sigma^{\mu\nu} \: (\tau^{A})_{b}^{\ c} \: D_{2c}\: 
 \widetilde{W}_{\mu\nu}^{A} \ + \ \mathrm{H.c.}\;,  
 \label{dipoles}
\end{align}     
%%%%%%%%%%%%%%%%%%%
where $B_{\mu\nu}$ and $W_{\mu\nu}^{A}$ are the $U(1)$ and $SU(2)_L$, field strength tensors,
respectively, and $\widetilde{B}_{\mu\nu} \equiv \epsilon_{\mu\nu}^{\ \ \rho\sigma}\, B_{\rho\sigma}$.
 These operators are electric and magnetic dipole moments for the DM particle. 
They arise directly at $d=5$ level,  whereas 
quark and/or lepton magnetic moments arise at $d=6$ level.

\end{itemize}
%%%%%%%%%%%%%%%%%%%%%%%
We have not found other than the above $d=5$ independent operators.

%%%%%%%%%%%%%%%%%%%%%%%%%%%%%%%%%
\subsection{$d=6$ non-renormalizable operators}
%%%%%%%%%%%%%%%%%%%%%%%%%%%%%%%%%
Focusing only in interactions between $f= D_{1}~\mathrm{or}~ D_{2}$ and the Higgs
 field\footnote{All others are identical to standard dimension-6 operators and can be found in
\cite{Grzadkowski:2010es}.} there are four Lorentz and gauge invariant categories:
$ff H^{3}$, $ff \mathscr{D} H^{2}$, $ff \mathscr{D}^{2} H$, $ff\mathscr{D}^{3}$, and of course
$ffff$.

\begin{itemize}

\item $ff H^{3}$ : There are no such operators which preserve the $Z_{2}$-symmetry, 
or, as a matter of fact, the charge conjugation or custodial symmetry or lepton number , e.g. there is 
$(H^{\dagger} D_{1} \bar{e}) (H^{\dagger}  H)$ and the one with triplets.

\item $ff \mathscr{D} H^{2}$ : There are quite a few invariant operators of this kind. The
independent ones  are %{\bf (please check!!)}
%%%%%%%%%%%%%%%
\begin{align}
&-\mathcal{L}_{\mathrm{dim}=6}  \ \supset \ 
\left (  \frac{a_{1}}{\Lambda^{2}} \: D_{1}^{\dagger a} \: \bar{\sigma}^{\mu}\: D_{1a} \ + \
\frac{a_{2}}{\Lambda^{2}} \: D_{2}^{\dagger a}\: \bar{\sigma}^{\mu}\: D_{2a} \right )\:
\left ( i\, H^{\dagger b}\:  (\overleftrightarrow{{\mathscr{D}}_{\mu}})_{b}^{\ c}\: H_{c} \right )
\nonumber \\[2mm]
& + \left (  \frac{a_{1}^{\prime}}{\Lambda^{2}} \: D_{1}^{\dagger a} \: (\tau^{A})_{a}^{\ b} \:
\bar{\sigma}^{\mu} \: D_{1b} \ + \
\frac{a_{2}^{\prime}}{\Lambda^{2}} \: D_{2}^{\dagger a}\: (\tau^{A})_{a}^{\ b}\:
 \bar{\sigma}^{\mu} D_{2b} \right )\:
\left (i\, H^{\dagger c}\:  (\overleftrightarrow{{\mathscr{D}}_{\mu}^{A}})_{c}^{\ d}\: H_{d} \right )
\nonumber \\[2mm]
%%%%%%%%%%%%%%%%%
&  + \frac{b_{1}}{\Lambda^{2}} \: (D_{2}^{\dagger a} \: \bar{\sigma}^{\mu}\: D_{1a} ) \:
 [\epsilon^{bc} H_{b}\:  (\overleftrightarrow{{\mathscr{D}}_{\mu}})_{c}^{\ d}\: H_{d}  ]
\ + \ \frac{b_{2}}{\Lambda^{2}} \: (D_{1}^{\dagger a} \: \bar{\sigma}^{\mu}\: D_{2a} )
\:( \epsilon_{bc}\: H^{\dagger b} \:  (\overleftrightarrow{{\mathscr{D}}_{\mu}}^{\dagger})^{c}_{\ d}\: 
H^{\dagger d} ) \ %+
%\nonumber \\[2mm]
%&  \frac{b_{1}^{\prime}}{\Lambda^{2}} \: (D_{2}^{\dagger a} \: (\tau^{A})_{a}^{\ b}
%\bar{\sigma}^{\mu}\: D_{1b} ) \:
 %[\epsilon^{bc} H_{b}\:  (\overleftrightarrow{{\mathscr{D}}_{\mu}^{A}} )_{c}^{\ d}\: H_{d}  ]
%\ + \ \frac{b_{2}^{\prime}}{\Lambda^{2}} \: (D_{1}^{\dagger a} \: (\tau^{A})_{a}^{\ b} \: 
%\bar{\sigma}^{\mu}\: D_{2b} )
%\:[ \epsilon_{cd}\: H^{\dagger c} \: ( \overleftrightarrow{{\mathscr{D}}_{\mu}^{A\, \dagger}} )_{\ e}^{d}\: 
%H^{\dagger e} ] \;,
%%%%%%%%%%%%%%%%%%%
\label{ffdh2}
\end{align}
%%%%%%%%%%%%%%
where $H^{\dagger}\: \overleftrightarrow{{\mathscr{D}}_{\mu}}\: H \equiv 
H^{\dagger}\: \overrightarrow{{\mathscr{D}}_{\mu}}\: H - 
H^{\dagger} \: \overleftarrow{{\mathscr{D}}_{\mu}}\: H$ and 
$H^{\dagger}\: \overleftrightarrow{{\mathscr{D}}_{\mu}^{A}}\: H \equiv 
H^{\dagger}\: \tau^{A} \: \overrightarrow{{\mathscr{D}}_{\mu}}\: H - 
H^{\dagger} \: \overleftarrow{{\mathscr{D}}_{\mu}}\: \tau^{A} \: H$, $a_{1,2}$ and $a_{1,2}^{\prime}$ are real numbers,
while $b_{2} = b_{1}^{*}$.
We can obtain new operators after changing $L\leftrightarrow D_{1}$ but
these would violate $Z_{2}$ or they would belong to existing SM operators given in \Ref{Grzadkowski:2010es}.

\item $ff \mathscr{D}^{2} H$ : Because $D_{1}$, $D_{2}$  and the $H$ are $SU(2)$-doublets 
only $Z_{2}$-breaking terms exist in this category,
e.g., $(D_{1} \sigma_{\mu\nu} \bar{e}) F^{\mu\nu} H^{\dagger}$ or when  the Higgs receives  
vev they reduce to $d=5$ operators already given in  \eqref{dipoles}.

\item $ff\mathscr{D}^{3}$ : We found no new operators. 
Lorentz invariance says that 
they exist only if $ff$ transforms as a vector e.g., 
$(D_{1}^{\dagger} \bar{\sigma}_{\mu} D_{1}) (\mathscr{D}_{\rho} B^{\rho \mu})$. 
By using equations of motion we get at most 
the operators of \eq{ffdh2}, or the four fermion operators, $ffff$, given  below
and/or  other like previously violating $Z_{2}$-symmetry. 
Acting with the covariant derivative to the left (on fermion current)  we obtain operators as in \eq{dipoles}.

\item $ffff$ :  we found the following independent operators: 
%%%%%%%%%%%%%%%%
\begin{align}
& -\mathcal{L}_{\mathrm{dim}=6}   \supset \  
\frac{c_{12}}{\Lambda^{2}}\: (\epsilon^{ab} D_{1a} D_{2b} ) \: (\epsilon^{cd} D_{1c} D_{2d} ) 
\ + \ \sum_{k,\ell=1}^{2} \frac{d_{k\ell}}{\Lambda^{2}}\: 
(D_{k}^{\dagger a} \:  \bar{\sigma}^{\mu} \: D_{k\, a}) \: 
(D_{\ell}^{\dagger b} \: \bar{\sigma}_{\mu}\: D_{\ell\, b}) \nonumber \\[2mm]
\ &+ \  \sum_{i,j=1}^{3}\sum_{k=1}^{2} \frac{1}{\Lambda^{2}}\:
(D_{k}^{\dagger a} \:  \bar{\sigma}^{\mu} \: D_{k\, a}) \:
\left [ f^{\ell}_{kij} (\ell^{\dagger}_{i} \bar{\sigma}_{\mu} \ell_{j} ) + f^{q}_{kij}(q^{\dagger}_{i} \bar{\sigma}_{\mu} 
q_{j} ) \right ]  \ +  \nonumber \\[2mm]
\ &+  \sum_{i,j=1}^{3}\sum_{k=1}^{2} \frac{1}{\Lambda^{2}}\:
(D_{k}^{\dagger a} \:  \bar{\sigma}^{\mu} \: (\tau^{A})_{a}^{\ b} D_{k\, b}) \:
\left [c^{L}_{kij} (L^{\dagger \; c}_{i} \bar{\sigma}_{\mu} (\tau^{A})_{c}^{\ d} L_{j \; d} ) + c^{Q}_{kij} (Q^{\dagger \; c}_{i} \bar{\sigma}_{\mu} (\tau^{A})_{c}^{\ d} 
Q_{j \; d} ) \right ]                                      \nonumber \\[2mm]
\ &+  \frac{c^{\prime}_{12}}{\Lambda^{2}}\:
(D_{1}^{\dagger a} \:  \bar{\sigma}^{\mu} \: (\tau^{A})_{a}^{\ b} D_{1\, b}) \:
(D_{2}^{\dagger c} \:  \bar{\sigma}^{\mu} \: (\tau^{A})_{c}^{\ d} D_{2\, d})      +          \mathrm{H.c.}\, ,   
 \label{ffff} 
\end{align}
%%%%%%%%%%%%%%%
not counting operators that violate $Z_{2}$. Note that $\ell \equiv L, \bar{e}$ and 
$q \equiv Q, \bar{u}, \bar{d}$  and $i,j$ indices stand for lepton or quark flavour.
Furthermore, there is only one scalar $d=6$ four-fermion operator, the one containing
 DM-self interactions proportional to $c_{12}$. 
In addition, there  are lepton number violating \emph{scalar} operators like:
 \begin{align}
 (D_{1a} D_{1b})(L^{\dagger \, a} L^{\dagger\, b}) \ +\
 \epsilon^{ac}   \epsilon^{bd}  \:(D_{2a} D_{2b})  (L_c L_d) \ + \ 
 \epsilon^{ab}  \epsilon^{cd} \:(D_{2a} L_b)   (D_{2c} L_d) \;.
 \end{align}
Other four-fermion \emph{scalar} operators between quarks/leptons and
 DM fields appear first at $d=7$  level and have the form 
 $D_{1} D_{2} (Q H \bar{u} + Q H^{\dagger} \bar{d} + L H^{\dagger} \bar{e} )$.
 All other operators in \eq{ffff} are vector-like, and, many of them  lead to spin-dependent 
 interactions in DM-nuclei collisions.
\end{itemize}

%%%%%%%%%%% BIBLIOGRAPHY %%%%%%%%%%%%%%%%%
\bibliography{DarkMatter-Biblio.bib}{}
\bibliographystyle{JHEP}

\end{document}